\documentclass{JHEP}
\usepackage{epsfig}
\usepackage{amssymb}

\newcommand{\be}{\begin{equation}}
\newcommand{\ee}{\end{equation}}
\newcommand{\bea}{\begin{eqnarray}}
\newcommand{\eea}{\end{eqnarray}}
\newcommand{\bean}{\begin{eqnarray*}}
\newcommand{\eean}{\end{eqnarray*}}

\def\IZ{\mathbb{Z}}

\def\IP{\mathbb{P}}

\def\beq{\begin{equation}}
\def\eeq{\end{equation}}

\newcommand{\fref}[1]{Figure~\ref{#1}}
\def\eref#1{(\ref{#1})}

\preprint{MIT-CTP-3487}

\title{Hidden Exceptional Global Symmetries in 4d CFTs}

\author{Sebasti\'{a}n Franco$^1$, Amihay Hanany$^{1,2}$ and Pavlos Kazakopoulos$^1$
\footnote{
Research supported in part by the CTP and the LNS
of MIT. Further support is
granted from the U.S. Department of Energy under cooperative
agreements $\#$DE-FC02-94ER40818 and $\#$DE-FG02-95ER40893.
A.~H.~is also supported by the Reed Fund Award
and a DOE OJI award.}

\\
1. \parbox[t]{6in}{Center for Theoretical Physics,
\\Massachusetts Institute of Technology,\\ Cambridge, MA 02139, USA}\\[.2cm]
 
2. \parbox[t]{6in}{Institute for Advanced Study, \\
Princeton NJ 08540 USA }\\[.4cm]
\email{sfranco,hanany,noablake@mit.edu}
}

\abstract{We study four dimensional ${\cal N}=1$ gauge theories that arise on the worldvolume of D3-branes probing
complex cones over del Pezzo surfaces. Global symmetries of the gauge theories are made explicit by using a correspondence
between bifundamental fields in the quivers and divisors in the underlying geometry.
These global symmetries are hidden, being unbroken when all inverse gauge couplings of the quiver theory vanish. In the broken phase, for finite gauge couplings, only the Cartan subalgebra is manifest as a global symmetry. Superpotentials for these models are 
constructed using global symmetry invariants as their building blocks. Higgsings connecting theories for different del 
Pezzos are immediately identified by performing
the appropriate higgsing of the global symmetry groups. The symmetric properties of the quivers
are also exploited to count the first few dibaryon operators in the gauge theories, matching their enumeration in the AdS duals.}

\begin{document}

\section{Introduction}

\label{section_introduction}

One of the possible ways in which gauge theories can be engineered in String Theory is by considering
the low energy limit of D-branes probes on singularities. When D3-branes are used to probe
local Calabi-Yau threefolds, the resulting theory on their worldvolume is a ${\cal N}=1$, 
$d=4$ quiver gauge theory.

A large class of interesting geometries is given by toric singularities. The gauge group, 
matter content and interaction superpotential of the gauge theory are dictated by the underlying 
geometry. A systematic procedure for obtaining the gauge theory consists in realizing the singularity 
as a partial resolution of an Abelian orbifold \cite{Beasley:1999uz,Feng:2000mi,Feng:2001xr}.
A remarkable fact is that the gauge theory living on the D3-brane probes is not unique. This phenomenon
is a manifestation of a full IR equivalence between different field theories and is called Toric 
Duality \cite{Feng:2000mi,Feng:2001xr}. It has been shown \cite{Feng:2001bn,Beasley:2001zp} that toric dual 
theories are non-trivial realizations of the well known Seiberg duality \cite{Seiberg:1994pq}.

A particular class of interesting singularities is the one of complex cones over del Pezzo surfaces.
Recently, a map between bifundamental fields in certain del Pezzo quivers and 2-cycles in the 
geometry has been established \cite{Intriligator:2003wr}. Furthermore, a set of $n$ conserved global $U(1)$ currents, 
together with corresponding charges of the bifundamental fields under these currents were identified. The existence of 
these currents is encouraging and actually fits the expectation for an enhanced global $E_n$ symmetry by identifying them 
with the Cartan elements of the $E_n$ symmetry.
The main objective of this paper is to exploit the map between geometric properties and matter fields in order 
to make the symmetries of the underlying geometry manifest at the level of the quiver gauge theory on the D-branes. 
This is not an obvious task since, as we will see in coming sections, a given irreducible representation of these
global $E_n$ symmetry groups may be formed by fields charged under different gauge groups.
What this implies is that quiver gauge quantum numbers actually break the global $E_n$ symmetry. The relevant 
deformations associated to each of these quiver gauge groups is their corresponding gauge couplings. As a result, 
we would expect that when all inverse gauge couplings vanish and the gauge group quantum numbers are absent, the 
$E_n$ symmetry will be restored. This leads to the natural identification that the gauge couplings are in fact Cartan 
elements of an adjoint valued complex scalar field of $E_n$. The gauge couplings are K\"ahler moduli and we expect 
an $E_n$ enhanced symmetry points at the origin of the K\"ahler moduli space.

The structure that is found here, i.e the grouping of matter fields into $E_n$ representations 
and the resulting successful reformulation of several properties of the gauge theories in the language of group
theory of these exceptional Lie algebras seems to point towards the existence of a fixed point with
enhanced exceptional global symmetry for each of the del Pezzo theories. The enhanced $E_n$ symmetry
is hidden in the sense that it does not appear in the perturbative Lagrangian definition of the theories, and one 
can only argue for the existence of a superconformal fixed point where the symmetry is realized. For $n=6,7,8$
this is of course to be expected, since there are no known Lagrangians that manifest this type of global
symmetry. Examples of hidden global symmetry 
enhancement have been discovered in three and five dimensional gauge theories. In three dimensions,
a $N=4$ $U(1)$ gauge theory with $2$ charged hypermultiplets flows to a fixed point with 
enhanced $SU(2)$ global symmetry and infinite gauge coupling \cite{Intriligator:1996ex}. More generally, 
the theories on D2 branes probing $ADE$ singularities of an ALE space are argued to possess a fixed point 
(at infinite gauge coupling) 
with hidden global symmetry of the corresponding $ADE$ type \cite{PZ}. In the $T$-dual picture one has three-branes
suspended between NS fivebranes and the global symmetry can be seen as the gauge symmetry living on the 
NS fivebranes \cite{Hanany:1996ie}. In five dimensions, the ${\cal N}=1$ gauge
theory on a D4 brane probing a certain type I$'$ background is shown to have a fixed point with enhanced
$E_n$ global symmetry, depending on the number of D8 branes in the background \cite{Seiberg:1996bd}. 
Here also the fixed point resides at infinite gauge coupling. More examples can be found in 
\cite{Intriligator:1997pq} (see also \cite{Feng:2000eq}). The study of hidden symmetry enhancement in 
five dimensions can also be approached through $(p,q)$ web techniques as in 
\cite{Aharony:1997ju,Aharony:1997bh} and the symmetry is made manifest in the string theory 
construction with the introduction of 7-branes \cite{DeWolfe:1999hj}. The theories we have at hand bear 
striking similarities to these examples, namely the 
appearance of the $E_n$ Lie groups as hidden global symmetries and the realization of the enhanced 
symmetry only at zero inverse gauge coupling. On the other hand, all the aforementioned examples are 
theories with eight supercharges in contrast to the four supercharges of the theories on D3-branes probing
cones over del Pezzo surfaces.
To the best of our knowledge, this is the first example of gauge theories with ${\cal N}=1$ supersymmetry   
in four dimensions that exhibit a hidden exceptional global symmetry.

There are several problems that can be addressed once quivers are classified using global symmetries. A specially
challenging task when deriving gauge theories that live on D-branes on singularities is the determination of the 
corresponding superpotentials. After organizing the matter content into representations, while ignoring their quiver gauge quantum numbers, the building blocks for 
superpotentials are given by invariant combinations of such irreducible representations. This is a key observation since, with this amount of 
supersymmetry, superpotentials are only affected by closed string complex moduli and will not be affected by changing K\"ahler moduli. As a result 
it is possible to compute the superpotentials at the enhanced point were the full $E_n$ symmetry is enhanced and restrict only to $E_n$ invariants. 
Once the symmetry is broken by turning on gauge interactions some of the terms in the superpotential become non-gauge invariant and are projected out. 
We are left then with the gauge invariant projection of the original $E_n$ symmetric terms.
In particular, the computation of superpotentials for 
non-toric del Pezzos has been elusive until recently. In \cite{Wijnholt:2002qz}, the superpotentials for some of 
the non-toric theories have been computed using exceptional collections. We will see along the 
paper how these superpotentials can also be derived by considering symmetric combinations and, in some cases, using 
simple inputs regarding the behavior of the theory under (un)higgsings. 

As we proceed with the study of del Pezzo quivers, we will encounter a further complication. The matter content 
of some quivers does not even seem to fit into irreducible representations of the corresponding $E_n$ group. We will see that it is 
still possible to treat all the examples within a unified framework, by introducing some new ideas such as the concept of 
{\it partial representations}. This idea simply states that the missing fields can be postulated to exist as massive fields,
thus completing the representation to its actual size. Our tools allow us to identify all the quantum numbers of such 
missing states. A crucial ingredient in such a construction is the possibility to add a global symmetry invariant mass 
term for these fields such that at energies smaller than this mass such fields will be integrated out and we will be left with what appears to be a ``partial representation." 

The outline of this paper is as follows. In Section \ref{section_En_and_geometry}, we review how $E_n$ symmetries arise in
del Pezzo surfaces, and we establish the general framework that will be used along the paper to make these symmetries 
explicit in the corresponding quiver theories. In Section 3, we follow this methodology and, starting 
from $\IP^2$, construct the divisors associated to bifundamental fields for all del Pezzo quivers up to $dP_6$ by performing 
successive blow-ups. We classify the bifundamental matter into irreducible representations of the global symmetry
group and show how invariance under the global symmetry group determines superpotentials. Section 
\ref{section_partial_representations} describes the concept of partial representations and shows in explicit examples how 
to determine which are the fields that are missing from them. In Section \ref{section_higgsing}
we identify the blow-downs that take from $dP_n$ to $dP_{n-1}$ with higgsing of the global symmetry group $E_n$ down to $E_{n-1}$ 
by a non-zero VEV in the fundamental representation. This offers a systematic approach to the connection among theories for 
different del Pezzos. Section \ref{section_symmetries_seiberg} presents a simple set of rules for transforming the $E_n$ 
representation content of a quiver under Seiberg duality. Sections \ref{section_dibaryon_operators} and \ref{section_dibaryon_counting}
explain how to use group theory to count and classify dibaryon operators in the gauge theory, matching the geometric enumeration of 
such states in the AdS dual. Finally, Section \ref{section_dP7_and_dP8} applies the decomposition into maximal subgroups 
of $E_n$ to the counting of dibaryons in $dP_7$ and $dP_8$.

\section{$E_n$ symmetries and del Pezzo surfaces}

\label{section_En_and_geometry}

Let us have a look at how exceptional symmetries appear in del Pezzo theories\footnote{There is a vast literature on the 
geometry of del Pezzo surfaces and their symmetries. Some recent papers that include nice discussions of the subject are 
\cite{Iqbal:2001ye,Intriligator:2003wr,Herzog:2003dj}.}. Del Pezzo surfaces $dP_n$ are manifolds of complex 
dimension $2$ constructed by blowing up $\mathbb{P}^2$ at $n$ generic points, $n=0,\ldots,8$. The lattice $H_2(dP_n,\mathbb{Z})$ 
is generated by the set $\{D,E_1,E_2,\ldots,E_n\}$. Here $D$ is the pullback of the generator of $H_2(\mathbb{P}^2,\mathbb{Z})$ under the 
projection that collapses the blown-up exceptional curves $E_1,E_2,\ldots,E_n$.
The intersection numbers for this basis are 

\beq
\begin{array}{ccccccc}
D\cdot D=1 & \ \ \ \ & D\cdot E_i=0 & \ \ \ \ & E_i\cdot E_j=-\delta_{ij} & \ \ \ \ & i,j=1,\ldots,n
\end{array}
\eeq
One can use a vector notation for the elements of $H_2(dP_n,\mathbb{Z})$ which will be useful later for counting
dibaryons. In this notation the basis elements read

\beq
\begin{array}{ccl}
D   & : & (1,0,0,...,0) \\
E_1 & : & (0,1,0,...,0) \\
E_2 & : & (0,0,1,...,0), \mbox{ etc}
\end{array}
\label{vector_notation}
\eeq
and the intersection numbers are computed by taking the scalar product between vectors using the Lorentzian metric $\mbox{diag } (1,-1,\ldots,-1)$.

The first Chern class for $dP_n$ is $c_1=3D-\sum_{i=1}^{n}E_i$. The canonical class is  $K_n=-c_1$. The orthogonal complement of $K_n$ according
to the above intersection product is a natural sublattice of $H_2(dP_n,\mathbb{Z})$, called the 
{\it normal sublattice}. There is an isomorphism between the normal sublattice and the root lattice of the $E_n$
Lie algebra for  $n \geq 3$. If we take as basis for this 
sublattice the set of vectors

\beq
\begin{array}{ll}
\alpha_i=E_i-E_{i+1} & i=1,...,n-1 \\
\alpha_n=D-E_1-E_2-E_3 & 
\end{array}
\label{simple_roots}
\eeq
then the intersection numbers for the $\alpha_i$ are

\beq
\alpha_i\cdot \alpha_j=-A_{ij}\;\;,\;\;i,j=1,\ldots,n,
\eeq
where $A_{ij}$ is the Cartan matrix of the Lie Algebra $E_n$. Thus, the $\alpha_i$ correspond to the simple roots of 
$E_n$. It is useful to keep in mind that $E_1=U(1)$, $E_2=SU(2) \times U(1)$, $E_3=SU(2) \times SU(3)$, $E_4=SU(5)$ and $E_5=SO(10)$.
Given an element $\mathcal{C}$ of $H_2(dP_n,\mathbb{Z})$, we can assign a weight vector of $E_n$ to it, with
Dynkin coefficients given by its projection on the normal sublattice

\beq
\lambda_i=- \mathcal{C} \cdot \alpha_i\;.
\label{Dynkin}
\eeq

Let us now consider the quiver gauge theories that appear on a stack of $N$ D3-branes probing complex cones over del 
Pezzo surfaces \footnote{Recently, the non-conformal theories resulting from the inclusion of fractional branes in these geometries and
the resulting RG flows have been investigated in \cite{Hanany:2003xh,Franco:2003ja,Franco:2003ea,Franco:2004jz}}. For $dP_n$, the gauge group 
is $\prod_{i=1}^k U(d_i N)$, where the $d_i$ are appropriate integers and 
$k=n+3$, the Euler characteristic of the del Pezzo. The near horizon geometry of this configuration will be $AdS_5\times H_5$ where $H_5$ is
a $U(1)$ fibration over the the del Pezzo surface $dP_n$. 
The AdS/CFT correspondence \cite{Maldacena:1997re,Gubser:1998bc,Witten:1998qj} conjectures a mapping between operators 
in the conformal gauge theory and states of the bulk string theory. Although this mapping is not known in its generality, it has been 
sufficiently explored for the special case of BPS operators.  
One such class of operators are dibaryons \cite{Gubser:1998fp,Gukov:1998kn}. These are generalizations of the usual baryons to theories with 
bifundamental matter. In the gravity
dual dibaryons correspond to D3-branes wrapping certain 3-cycles in $H_5$. These 3-cycles are holomorphic 2-cycles of $dP_n$ together with the $U(1)$ fiber of $H_5$. Therefore, it is possible to assign a curve in $H_2(dP_n,\mathbb{Z})$ to every 
dibaryon \cite{Intriligator:2003wr, Herzog:2003dj} (more details 
of this correspondence later). In the special case where the quiver theory is in the so-called toric phase\footnote{We want to bring to the reader's attention
the particular use we are making here of the concept of a toric phase. It follows the use given to it in \cite{Feng:2002zw,Feng:2002fv} 
and it simply refers to a quiver in which all $d_i=1$, i.e. all the gauge groups are equal to $U(N)$. In particular, there can be toric quivers 
for non-toric del Pezzos, as we will see along the paper.}, i.e. when all the gauge group factors are $U(N)$, there are some dibaryons which are 
formed by the anti-symmetrization of $N$ copies of a single bifundamental field

\beq
\epsilon_{i_1i_2\cdots i_N}\epsilon^{j_1j_2\cdots j_N}X^{i_1}_{j_1}X^{i_2}_{j_2}\cdots X^{i_N}_{j_N}.
\label{simple_dibaryons}
\eeq

This can be repeated for every bifundamental field, allowing us to extend the correspondence between holomorphic 2-cycles (also called divisors)
and dibaryons and assign an element of $H_2(dP_n,\mathbb{Z})$ to each bifundamental matter field in the quiver \cite{Intriligator:2003wr}.
Thus, if $X_{\alpha\beta}$ is a bifundamental field extending from node $\alpha$ to node $\beta$ in the
quiver representation then we can associate to it an element $L_{\alpha\beta}$ of $H_2(dP_n,\mathbb{Z})$.
In fact, the $L_{\alpha\beta}$ of toric quivers can be written as differences of divisors $L_{\alpha}$ associated to the nodes. 
The precise nature of the node divisors was clarified in \cite{Herzog:2003dj}, where they were identified with the first Chern 
class of the sheaves in the dual exceptional collection associated to the quiver. This result has been generalized in 
\cite{Herzog:2003dj} to the case in which the ranks of the gauge groups are not necessarily equal, yielding the following 
expressions

\beq
\begin{array}{l}
L_{\alpha\beta}={L_{\beta}\over d_{\beta}}-{L_{\alpha}\over d_{\alpha}}\;\;\;\;\;\;\;\;\;\;\,\mbox{if}\;\;\;{L_{\beta}\over d_{\beta}}-{L_{\alpha}\over d_{\alpha}}\geq0\\\\
L_{\alpha\beta}={L_{\beta}\over d_{\beta}}-{L_{\alpha}\over d_{\alpha}}+c_1\;\;\;\mbox{if}\;\;\;{L_{\beta}\over d_{\beta}}-{L_{\alpha}\over d_{\alpha}}<0.
\label{nodes_divisors}
\end{array}
\eeq
where the sign refers to the sign of $(L_{\beta}/ d_{\beta}-L_{\alpha}/ d_{\alpha})\cdot c_1$. 
The supersymmetric gauge theories living on the stack of D3-branes probing these geometries are invariant 
under a set of global $U(1)$ symmetries. One of these is  the $U(1)_R$ symmetry which is part of the 
superconformal algebra. There are also $n$  flavor $U(1)$ symmetries under which the bifundamentals
are charged. Dibaryons are correspondingly charged under these symmetries, thus
we refer to them as baryonic $U(1)$'s.
The aforementioned correspondence allows us to readily calculate the charges of bifundamentals under the 
{\it baryonic} $U(1)$'s and $U(1)_R$ contained in
the global symmetry group of the quivers. In particular, the $R$ charge, being proportional to the volume of the 
3-cycle wrapped by the D3-brane in the dual geometry, is given by

\beq
R(X_{\alpha\beta})=-2\frac{K_n\cdot L_{\alpha\beta}}{K_n\cdot K_n}.
\label{R-charge_formula}
\eeq

The global baryonic $U(1)$ symmetries are gauge symmetries in the $AdS_5$ bulk, with the $U(1)$
gauge fields coming from the reduction of the RR gauge field $C_4$ on $n$ independent 3-cycles of $H_5$.
The flavor currents $J_i$ of these $U(1)$'s must be neutral under the R-symmetry, which translates in the dual 
geometry as $J_i\cdot K_n=0$. Therefore, the divisors $J_i$ corresponding to these are elements of the normal 
sublattice and can be chosen to be the basis vectors $\alpha_i$ defined in  \eref{simple_roots}.
The vector of $U(1)$ charges for each bifundamental $X_{\alpha\beta}$ is then 

\beq
q_i=L_{\alpha\beta}\cdot J_i.
\label{charge_vector}
\eeq   

According to \eref{Dynkin}, these are (modulo an unimportant overall minus sign) the Dynkin coefficients of the weight 
vector $L_{\alpha\beta}$. We can indeed compute the weight vectors for all the toric phases of the del Pezzos (and will in fact 
do so in Section 3). What one finds using these weight vectors is that the bifundamental matter fields can be accommodated into irreducible 
representations of the $E_n$ Lie algebra for each of these theories. The matter fields within a representation have the same $R$ 
charge, which is characteristic of the representation.



These theories also have superpotentials. Each term in the superpotential must be invariant under the
$U(1)$ flavor symmetries and have R-charge equal to two.
The superpotential for all these models can actually be written as the gauge invariant part of singlets of the Weyl
group of $E_n$ formed by products of these irreducible representations. As we will see, this description makes it 
possible to recast most of what is known  
about the del Pezzo theories, including superpotentials, Seiberg duality relations and higgsing relations, in an elegant 
group theoretic language. Although the global symmetry of these models at a generic point in the moduli space 
is just the $U(1)^n\times U(1)_R$ symmetry, in the limit where all the gauge couplings 
$g_i\rightarrow\infty$ the full $E_n$ symmetry is restored. This enhancement of the symmetry
leaves its mark on the theory even for finite $g_i$, if appropriately combined with the principle of gauge 
invariance. In fact, the $U(1)^n\times U(1)_R$ global symmetry algebra forms the Cartan sub-algebra of the affine algebra, ${\hat E}_n$, with $U(1)_R$ being the Cartan element associated with the imaginary root of the affine algebra. It is important to note that the sub-algebras which can be enhanced by tuning the inverse gauge coupling are always finite dimensional and, as usual, the affine algebra is never completely enhanced. It will be interesting to study the signature of the affine algebras on these quiver theories.

To conclude this section, let us stress that for non-toric quivers it is no longer possible to arrange individual bifundamental fields 
into representations. In fact, if we apply \eref{charge_vector} to the divisors computed using \eref{nodes_divisors}, we obtain in 
general a set of fractional charges that cannot be interpreted as Dynkin coefficients defining a representation of the Weyl group of 
the corresponding $E_n$. It is only when various bifundamental fields are combined into dibaryons that the resulting objects
have integer $U(1)$ charges and, accordingly, well defined transformation properties under the global symmetry group. 
This is not surprising, since the only operators in the CFT that are mapped to the gravity side by the AdS/CFT are 
gauge invariant. We will discuss in Section \ref{section_dP7_and_dP8} how a classification of bifundamentals into subgroups 
of $E_n$  is still possible and turns out to be useful for the enumeration of dibaryons.

\subsection{The Weyl group and dibaryons}

\label{subsection_Weyl_group}

As mentioned above, there is an interesting relation between the Weyl group of $E_n$ and dibaryons in del Pezzo gauge theories.
Let $\mathcal{C}$ be an element of $H_2(dP_n,\mathbb{Z})$ corresponding to a dibaryon state in $dP_n$.
The degree of this curve is defined as:

\beq
k=-(K_n\cdot \mathcal{C}).
\label{equation_degree}
\eeq
Now, there is a natural action of the Weyl group of $E_n$ on these curves that
preserves their degree. If $\alpha_i\in H_2(dP_n,\mathbb{Z})$, $i=1,\ldots,n$ is any of the simple roots 
in \eref{simple_roots} then the corresponding Weyl group element acts on $\mathcal{C}$ as 

\beq
w_{\alpha_i} \;\;:\;\; \mathcal{C}\rightarrow  \mathcal{C}+(\mathcal{C}\cdot \alpha_i) \ \alpha_i
\eeq
and the curve produced by this action has the same degree $k$, because $K_n\cdot \alpha_i=0$. 
Thus, the curves of a given degree form a representation of the Weyl group of $E_n$. 
So there is some  $E_n$ related structure for dibaryon states at a generic point in the moduli space, even
though they do not form complete $E_n$ representations because of the requirement of gauge invariance in 
their construction.  
On the other hand, every representation of $E_n$ is the union of irreducible representations of the Weyl group 
(Weyl orbits). For basic representations, i.e. representations whose highest weight vector has only one nonzero 
element, equal to one, it can be shown that they consist of a single nontrivial Weyl orbit (plus $n$ Weyl singlets 
in the case of the adjoint). This means that for low levels $k$ the dimensions of Weyl orbits and irreducible 
$E_n$  representations coincide (modulo a difference of $n$ for the adjoint) and we shall see in later sections
how this can be used for performing an algebraic counting of dibaryon states.







\section{Global symmetry classification of quivers}

\label{section_symmetries}

The discussion in Section \ref{section_En_and_geometry} provides us with a systematic procedure to classify del 
Pezzo quivers according to the transformation properties of bifundamental fields under the corresponding $E_n$ 
groups, which can be summarized as follows. The divisors associated to bifundamental fields are computed from the 
divisors assigned to the quiver nodes using \eref{nodes_divisors}. The baryonic $U(1)$ and $R$ charges are 
calculated from the  intersection numbers of these divisors with the normal sublattice and the canonical class 
according to \eref{charge_vector} and \eref{R-charge_formula}. 
The vector of $U(1)$ charges for each of the matter fields in $dP_n$ is actually a weight vector of the $E_n$ 
Lie algebra and, as we will see by computing them, these weight vectors form irreducible representations of $E_n$. 

In this section we will summarize the transformation properties under global symmetries of bifundamental fields 
in different phases of gauge theories on D3-branes probing complex cones over del Pezzo surfaces.
This information will be used in the subsequent sections of the paper. 
We will closely examine the toric phases of the del Pezzo theories, $dP_n$ for $2 \leq n \leq 6$
\footnote{We skip $dP_1$ because its global symmetry $E_1=U(1)$ makes it rather trivial 
from the point of view of grouping the fields in irreducible representations (all irreducible representations
are one-dimensional). For $n \leq 3$, these models have been extensively studied and the information we will
use here can be found, for example, in \cite{Feng:2002zw}. Several aspects of the gauge theories for 
$4 \leq n \leq 6$ have been studied in \cite{Hanany:2001py,Feng:2002fv,Wijnholt:2002qz,Intriligator:2003wr,Herzog:2003wt}.}. 

The corresponding superpotentials of these theories are singlets under the global symmetry transformations. 
Thus, they can be written as a sum of products of irreducible 
representations, with the understanding that from each of these products we consider the $E_n$ singlet 
included in it and that only the gauge invariant terms in this singlet actually contribute to the 
superpotential. We will only write down those superpotential terms from which some contributions survive 
the projection onto gauge invariants, although sometimes more terms that are invariant under global transformations 
can be written down.

The only subtlety we encounter in using this description is the appearance of {\it partial representations}. 
This is just a name for groups of fields that do not completely fill representations of $E_n$. 
A detailed discussion of partial representations is deferred to Section \ref{section_partial_representations}. For 
now it will suffice to say that the missing components in these representations are actually massive matter 
fields that do not appear in the low energy limit of the theory.   

In writing down these theories, we have decided to number the gauge groups following the order
of the corresponding external legs of the associated $(p,q)$ webs (we refer the reader to \cite{Hanany:2001py,Franco:2002ae} 
for a description of the connection between $(p,q)$ webs and 4d gauge theories on D3-branes probing toric
singularities \footnote{See also \cite{Feng:2004uq}, for a recent exploration of the correspondence.}). This ordering is
closely related to the one of the associated dual exceptional collection. In fact, both ordering prescriptions 
are almost identical, differing at most by a possible reordering of nodes within each block (set of parallel 
external legs in the $(p,q)$ web description).

There are different ways to go about calculating the divisors associated to nodes and bifundamental
fields in these quiver theories. According to \cite{Herzog:2003dj}, the divisors corresponding to the nodes are 
the elements of a dual exceptional collection, obtained through a certain braiding operation from the 
exceptional collection used to construct the quiver theory. Some recent progress has also been made toward
reading off this dual exceptional collection from the quiver diagram of the theory \cite{Herzog:2003zc}. 

We will follow here another procedure, also used in \cite{Intriligator:2003wr}, 
which makes use of the fact that these phases are related to one another by Seiberg dualities and/or 
higgsing. We can compute the divisor configurations starting from any of these models and using Seiberg
duality and blowing down or blowing up cycles in the geometry, which means higgsing or unhiggsing the 
quiver theory respectively. We will primarily focus on toric phases. By the use of Seiberg dualities on 
selfdual nodes (i.e. nodes whose rank does not change upon dualisation) we can `move' among different 
toric phases of the theory, while the operation of blowing cycles up or down takes us from the $dP_n$ 
theory to $dP_{n+1}$ or $dP_{n-1}$ respectively. The ways these operations act on the divisors are 
described in \cite{Intriligator:2003wr}. Here is a quick review of the rules for quivers in which all 
the gauge groups are equal to $U(N)$:

\begin{itemize}

\item
{\bf \underline{Seiberg duality}:} when a self-dual node $\alpha$ (i.e., for the class of quivers under 
consideration, a node with $2N$ flavors) is dualised, the divisor $L_{\alpha}$ changes to 
$L_{\alpha}'=L_{\beta}+L_{\gamma}-L_{\alpha}$ where $\beta$ and $\gamma$ are the nodes where arrows 
starting from $\alpha$ end or, equivalently, where arrows that end at $\alpha$ begin. In the case of a 
double arrow, $\beta$ and $\gamma$ can be the same.

\item
{\bf \underline{Blow-down}:} to blow down from $dP_n$ to $dP_{n-1}$ we eliminate $E_n$ from the divisors and 
identify the nodes that have the same divisor after the elimination.

\item
{\bf \underline{Blow-up}:} to go from $dP_{n-1}$ to $dP_n$ we add a new node and attribute to it a divisor 
such that the field that is unhiggsed in the quiver corresponds to the divisor $E_n$ that we blow up. Moreover, 
all other divisors can differ from their blown-down counterparts only by $E_n$.           

\end{itemize}

We now present the results of this classification for the toric phases of the del Pezzo theories, 
$dP_n$ for $2 \leq n \leq 6$.


\subsection{del Pezzo 2}

The first quiver theory we examine is $dP_2$.
The exceptional Lie algebra $E_2$ is $SU(2)\times U(1)$ and of the two $U(1)$ charges used below, 
$J_1=E_1-E_2$ corresponds to the Cartan generator of $SU(2)$ and  $J_2=2D-3E_1-3E_2$ is
the $U(1)$ factor. This is a somewhat irregular case because $E_2$ is not semisimple. However with the choice 
of currents given above the fields are still organized in representations of $SU(2)$  with each representation
carrying a charge under the $U(1)$ factor of $E_2$.
Note that the $J_2$ current that is used here is different from the one used in \cite{Intriligator:2003wr}. There are two 
toric phases for $dP_2$ \cite{Feng:2002zw}, we proceed now to study them.

\subsubsection*{Model I}

This phase has 13 fields. The divisors for the nodes and the fields of this model are listed below together
with the global $U(1)$ and $R$ charges.

{\footnotesize

\beq
\begin{array}{ccc}

\begin{array}{cc}
Node & L_{\alpha} \\
& \\
1 & E_1 \\
2 & E_2 \\
3 & D \\
4 & 0 \\
5 & 2D 
\end{array}

& \ \ \ \ \ \ \ \ \ \ &

\begin{array}{lcrrc}
& L_{\alpha\beta} & J_1 & J_2 & R \\
&&&&\\
X_{41} & E_1 & -1 & 3 & 2/7 \\
X_{42} & E_2 & 1 & 3 & 2/7 \\
X_{54} & D-E_1-E_2 & 0 & -4 & 2/7 \\
X_{52} & D-E_1 & 1 & -1 & 4/7 \\
X_{51} & D-E_2 & -1 & -1 & 4/7 \\
X_{13} & D-E_1 & 1 & -1 & 4/7 \\
X_{23} & D-E_2 & -1 & -1 & 4/7 \\
X_{35} & D & 0 & 2 & 6/7 \\ 
X_{34} & 2D-E_1-E_2 & 0 & -2 & 8/7 
\end{array} 

\end{array}
\eeq
}

\begin{figure}[htb]
\begin{minipage}[b]{0.3\textwidth}
\centering
\includegraphics[width=0.7\textwidth]{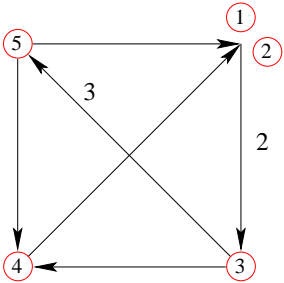}
\caption{Quiver diagram for Model I of $dP_2$.}
\label{quiver_dP2_1}
\end{minipage}%
\begin{minipage}[b]{0.7\textwidth}
\centering
{\footnotesize
\beq
\begin{array}{c|c}
 \ \ \ \ \ \mbox{Fields} \ \ \ \ \ & \ \ \ \ \ SU(2) \times U(1) \ \ \ \ \ \\
\hline
\hline
(X_{41},X_{42}) & 2_3 \\
\hline
(X_{52},X_{51}) & 2_{-1}^{(a)} \\
\hline
(X_{13},X_{23}) & 2_{-1}^{(b)} \\
\hline
(Y_{13},Y_{23}) & 2_{-1}^{(c)} \\
\hline
X_{35} & 1_2^{(a)} \\
\hline
Y_{35} & 1_2^{(b)} \\
\hline
Z_{35} & 1_2^{(c)} \\
\hline
X_{54} & 1_{-4} \\
\hline
X_{34} & 1_{-2} 
\end{array}
\label{symmetries_dP2_1}
\eeq
}
\end{minipage}
\end{figure}

The quiver diagram for this model is shown in \fref{quiver_dP2_1}. Throughout the paper, we will present quivers in compact
block form. Every circle represents a $U(N)$ gauge group. A block of circles at a given node of the quiver represents 
a set of gauge groups such that there is no bifundamental matter charged under any pair of them, and that have identical
intersection numbers with the other gauge groups in the theory. The fields can be assigned to irreducible representations 
of $SU(2)$ as in \eref{symmetries_dP2_1}. The $U(1)$ charge of each of these representations is indicated with a subscript.
When the same representation appears more than once, as $2_{-1}$ and $1_2$ do in this case, we distinguish them 
using a lowercase superscript (a, b, c, etc) in parentheses, so as not to confuse it with
the $U(1)$ charge. This notation will be simplified for $dP_n$ with $n \geq 3$, where no $U(1)$ subscripts will be present. 
Note that all fields within a representation have the same $R$ charge.


Now we can write the superpotential of this theory as singlets of products of these representations. As discussed
in Section \ref{section_symmetries}, only the gauge invariant terms of those singlets are actually contained in the superpotential.

\beq
\begin{array}{ccl}
W_{I}&=& 2_3 \otimes 2_{-1}^{(a)} \otimes 1_{-2}+ 2_3 \otimes 2_{-1}^{(b)} \otimes 1_{-2}+2_3 \otimes 2_{-1}^{(c)} \otimes 1_2^{(a)} \otimes 1_{-4}+2_3 \otimes 2_{-1}^{(c)} \otimes 1_2^{(b)} \otimes 1_{-4}+ \\
     & & +2_{-1}^{(a)} \otimes 2_{-1}^{(b)} \otimes 1_2^{(a)}+2_{-1}^{(a)} \otimes 2_{-1}^{(b)} \otimes 1_2^{(b)}+2_{-1}^{(a)} \otimes 2_{-1}^{(c)} \otimes 1_2^{(c)}
\end{array}
\eeq




\subsubsection*{Model II}

This phase has 11 fields. It is obtained by dualising node 2 of Model I. The divisors, global charges 
and representation assignments are listed below.

{\footnotesize

\beq
\begin{array}{ccc}
\begin{array}{cc}
Node & L_{\alpha} \\
& \\
1 & D \\
2 & 2D-E_2 \\
3 & 2D \\
4 & 0 \\
5 & E_1 
\end{array}

& \ \ \ \ \ \ \ \ \ \ & 

\begin{array}{lcrrc}
& L_{\alpha\beta} & J_1 & J_2 & R \\
&&&&\\
X_{45} & E_1 & -1 & 3 & 2/7 \\
X_{23} & E_2 & 1 & 3 & 2/7 \\
X_{34} & D-E_1-E_2 & 0 & -4 & 2/7 \\
X_{24} & D-E_1 & 1 & -1 & 4/7 \\
X_{35} & D-E_2 & -1 & -1 & 4/7 \\
X_{51} & D-E_1 & 1 & -1 & 4/7 \\
X_{12} & D-E_2 & -1 & -1 & 4/7 \\
X_{41} & D & 0 & 2 & 6/7 \\ 
X_{13} & D & 0 & 2 & 6/7  
\end{array} 

\end{array}
\label{divisors_dP2_2}
\eeq

}

\begin{figure}[htb]
\begin{minipage}[b]{0.3\textwidth}
\centering
\includegraphics[width=0.7\textwidth]{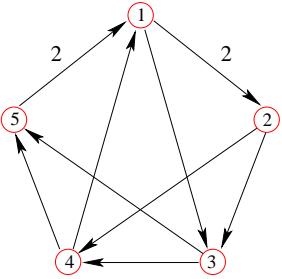}
\caption{Quiver diagram for Model II of $dP_2$.}
\label{quiver_dP2_2}
\end{minipage}%
\begin{minipage}[b]{0.7\textwidth}
\centering
{\footnotesize
\beq
\begin{array}{c|c}
 \ \ \ \ \ \mbox{Fields} \ \ \ \ \ & \ \ \ \ \ SU(2) \times U(1) \ \ \ \ \ \\
\hline
\hline
(X_{23},X_{45}) & 2_3 \\
\hline
(X_{35},X_{24}) & 2_{-1}^{(a)} \\
\hline
(X_{12},X_{51}) & 2_{-1}^{(b)} \\
\hline
(Y_{12},Y_{51}) & 2_{-1}^{(c)} \\
\hline
X_{13} & 1_2^{(a)} \\
\hline
X_{41} & 1_2^{(b)} \\
\hline
X_{34} & 1_{-4} 
\end{array}
\label{symmetries_dP2_2}
\eeq
}
\bigskip
\end{minipage}
\end{figure}

\fref{quiver_dP2_2} shows the quiver for this model. There are two double arrows in this theory, given by the fields
$X_{12}$, $Y_{12}$ $X_{51}$ and $Y_{51}$. Identical divisors are assigned to fields connecting the same pair of nodes. 
In \eref{divisors_dP2_2}, we list only one field for each double arrow. This convention will be followed for all other 
models with multiple arrows in the paper. As above, $SU(2)$ representations and $U(1)$ charges are presented in Table 
\eref{symmetries_dP2_2}. 


The superpotential for this theory becomes

\beq
\begin{array}{ccl}
W_{II}&=& 1_2^{(a)} \otimes 1_{-4} \otimes 1_2^{(b)} + 1_2^{(b)} \otimes 2_{-1}^{(c)} \otimes 2_{-1}^{(a)} + 1_2^{(a)} \otimes 2_{-1}^{(a)} \otimes 2_{-1}^{(c)} + \\
      & & + 2_{-1}^{(c)} \otimes 2_{-1}^{(b)} \otimes 2_{-1}^{(a)} \otimes 2_3 + 2_{-1}^{(b)} \otimes 2_{-1}^{(b)} \otimes 2_3 \otimes 1_{-4} \otimes 2_3 
\end{array}
\eeq

The terms $1_2^{(a)} \otimes 1_{-4} \otimes 2_3 \otimes 2_{-1}^{(b)}$ and $1_2^{(a)} \otimes 1_{-4} \otimes 2_3 \otimes 2_{-1}^{(c)}$ are globally symmetric and 
gauge invariant, but nevertheless are not present in the superpotential. We can check that this is so by looking at how it is generated by 
higgsing $dP_3$.

\subsection{del Pezzo 3}

There are four toric phases for $dP_3$, related to one another by Seiberg dualities \cite{Feng:2002zw}. 
For each of them we list the divisors corresponding to the nodes and fields, the assignment of fields 
to representations and the superpotential written as an $E_3$ singlet. We obtain $dP_3$ by blowing up 
a 2-cycle in $dP_2$.

\subsubsection*{Model I}

This model has 12 fields.


{\footnotesize

\beq
\begin{array}{ccc}

\begin{array}{cc}
Node & L_{\alpha} \\
& \\
1 & D+E_3 \\
2 & 2D-E_2 \\
3 & 2D \\
4 & E_3 \\
5 & E_1+E_3 \\
6 & D 
\end{array}

& \ \ \ \ \ \ \ \ \ \ & 

\begin{array}{lcrrrc}
& L_{\alpha\beta} & J_1 & J_2 & J_3 & R \\
&&&&&\\
X_{45} & E_1 & -1 & 0 & 1 & 1/3 \\
X_{23} & E_2 & 1 & -1 & 1 & 1/3 \\
X_{61} & E_3 & 0 & 1 & 1 & 1/3 \\
X_{34} & D-E_1-E_2 & 0 & 1 & -1 & 1/3 \\
X_{56} & D-E_1-E_3 & 1 & -1 & -1 & 1/3 \\
X_{12} & D-E_2-E_3 & -1 & 0 & -1 & 1/3 \\
X_{51} & D-E_1 & 1 & 0 & 0 & 2/3 \\
X_{35} & D-E_2 & -1 & 1 & 0 & 2/3 \\
X_{13} & D-E_3 & 0 & -1 & 0 & 2/3 \\
X_{24} & D-E_1 & 1 & 0 & 0 & 2/3 \\
X_{62} & D-E_2 & -1 & 1 & 0 & 2/3 \\
X_{46} & D-E_3 & 0 & -1 & 0 & 2/3 
\end{array}  

\end{array}
\eeq

}

\begin{figure}[htb]
\begin{minipage}[b]{0.3\textwidth}
\centering
\includegraphics[width=0.7\textwidth]{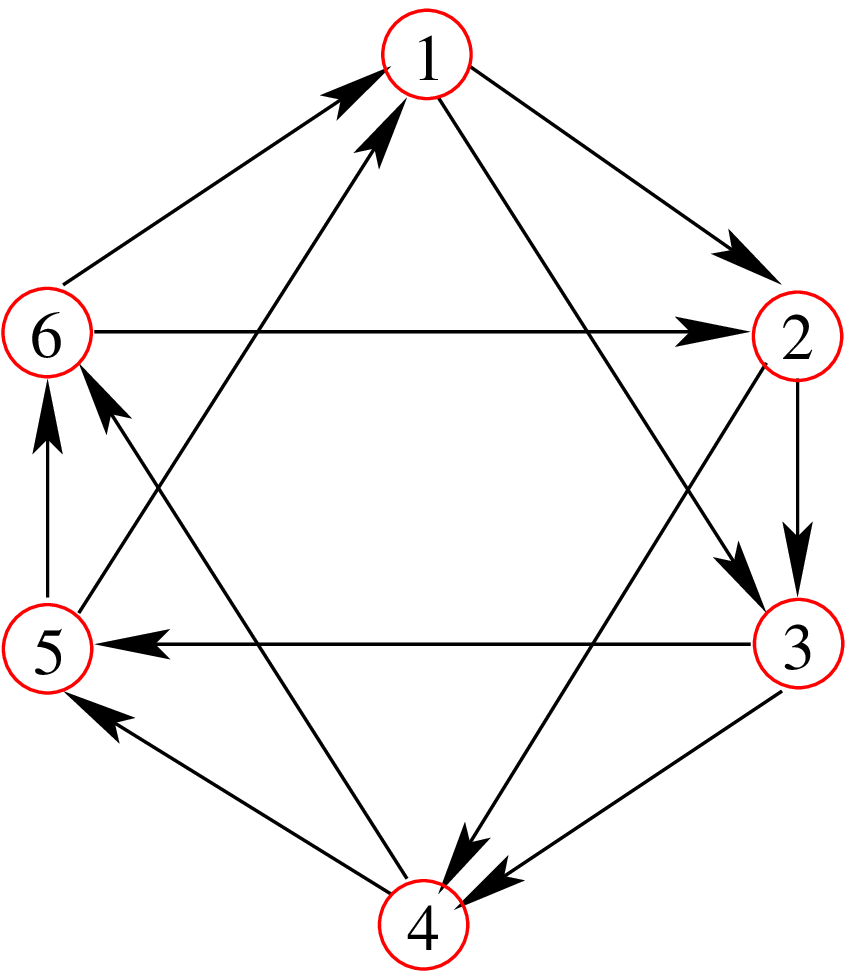}
\caption{Quiver diagram for Model I of $dP_3$.}
\label{quiver_dP3_1}
\end{minipage}%
\begin{minipage}[b]{0.7\textwidth}
\centering
{\footnotesize
\beq
\begin{array}{c|c}
 \ \ \ \ \ \mbox{Fields} \ \ \ \ \ & \ \ \ \ \ SU(2) \times SU(3) \ \ \ \ \ \\
\hline
\hline
(X_{12},X_{23},X_{34},X_{45},X_{56},X_{61}) & (2,3) \\
\hline
(X_{13},X_{35},X_{51}) & (1,\bar{3})_a \\
\hline
(X_{24},X_{46},X_{62}) & (1,\bar{3})_b
\end{array}
\label{symmetries_dP3_1}
\eeq
}
\bigskip
\bigskip
\bigskip
\bigskip
\end{minipage}
\end{figure}

The quiver diagram for this model is displayed in \fref{quiver_dP3_1}.
The fields are arranged in irreducible representations of $E_3=SU(2) \times SU(3)$ as shown in 
\eref{symmetries_dP3_1}. The assignment is done by comparing the $U(1)$ charges above with the Dynkin 
labels of the weight vectors of an $E_3$ irreducible representation. We remind the reader that because of 
the difference in their definitions there is an overall minus sign difference that must be taken into 
account in this comparison. From now on, repeated representations will be identified with lowercase subscript 
(a, b, c, etc). 


The superpotential can then be written as

{\footnotesize
\beq
W_I=(2,3)^6+(2,3)^2 \otimes (1,\bar{3})_a \otimes (1,\bar{3})_b+(1,\bar{3})^3_a+(1,\bar{3})^3_b 
\eeq
}

\subsubsection*{Model II}

There are 14 fields in this phase. We can get it by dualising node 1 of the previous model. Again, we calculate
the divisors and charges and compare them with $E_3$ weight vectors in order to assign the fields to 
representations. The results are shown in the tables below. 

{\footnotesize

\beq
\begin{array}{ccc}

\begin{array}{cc}
Node & L_{\alpha} \\
& \\
1 & D \\
2 & E_1+E_3 \\
3 & E_3 \\
4 & E_1 \\
5 & 2D \\
6 & 2D-E_2 
\end{array} 

& \ \ \ \ \ \ \ \ \ \ & 

\begin{array}{lcrrrc}
& L_{\alpha\beta} & J_1 & J_2 & J_3 & R \\
&&&&&\\
X_{32} & E_1 & -1 & 0 & 1 & 1/3 \\
X_{65} & E_2 & 1 & -1 & 1 & 1/3 \\
X_{42} & E_3 & 0 & 1 & 1 & 1/3 \\
X_{53} & D-E_1-E_2 & 0 & 1 & -1 & 1/3 \\
X_{21} & D-E_1-E_3 & 1 & -1 & -1 & 1/3 \\
X_{54} & D-E_2-E_3 & -1 & 0 & -1 & 1/3 \\
X_{41} & D-E_1 & 1 & 0 & 0 & 2/3 \\
X_{16} & D-E_2 & -1 & 1 & 0 & 2/3 \\
X_{64} & D-E_3 & 0 & -1 & 0 & 2/3 \\
X_{63} & D-E_1 & 1 & 0 & 0 & 2/3 \\
Y_{16} & D-E_2 & -1 & 1 & 0 & 2/3 \\
X_{31} & D-E_3 & 0 & -1 & 0 & 2/3 \\
X_{15} & D & 0 & 0 & 1 & 1 \\ 
X_{26} & 2D-E_1-E_2-E_3 & 0 & 0 & -1 & 1 
\end{array} 

\end{array}
\eeq

}

\begin{figure}[htb]
\begin{minipage}[b]{0.3\textwidth}
\centering
\includegraphics[width=0.7\textwidth]{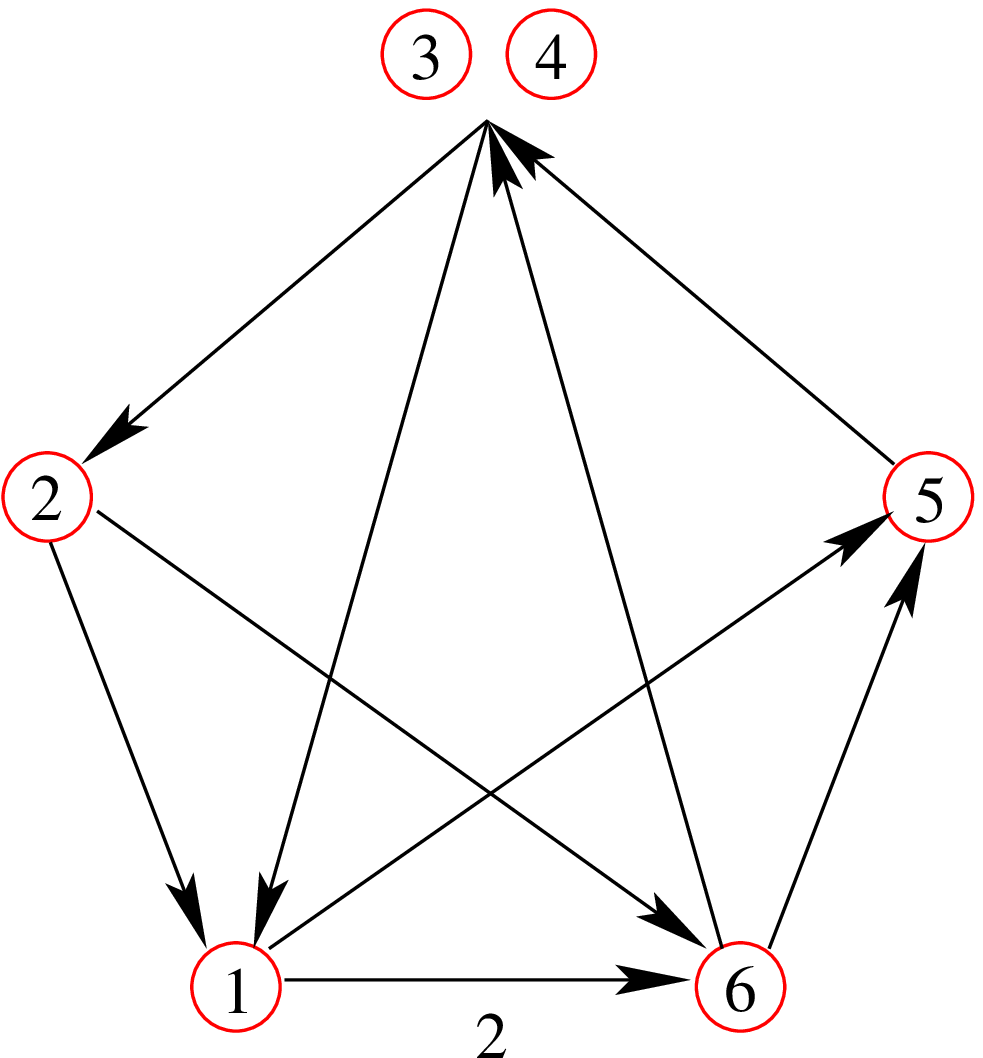}
\caption{Quiver diagram for Model II of $dP_3$.}
\label{quiver_dP3_2}
\end{minipage}%
\begin{minipage}[b]{0.7\textwidth}
\centering
{\footnotesize
\beq
\begin{array}{c|c}
 \ \ \ \ \ \mbox{Fields} \ \ \ \ \ & \ \ \ \ \ SU(2) \times SU(3) \ \ \ \ \ \\
\hline
\hline
(X_{54},X_{65},X_{53},X_{32},X_{21},X_{42}) & (2,3) \\
\hline
(X_{63},X_{31},X_{16}) & (1,\bar{3})_a \\
\hline
(X_{64},X_{41},Y_{16}) & (1,\bar{3})_b \\
\hline
(X_{26},X_{15}) & (2,1) 
\end{array}
\label{symmetries_dP3_2}
\eeq
}
\bigskip
\bigskip
\bigskip
\bigskip
\end{minipage}
\end{figure}

The quiver for this model is shown in \fref{quiver_dP3_2}. The representation structure is shown in \eref{symmetries_dP3_2}.


Using this, we can write the superpotential as
{\footnotesize
\beq
\begin{array}{ccl}
W_{II}&=&(2,3)^4 \otimes (1,\bar{3})_a+(2,3)^4 \otimes (1,\bar{3})_b \\
& & + (2,3) \otimes (2,1) \otimes (1,\bar{3})_a+(2,3) \otimes (2,1) \otimes (1,\bar{3})_b \\
& & +(1,\bar{3})^2_a \otimes (1,\bar{3})_b+(1,\bar{3})_a \otimes (1,\bar{3})^2_b
\end{array}
\eeq
}

\subsubsection*{Model III}

Dualising node 2 of Model II we obtain Model III. This phase has 14 fields.

{\footnotesize

\beq
\begin{array}{ccc}

\begin{array}{cc}
Node & L_{\alpha} \\
& \\
1 & D\\
2 & E_1 \\
3 & E_3 \\
4 & 0\\
5 & 2D \\
6 & 2D-E_2
\end{array}

& \ \ \ \ \ \ \ \ \ \ &

\begin{array}{lcrrrc}
& L_{\alpha\beta} & J_1 & J_2 & J_3 & R \\
&&&&&\\
X_{42} & E_1 & -1 & 0 & 1 & 1/3 \\
X_{65} & E_2 & 1 & -1 & 1 & 1/3 \\
X_{43} & E_3 & 0 & 1 & 1 & 1/3 \\
X_{53} & D-E_1-E_2 & 0 & 1 & -1 & 1/3 \\
X_{64} & D-E_1-E_3 & 1 & -1 & -1 & 1/3 \\
X_{52} & D-E_2-E_3 & -1 & 0 & -1 & 1/3 \\
X_{21} & D-E_1 & 1 & 0 & 0 & 2/3 \\
X_{16} & D-E_2 & -1 & 1 & 0 & 2/3 \\
X_{31} & D-E_3 & 0 & -1 & 0 & 2/3 \\
X_{15} & D & 0 & 0 & 1 & 1 \\ 
X_{14} & 2D-E_1-E_2-E_3 & 0 & 0 & -1 & 1 
\end{array} 

\end{array}
\eeq

}

\begin{figure}[htb]
\begin{minipage}[b]{0.3\textwidth}
\centering
\includegraphics[width=0.7\textwidth]{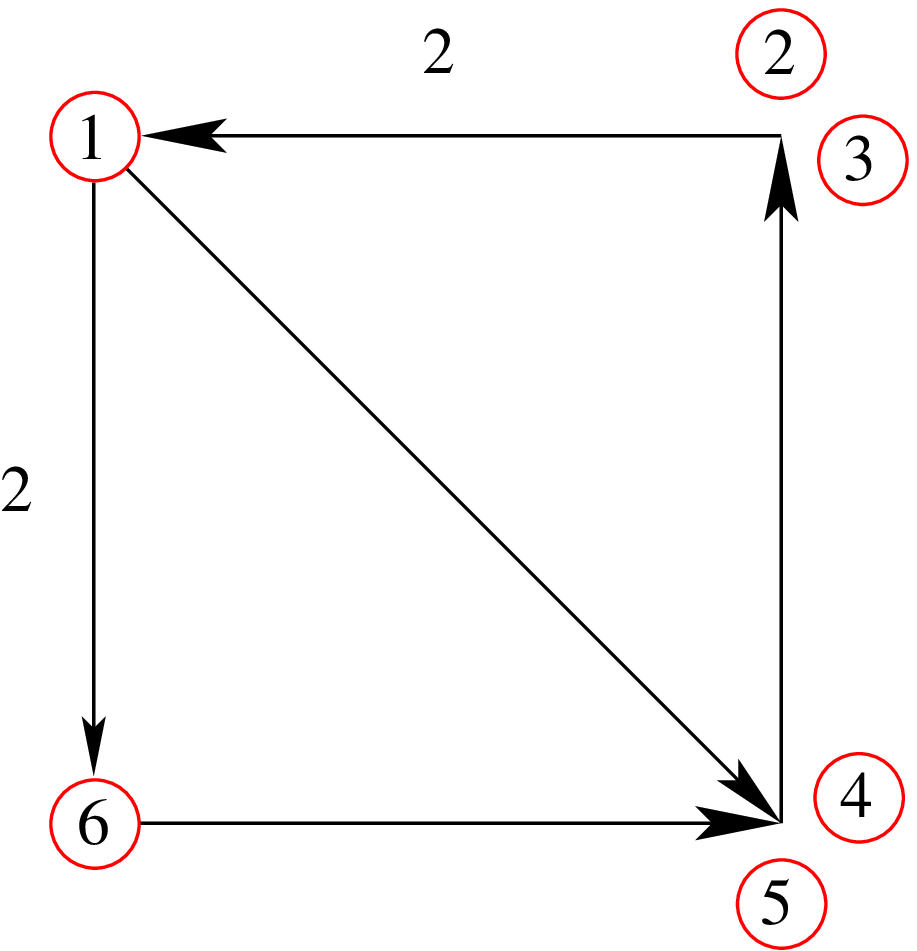}
\caption{Quiver diagram for Model III of $dP_3$.}
\label{quiver_dP3_3}
\end{minipage}%
\begin{minipage}[b]{0.7\textwidth}
\centering
{\footnotesize
\beq
\begin{array}{c|c}
 \ \ \ \ \ \mbox{Fields} \ \ \ \ \ & \ \ \ \ \ SU(2) \times SU(3) \ \ \ \ \ \\
\hline
\hline
(X_{21},X_{64},X_{23},X_{43},X_{62},X_{41}) & (2,3) \\
\hline
(X_{15},X_{56},X_{35}) & (1,\bar{3})_a \\
\hline
(Y_{15},Y_{56},Y_{35}) & (1,\bar{3})_b \\
\hline
(X_{54},X_{52}) & (2,1) 
\end{array}
\label{symmetries_dP3_3}
\eeq
}
\bigskip
\bigskip
\bigskip
\bigskip
\end{minipage}
\end{figure}

\fref{quiver_dP3_3} shows the quiver diagram and Table \eref{symmetries_dP3_3} summarizes how the fields fall into 
representations. 


The superpotential for this theory can be written in an invariant form as
\beq
\begin{array}{ccl}
W_{III}&=&(2,3)^2 \otimes (1,\bar{3})^2_a+(2,3)^2 \otimes (1,\bar{3})_a \otimes (1,\bar{3})_b+(2,3)^2 \otimes (1,\bar{3})^2_b \\
   & & +(2,3) \otimes (2,1) \otimes (1,\bar{3})_a+(2,3) \otimes (2,1) \otimes (1,\bar{3})_b
\end{array}
\eeq

\subsubsection*{Model IV}

There are 18 fields in this phase, which is produced by dualising node 6 of Model III. 

{\footnotesize

\beq
\begin{array}{ccc}

\begin{array}{cc}
Node & L_{\alpha} \\
& \\
1 & E_1 \\
2 & E_2 \\
3 & E_3 \\
4 & 2D \\
5 & 0 \\
6 & D 
\end{array}

& \ \ \ \ \ \ \ \ \ \ &

\begin{array}{lcrrrc}
& L_{\alpha\beta} & J_1 & J_2 & J_3 & R \\
&&&&&\\
X_{51} & E_1 & -1 & 0 & 1 & 1/3 \\
X_{52} & E_2 & 1 & -1 & 1 & 1/3 \\
X_{53} & E_3 & 0 & 1 & 1 & 1/3 \\
X_{43} & D-E_1-E_2 & 0 & 1 & -1 & 1/3 \\
X_{42} & D-E_1-E_3 & 1 & -1 & -1 & 1/3 \\
X_{41} & D-E_2-E_3 & -1 & 0 & -1 & 1/3 \\
X_{16} & D-E_1 & 1 & 0 & 0 & 2/3 \\
X_{26} & D-E_2 & -1 & 1 & 0 & 2/3 \\
X_{36} & D-E_3 & 0 & -1 & 0 & 2/3 \\
X_{64} & D & 0 & 0 & 1 & 1 \\ 
X_{65} & 2D-E_1-E_2-E_3 & 0 & 0 & -1 & 1 
\end{array} 

\end{array}
\eeq

}

The table below shows how the fields are organized in representations. 



\begin{figure}[htb]
\begin{minipage}[b]{0.3\textwidth}
\centering
\includegraphics[width=0.7\textwidth]{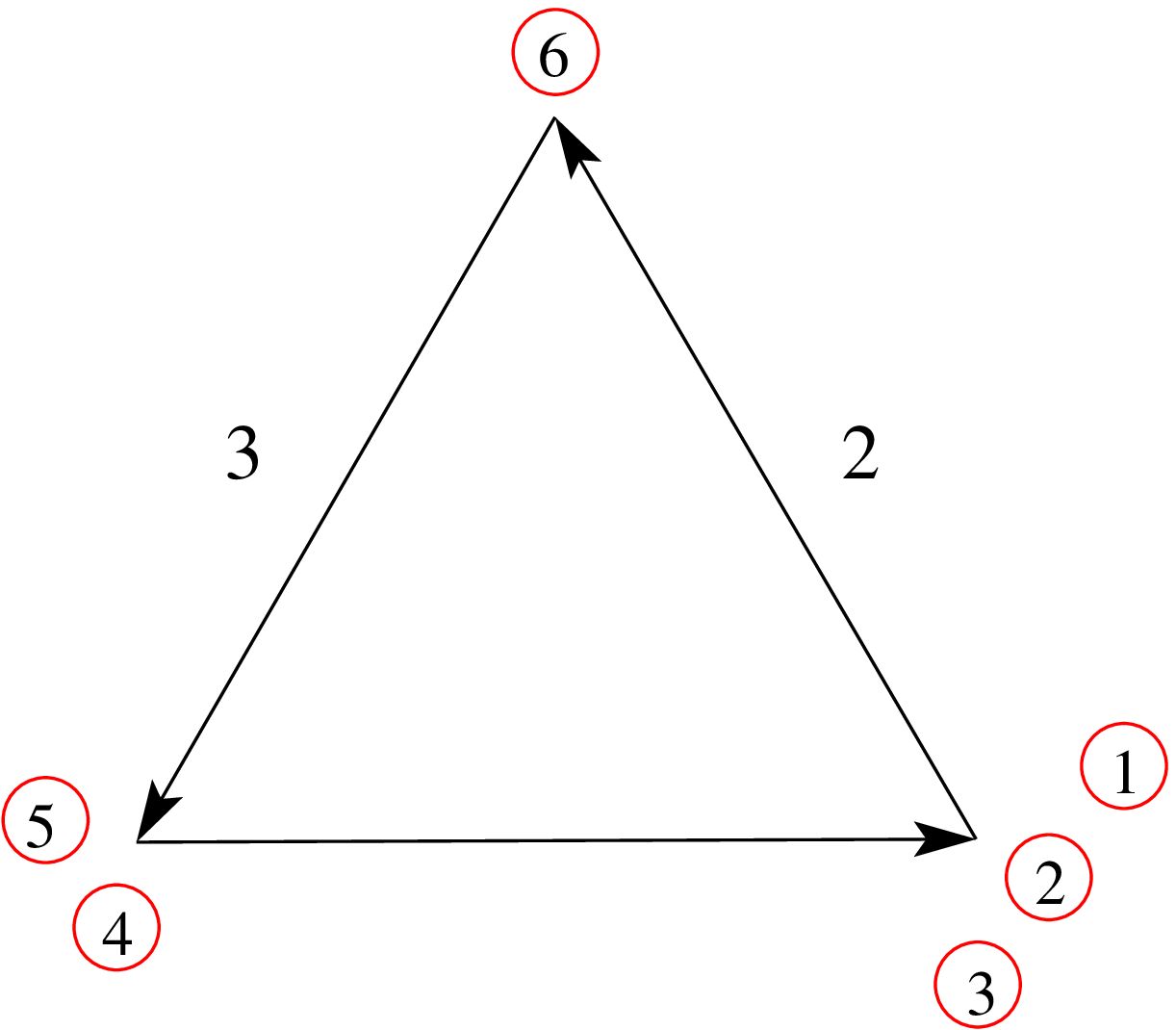}
\caption{Quiver diagram for Model IV of $dP_3$.}
\label{quiver_dP3_4}
\end{minipage}%
\begin{minipage}[b]{0.7\textwidth}
\centering
{\footnotesize
\beq
\begin{array}{c|c}
 \ \ \ \ \ \mbox{Fields} \ \ \ \ \ & \ \ \ \ \ SU(2) \times SU(3) \ \ \ \ \ \\
\hline
\hline
(X_{51},X_{52},X_{53},X_{41},X_{42},X_{43}) & (2,3) \\
\hline
(X_{16},X_{26},X_{36}) & (1,\bar{3})_a \\
\hline
(Y_{16},Y_{26},Y_{36}) & (1,\bar{3})_b \\
\hline
(X_{64},X_{65}) & (2,1)_a \\
\hline
(Y_{64},Y_{65}) & (2,1)_b \\
\hline
(Z_{64},Z_{65}) & (2,1)_c 
\end{array}
\label{symmetries_dP3_4}
\eeq
}
\bigskip
\bigskip
\bigskip
\bigskip
\end{minipage}
\end{figure}


The superpotential is
\beq
\begin{array}{ccl}
W_{IV}=(2,3) \otimes [(1,\bar{3})_a + (1,\bar{3})_b] \otimes [(2,1)_a+(2,1)_b+(2,1)_c] 
\end{array}
\eeq

\subsection{del Pezzo 4}

There are two toric phases for $dP_4$. The organization of matter fields into $E_4$ representations 
is in this case more subtle than in the 
preceding examples. The classification of these theories can be achieved with the same reasoning as 
before, by introducing the idea of partial representations. Partial representations are ordinary
representations in which some of the fields are massive, being integrated out in the
low energy limit. We will summarize the results in this section, and postpone a detailed 
explanation of partial representations to Sections \ref{section_partial_representations} 
and \ref{section_symmetries_seiberg}. 



\subsubsection*{Model I}

This theory has 15 fields.

{\footnotesize

\beq
\begin{array}{ccc}

\begin{array}{cc}
Node & L_{\alpha} \\
& \\
1 & D \\
2 & E_1 \\
3 & E_3 \\
4 & 2D-E_4 \\
5 & 0 \\
6 & D-E_4 \\
7 & 2D-E_2-E_4 
\end{array}

& \ \ \ \ \ \ \ \ \ \ &

\begin{array}{lcrrrrc}
& L_{\alpha\beta} & J_1 & J_2 & J_3 & J_4 & R \\
&&&&&& \\
X_{45} & E_1 & -1 & 0 & 0 & 1 & 2/5 \\
X_{23} & E_2 & 1 & -1 & 0 & 1 & 2/5 \\
X_{46} & E_3 & 0 & 1 & -1 & 1 & 2/5 \\
X_{71} & E_4 & 0 & 0 & 1 & 0 & 2/5 \\
X_{36} & D-E_1-E_2 & 0 & 1 & 0 & -1 & 2/5 \\
X_{24} & D-E_1-E_3 & 1 & -1 & 1 & -1 & 2/5 \\
X_{57} & D-E_1-E_4 & 1 & 0 & -1 & 0 & 2/5 \\
X_{35} & D-E_2-E_3 & -1 & 0 & 1 & -1 & 2/5 \\
X_{12} & D-E_2-E_4 & -1 & 1 & -1 & 0 & 2/5 \\
X_{67} & D-E_3-E_4 & 0 & -1 & 0 & 0 & 2/5 \\
X_{51} & D-E_1 & 1 & 0 & 0 & 0 & 4/5 \\
X_{72} & D-E_2 & -1 & 1 & 0 & 0 & 4/5 \\
X_{61} & D-E_3 & 0 & -1 & 1 & 0 & 4/5 \\
X_{13} & D-E_4 & 0 & 0 & -1 & 1 & 4/5 \\
X_{14} & 2D-E_1-E_2-E_3-E_4 & 0 & 0 & 0 & -1 & 4/5 
\end{array} 

\end{array}
\eeq

}

\begin{figure}[htb]
\begin{minipage}[b]{0.3\textwidth}
\centering
\includegraphics[width=0.7\textwidth]{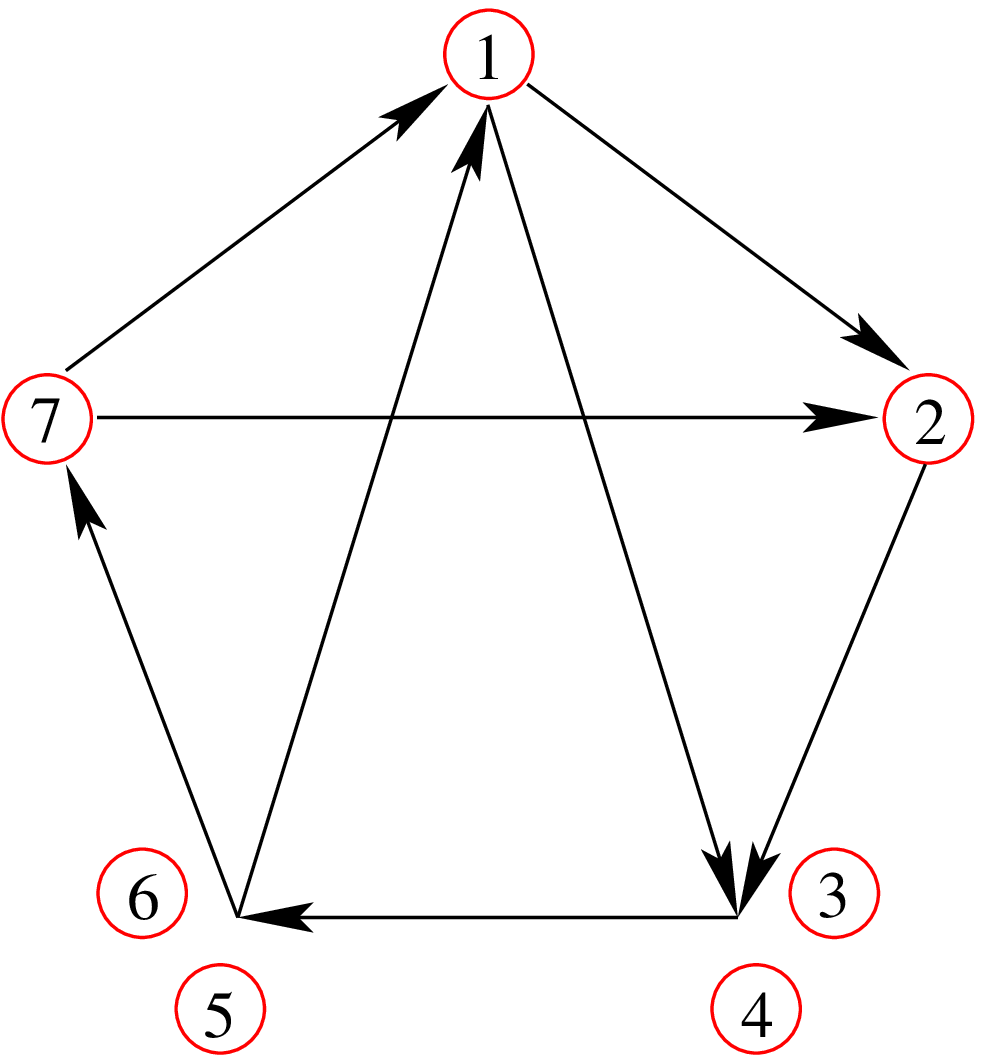}
\caption{Quiver diagram for Model I of $dP_4$.}
\label{quiver_dP4_1}
\end{minipage}%
\begin{minipage}[b]{0.7\textwidth}
\centering
{\footnotesize
\beq
\begin{array}{c|c}
 \ \ \ \ \ \mbox{Fields} \ \ \ \ \ & \ \ \ \ \ SU(5) \ \ \ \ \ \\
\hline
\hline
(X_{45},X_{23},X_{46},X_{71},X_{36},X_{24},X_{57},X_{35},X_{12},X_{67}) & 10 \\
\hline
(X_{51},X_{72},X_{67},X_{13},X_{14}) & \bar{5}
\label{symmetries_dP4_1}
\end{array}
\eeq
}
\bigskip
\bigskip
\bigskip
\bigskip
\end{minipage}
\end{figure}

Comparing the charges with weight vectors of $SU(5)$ representations we find the assignment tabulated in 
\eref{symmetries_dP4_1}.


The superpotential is written in terms of singlets as
\beq
W_I=10 \otimes \bar{5}^2+10^3 \otimes \bar{5}
\eeq

One might wonder whether a ${\bf 10}^5$ term should be present in $W_{dP_{4,I}}$. At first sight
it appears as a valid contribution, since this product of representations contains an $E_4$ singlet 
and we see from \fref{quiver_dP4_1} that it would survive the projection onto gauge invariant states. 
As we shall discuss in Section \ref{section_higgsing}, this model can be obtained from Model I of $dP_5$ by higgsing. All 
the gauge invariants in that theory are quartic. In particular, since there are no cubic terms, masses are 
not generated when turning on a non-zero vev for a bifundamental field. Then, we conclude that any fifth order 
term in $W_{dP_{4,I}}$ should have its origin either in a fifth or sixth order term in $W_{dP_{5,I}}$. Since 
$W_{dP_{5,I}}$ is purely quartic, we conclude that the ${\bf 10}^5$ is not present in $W_{dP_{4,I}}$.

\subsubsection*{Model II}
Upon dualisation of node 7 of Model I we get Model II. There are 19 fields in this model.

{\footnotesize

\beq
\begin{array}{ccc}

\begin{array}{cc}
Node & L_{\alpha} \\
& \\
1 & D \\
2 & E_1 \\
3 & E_3 \\
4 & 2D-E_4 \\
5 & 0 \\
6 & 2D-E_2 \\
7 & 2D-E_2-E_4 
\end{array}

& \ \ \ \ \ \ \ \ \ \ &

\begin{array}{lcrrrrc}
& L_{\alpha\beta} & J_1 & J_2 & J_3 & J_4 & R \\
&&&&&& \\
X_{52} & E_1 & -1 & 0 & 0 & 1 & 2/5 \\
X_{74} & E_2 & 1 & -1 & 0 & 1 & 2/5 \\
X_{53} & E_3 & 0 & 1 & -1 & 1 & 2/5 \\
X_{76} & E_4 & 0 & 0 & 1 & 0 & 2/5 \\
X_{43} & D-E_1-E_2 & 0 & 1 & 0 & -1 & 2/5 \\
X_{75} & D-E_1-E_3 & 1 & -1 & 1 & -1 & 2/5 \\
X_{63} & D-E_1-E_4 & 1 & 0 & -1 & 0 & 2/5 \\
X_{42} & D-E_2-E_3 & -1 & 0 & 1 & -1 & 2/5 \\
X_{17} & D-E_2-E_4 & -1 & 1 & -1 & 0 & 2/5 \\
X_{62} & D-E_3-E_4 & 0 & -1 & 0 & 0 & 2/5 \\
X_{21} & D-E_1 & 1 & 0 & 0 & 0 & 4/5 \\
X_{16} & D-E_2 & -1 & 1 & 0 & 0 & 4/5 \\
X_{31} & D-E_3 & 0 & -1 & 1 & 0 & 4/5 \\
X_{14} & D-E_4 & 0 & 0 & -1 & 1 & 4/5 \\
X_{15} & 2D-E_1-E_2-E_3-E_4 & 0 & 0 & 0 & -1 & 4/5 \\ 
X_{27} & 2D-E_1-E_2-E_4 & 0 & 1 & -1 & 0 & 6/5 \\
X_{37} & 2D-E_2-E_3-E_4 & -1 & 0 & 0 & 0 & 6/5 
\end{array} 

\end{array}
\label{divisors_dP4_II}
\eeq

}

\begin{figure}[htb]
\begin{minipage}[b]{0.3\textwidth}
\centering
\includegraphics[width=0.7\textwidth]{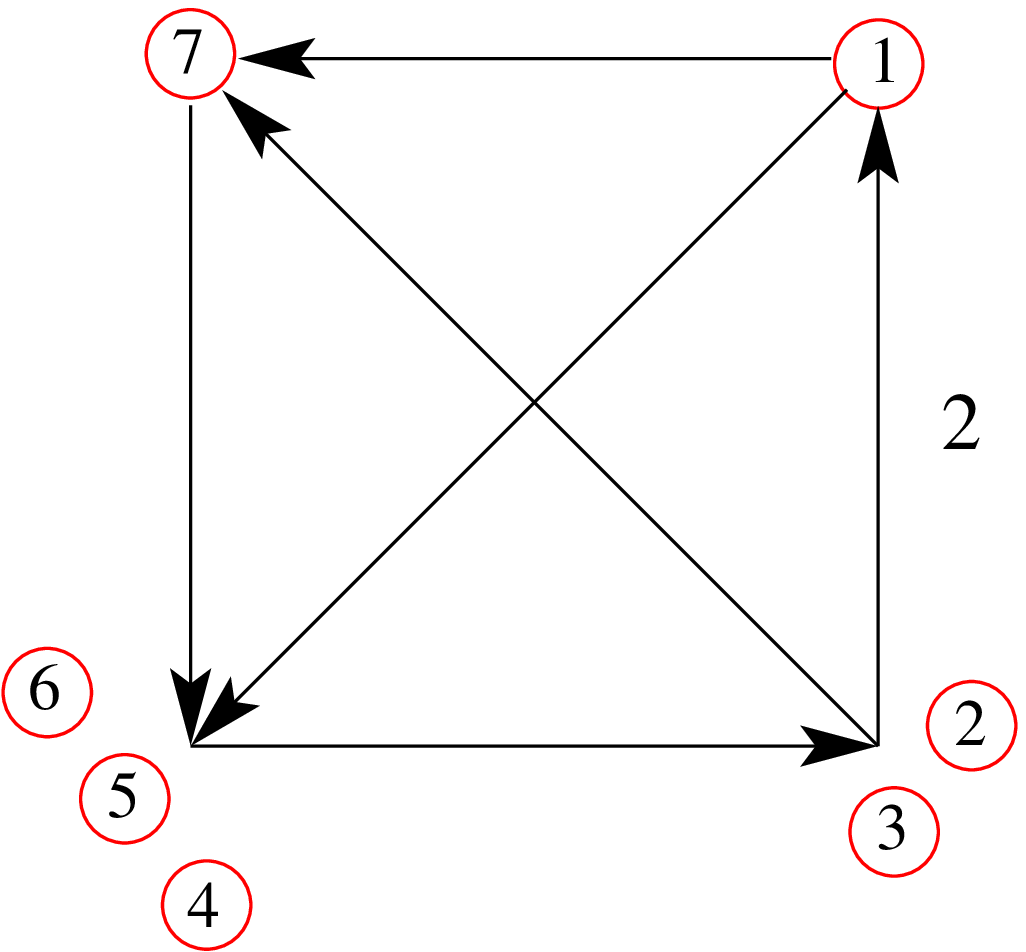}
\caption{Quiver diagram for Model II of $dP_4$.}
\label{quiver_dP4_2}
\end{minipage}%
\begin{minipage}[b]{0.7\textwidth}
\centering
{\footnotesize
\beq
\begin{array}{c|c}
 \ \ \ \ \ \mbox{Fields} \ \ \ \ \ & \ \ \ \ \ SU(5) \ \ \ \ \ \\
\hline
\hline
(X_{52},X_{74},X_{53},X_{76},X_{43},X_{75},X_{63},X_{42},X_{17},X_{62}) & 10 \\
\hline
(X_{21},X_{16},X_{31},X_{14},X_{15}) & \bar{5}_a \\
\hline
(Y_{21},Y_{31},*,*,*) & \mbox{partial } \bar{5}_b \\
\hline
(X_{27},X_{37},*,*,*) & \mbox{partial } 5
\end{array}
\label{symmetries_dP4_2}
\eeq
}
\bigskip
\bigskip
\bigskip
\bigskip
\end{minipage}
\end{figure}

This is the first example where partial representations appear. We will study this further in Section  
\ref{section_partial_representations}, and for the moment it will suffice to say that the missing fields 
(indicated by asterisks in Table \eref{symmetries_dP4_2}) are massive and the terms containing those fields 
in the singlets forming the superpotential should be thrown out, as 
they are not a part of the low energy theory.


The superpotential for this theory is

\beq
\begin{array}{ccl}
W_{II}&=& 10 \otimes \bar{5}_a^2+10 \otimes \bar{5}_a \otimes \bar{5}_b \\
      & &+10^3 \otimes \bar{5}_a+10^3 \otimes \bar{5}_b+10^2 \otimes 5 
\label{W_dP4_II}
\end{array}
\eeq
where the superpotential corresponds only to those $E_4$ invariant contributions that survive the 
projection onto gauge invariant terms. The terms including the partial ${\bf \bar{5}_b}$ representation 
are naturally understood to be truncated to fields actually appearing in the quiver.

\subsection{del Pezzo 5}

Let us study the three toric phases of $dP_5$. Models II and III exhibit again the phenomenon of partial representations. 


\subsubsection*{Model I}

This model has 16 fields.

{\footnotesize

\beq
\begin{array}{ccc}

\begin{array}{cc}
Node & L_{\alpha} \\
& \\
1 & E_1 \\
2 & E_3 \\
3 & D-E_4 \\
4 & D-E_5 \\
5 & 2D-E_2-E_4-E_5 \\
6 & D \\
7 & 2D-E_4-E_5 \\
8 & 0 
\end{array}

& \ \ \ \ \ \ \ \ \ \ &

\begin{array}{lcrrrrrc}
& L_{\alpha\beta} & J_1 & J_2 & J_3 & J_4 & J_5 & R \\
&&&&&&&\\
X_{81} & E_1 & -1 & 0 & 0 & 0 & 1 & 1/2 \\
X_{57} & E_2 & 1 & -1 & 0 & 0 & 1 & 1/2 \\
X_{82} & E_3 & 0 & 1 & -1 & 0 & 1 & 1/2 \\
X_{36} & E_4 & 0 & 0 & 1 & -1 & 0 & 1/2 \\
X_{46} & E_5 & 0 & 0 & 0 & 1 & 0 & 1/2 \\
X_{72} & D-E_1-E_2 & 0 & 1 & 0 & 0 & -1 & 1/2 \\
X_{58} & D-E_1-E_3 & 1 & -1 & 1 & 0 & -1 & 1/2 \\
X_{13} & D-E_1-E_4 & 1 & 0 & -1 & 1 & 0 & 1/2 \\
X_{14} & D-E_1-E_5 & 1 & 0 & 0 & -1 & 0 & 1/2 \\
X_{71} & D-E_2-E_3 & -1 & 0 & 1 & 0 & -1 & 1/2 \\
X_{45} & D-E_2-E_4 & -1 & 1 & -1 & 1 & 0 & 1/2 \\
X_{35} & D-E_2-E_5 & -1 & 1 & 0 & -1 & 0 & 1/2 \\
X_{23} & D-E_3-E_4 & 0 & -1 & 0 & 1 & 0 & 1/2 \\
X_{24} & D-E_3-E_5 & 0 & -1 & 1 & -1 & 0 & 1/2 \\
X_{67} & D-E_4-E_5 & 0 & 0 & -1 & 0 & 1 & 1/2 \\
X_{68} & 2D-E_1-E_2-E_3-E_4-E_5 & 0 & 0 & 0 & 0 & -1 & 1/2
\end{array} 

\end{array}
\eeq

}

All sixteen fields here are accommodated in a single ${\bf 16}$ representation of $E_5=SO(10)$.


\begin{figure}[htb]
\begin{minipage}[b]{0.3\textwidth}
\centering
\includegraphics[width=0.7\textwidth]{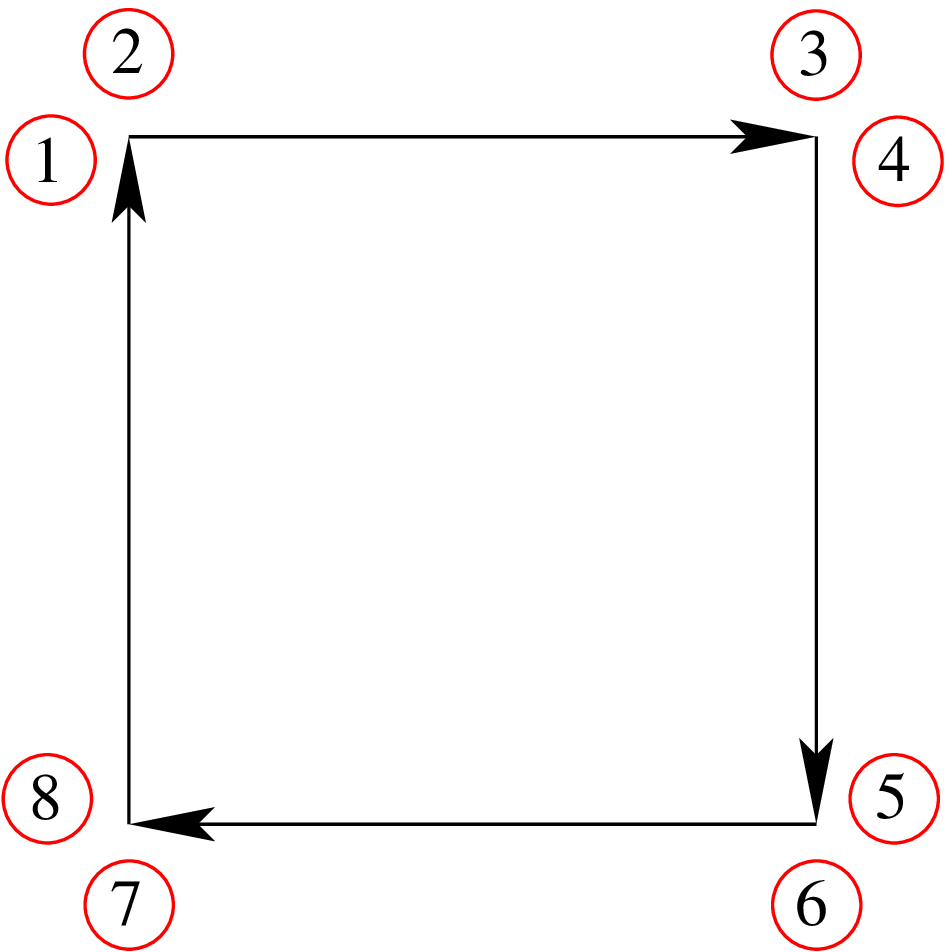}
\caption{Quiver diagram for Model I of $dP_5$.}
\label{quiver_dP5_1}
\end{minipage}%
\begin{minipage}[b]{0.7\textwidth}
\centering
{\footnotesize
\beq
\begin{array}{c|c}
 \ \ \ \ \ \mbox{Fields} \ \ \ \ \ & \ \ \ \ \ S0(10) \ \ \ \ \ \\
\hline
\hline
\begin{array}{l}(X_{81},X_{57},X_{82},X_{36},X_{46},X_{72},X_{58},X_{13}, \\
                \ X_{14},X_{71},X_{45},X_{35},X_{23},X_{24},X_{67},X_{68}) 
\end{array}& 16 
\end{array}
\label{symmetries_dP5_1}
\eeq
}
\bigskip
\bigskip
\bigskip
\bigskip
\end{minipage}
\end{figure}


The superpotential consists simply of all quartic gauge invariants and can be written as
\beq
W_I=16^4
\eeq

\subsubsection*{Model II}

Model II is can be obtained by dualising node 5 of model I. The divisors and charges are as shown in the tables.

{\footnotesize

\beq
\begin{array}{ccc}

\begin{array}{cc}
Node & L_{\alpha} \\
& \\
1 & D \\
2 & 2D-E_4-E_5 \\
3 & 0 \\
4 & E_1 \\
5 & E_2 \\
6 & E_3 \\
7 & D-E_4 \\
8 & D-E_5 
\end{array}

& \ \ \ \ \ \ \ \ \ \ &

\begin{array}{lcrrrrrc}
& L_{\alpha\beta} & J_1 & J_2 & J_3 & J_4 & J_5 & R \\
&&&&&&&\\
X_{34} & E_1 & -1 & 0 & 0 & 0 & 1 & 1/2 \\
X_{35} & E_2 & 1 & -1 & 0 & 0 & 1 & 1/2 \\
X_{36} & E_3 & 0 & 1 & -1 & 0 & 1 & 1/2 \\
X_{71} & E_4 & 0 & 0 & 1 & -1 & 0 & 1/2 \\
X_{81} & E_5 & 0 & 0 & 0 & 1 & 0 & 1/2 \\
X_{26} & D-E_1-E_2 & 0 & 1 & 0 & 0 & -1 & 1/2 \\
X_{25} & D-E_1-E_3 & 1 & -1 & 1 & 0 & -1 & 1/2 \\
X_{47} & D-E_1-E_4 & 1 & 0 & -1 & 1 & 0 & 1/2 \\
X_{48} & D-E_1-E_5 & 1 & 0 & 0 & -1 & 0 & 1/2 \\
X_{24} & D-E_2-E_3 & -1 & 0 & 1 & 0 & -1 & 1/2 \\
X_{57} & D-E_2-E_4 & -1 & 1 & -1 & 1 & 0 & 1/2 \\
X_{58} & D-E_2-E_5 & -1 & 1 & 0 & -1 & 0 & 1/2 \\
X_{67} & D-E_3-E_4 & 0 & -1 & 0 & 1 & 0 & 1/2 \\
X_{68} & D-E_3-E_5 & 0 & -1 & 1 & -1 & 0 & 1/2 \\
X_{12} & D-E_4-E_5 & 0 & 0 & -1 & 0 & 1 & 1/2 \\
X_{13} & 2D-E_1-E_2-E_3-E_4-E_5 & 0 & 0 & 0 & 0 & 1 & 1/2 \\ 
X_{82} & D-E_4 & 0 & 0 & -1 & 1 & 1 & 1 \\
X_{72} & D-E_5 & 0 & 0 & 0 & -1 & 1 & 1 \\
X_{83} & 2D-E_1-E_2-E_3-E_4 & 0 & 0 & 0 & 1 & -1 & 1 \\
X_{73} & 2D-E_1-E_2-E_3-E_5 & 0 & 0 & 1 & -1 & -1 & 1 
\end{array}

\end{array}
\eeq

}

The meson fields created by the dualization form a partial ${\bf 10}$ representation of $SO(10)$.


\begin{figure}[htb]
\begin{minipage}[b]{0.3\textwidth}
\centering
\includegraphics[width=0.7\textwidth]{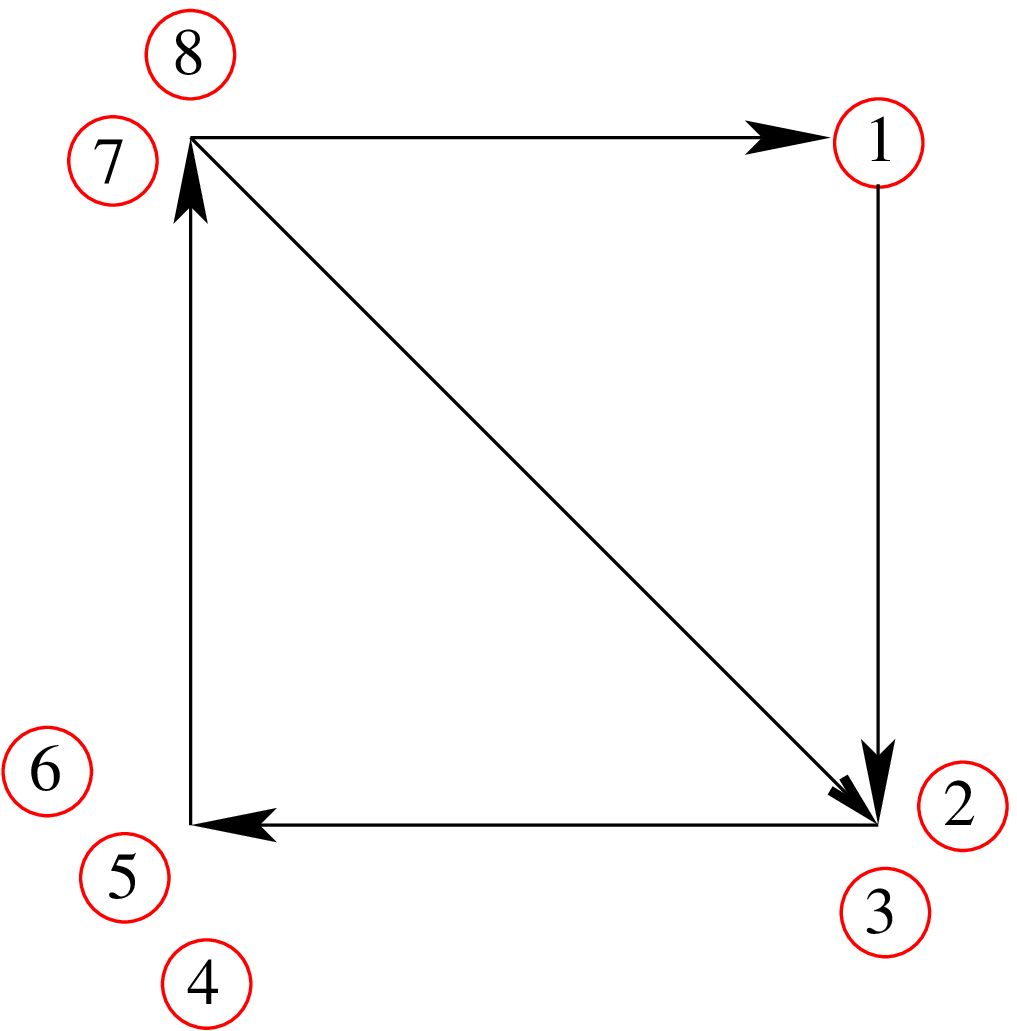}
\caption{Quiver diagram for Model II of $dP_5$.}
\label{quiver_dP5_2}
\end{minipage}%
\begin{minipage}[b]{0.7\textwidth}
\centering
{\footnotesize
\beq
\begin{array}{c|c}
 \ \ \ \ \ \mbox{Fields} \ \ \ \ \ & \ \ \ \ \ S0(10) \ \ \ \ \ \\
\hline
\hline
\begin{array}{l}(X_{34},X_{35},X_{36},X_{71},X_{81},X_{26},X_{25},X_{47}, \\
               \ X_{48},X_{24},X_{57},X_{58},X_{67},X_{68},X_{12},X_{13}) 
\end{array}& 16 \\
\hline
(X_{82},X_{72},X_{83},X_{73},*,*,*,*,*,*) & \mbox{partial } 10 
\end{array}
\label{symmetries_dP5_2}
\eeq
}
\bigskip
\bigskip
\bigskip
\bigskip
\end{minipage}
\end{figure}


The superpotential is
\beq
W_{II}=16^4+16^2 \otimes 10
\label{W_dP5_II}
\eeq

\subsubsection*{Model III}
The last toric phase of $dP_5$ is produced by dualising on node 6 of model II and has 24 fields, 
arranged as follows.
{\footnotesize

\beq
\begin{array}{ccc}

\begin{array}{cc}
Node & L_{\alpha} \\
& \\
1 & E_1 \\
2 & E_3 \\
3 & E_2 \\
4 & D-E_4-E_5 \\
5 & 2D-E_4-E_5 \\
6 & 0 \\
7 & D-E_4 \\
8 & D-E_5 \\

\end{array}

& \ \ \ \ \ \ \ \ \ \ &

\begin{array}{lcrrrrrc}
& L_{\alpha\beta} & J_1 & J_2 & J_3 & J_4 & J_5 & R \\
&&&&&&&\\
X_{61} & E_1 & -1 & 0 & 0 & 0 & 1 & 1/2 \\
X_{63} & E_2 & 1 & -1 & 0 & 0 & 1 & 1/2 \\
X_{62} & E_3 & 0 & 1 & -1 & 0 & 1 & 1/2 \\
X_{48} & E_4 & 0 & 0 & 1 & -1 & 0 & 1/2 \\
X_{47} & E_5 & 0 & 0 & 0 & 1 & 0 & 1/2 \\
X_{52} & D-E_1-E_2 & 0 & 1 & 0 & 0 & -1 & 1/2 \\
X_{53} & D-E_1-E_3 & 1 & -1 & 1 & 0 & -1 & 1/2 \\
X_{17} & D-E_1-E_4 & 1 & 0 & -1 & 1 & 0 & 1/2 \\
X_{18} & D-E_1-E_5 & 1 & 0 & 0 & -1 & 0 & 1/2 \\
X_{51} & D-E_2-E_3 & -1 & 0 & 1 & 0 & -1 & 1/2 \\
X_{37} & D-E_2-E_4 & -1 & 1 & -1 & 1 & 0 & 1/2 \\
X_{38} & D-E_2-E_5 & -1 & 1 & 0 & -1 & 0 & 1/2 \\
X_{27} & D-E_3-E_4 & 0 & -1 & 0 & 1 & 0 & 1/2 \\
X_{28} & D-E_3-E_5 & 0 & -1 & 1 & -1 & 0 & 1/2 \\
X_{64} & D-E_4-E_5 & 0 & 0 & -1 & 0 & 1 & 1/2 \\
X_{54} & 2D-E_1-E_2-E_3-E_4-E_5 & 0 & 0 & 0 & 0 & 1 & 1/2 \\ 
X_{85} & D-E_4 & 0 & 0 & -1 & 1 & 1 & 1 \\
X_{75} & D-E_5 & 0 & 0 & 0 & -1 & 1 & 1 \\
X_{86} & 2D-E_1-E_2-E_3-E_4 & 0 & 0 & 0 & 1 & -1 & 1 \\
X_{76} & 2D-E_1-E_2-E_3-E_5 & 0 & 0 & 1 & -1 & -1 & 1 
\end{array} 

\end{array}
\eeq

}

A second copy of the partial ${\bf 10}$ representation appears here.


\begin{figure}[htb]
\begin{minipage}[b]{0.3\textwidth}
\centering
\includegraphics[width=0.7\textwidth]{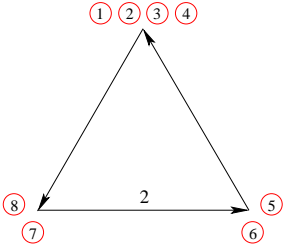}
\caption{Quiver diagram for Model III of $dP_5$.}
\label{quiver_dP5_3}
\end{minipage}%
\begin{minipage}[b]{0.7\textwidth}
\centering
{\footnotesize
\beq
\begin{array}{c|c}
 \ \ \ \ \ \mbox{Fields} \ \ \ \ \ & \ \ \ \ \ S0(10) \ \ \ \ \ \\
\hline
\hline
\begin{array}{l}(X_{61},X_{63},X_{62},X_{48},X_{47},X_{52},X_{53},X_{17}, \\
                \ X_{18},X_{51},X_{37},X_{38},X_{27},X_{28},X_{64},X_{54}) 
\end{array}& 16 \\
\hline
(X_{85},X_{75},X_{86},X_{76},*,*,*,*,*,*) & \mbox{partial } 10_a \\
\hline
(Y_{85},Y_{75},Y_{86},Y_{76},*,*,*,*,*,*) & \mbox{partial } 10_b
\end{array}
\label{symmetries_dP5_3}
\eeq
}
\bigskip
\bigskip
\bigskip
\bigskip
\end{minipage}
\end{figure}


The superpotential for this theory is
\beq
W_{III}=16^2 \otimes 10_a+16^2 \otimes 10_b
\eeq
Note that the global symmetry invariant term ${\bf 16}^4$ does not appear, since none of its components survives 
the projection onto gauge invariants. It is clear from looking at the 3-block quiver in \fref{quiver_dP5_3} that
there are no quartic gauge invariants in this case.
 
\subsection{del Pezzo 6}

The final model we study is the toric phase of $dP_6$. There are some indications suggesting that this theory 
completes the list of toric phases of del Pezzo theories. In particular, the geometric computation of dibaryon 
$R$ charges (the charges of bifundamental fields can be derived from them) determines that, if all gauge groups were
$U(N)$, the least possible R-charge for a bifundamental field is one for $dP_7$ and two for $dP_8$.
In this case, the superpotential for $dP_7$ could only consist of quadratic mass terms, while it would be impossible
to construct a superpotential for $dP_8$ \cite{Herzog:2003wt}. Since we expect all del Pezzo theories to have 
nontrivial superpotentials, this seems to rule out such models. Some other particular features, that might be related to 
the previous one, and differentiate $dP_7$ and $dP_8$ from the rest of the del Pezzos arise in the context of $(p,q)$ webs 
\cite{Hanany:2001py,Franco:2002ae}, where webs without crossing external legs cannot be constructed beyond $dP_6$. 

The toric phase of $dP_6$ has 27 fields. The associated divisors are listed in \eref{divisors_dP6}.

\bigskip

{\footnotesize

\beq
\begin{array}{ccc}

\begin{array}{cc}
Node & L_{\alpha} \\
& \\
1 & E_1 \\
2 & E_2 \\
3 & E_3 \\
4 & D-E_4 \\
5 & D-E_5 \\
6 & D-E_6 \\
7 & D \\
8 & 0 \\
9 & 2D-E_4-E_5-E_6 
\end{array}

& \ \ \ \ \ \ \ \ \ \ &

\begin{array}{lcrrrrrrc}
& L_{\alpha\beta} & J_1 & J_2 & J_3 & J_4 & J_5 & J_6 & R \\
&&&&&&&&\\
X_{81} & E_1 & -1 & 0 & 0 & 0 & 0 & 1 & 2/3 \\
X_{82} & E_2 & 1 & -1 & 0 & 0 & 0 & 1 & 2/3 \\
X_{83} & E_3 & 0 & 1 & -1 & 0 & 0 & 1 & 2/3 \\
X_{47} & E_4 & 0 & 0 & 1 & -1 & 0 & 0 & 2/3 \\
X_{57} & E_5 & 0 & 0 & 0 & 1 & -1 & 0 & 2/3 \\
X_{67} & E_6 & 0 & 0 & 0 & 0 & 1 & 0 & 2/3 \\ 
X_{93} & D-E_1-E_2 & 0 & 1 & 0 & 0 & 0 & -1 & 2/3 \\
X_{92} & D-E_1-E_3 & 1 & -1 & 1 & 0 & 0 & -1 & 2/3 \\
X_{14} & D-E_1-E_4 & 1 & 0 & -1 & 1 & 0 & 0 & 2/3 \\
X_{15} & D-E_1-E_5 & 1 & 0 & 0 & -1 & 1 &  0 & 2/3 \\
X_{16} & D-E_1-E_6 & 1 & 0 & 0 & 0 & -1 & 0 & 2/3 \\
X_{91} & D-E_2-E_3 & -1 & 0 & 1 & 0 & 0 & -1 & 2/3 \\
X_{24} & D-E_2-E_4 & -1 & 1 & -1 & 1 & 0 & 0 & 2/3 \\
X_{25} & D-E_2-E_5 & -1 & 1 & 0 & -1 & 1 & 0 & 2/3 \\
X_{26} & D-E_2-E_6 & -1 & 1 & 0 & 0 & -1 & 0 & 2/3 \\
X_{34} & D-E_3-E_4 & 0 & -1 & 0 & 1 & 0 & 0 & 2/3 \\
X_{35} & D-E_3-E_5 & 0 & -1 & 1 & -1 & 1 & 0 & 2/3 \\
X_{36} & D-E_3-E_6 & 0 & -1 & 1 & 0 & -1 & 0 & 2/3 \\
X_{69} & D-E_4-E_5 & 0 & 0 & -1 & 0 & 1 & 1 & 2/3 \\
X_{59} & D-E_4-E_6 & 0 & 0 & -1 & 1 & -1 & 1 & 2/3 \\
X_{49} & D-E_5-E_6 & 0 & 0 & 0 & -1 & 0 & 1 & 2/3 \\
X_{71} & 2D-E_2-E_3-E_4-E_5-E_6 & -1 & 0 & 0 & 0 & 0 & 0 & 2/3 \\ 
X_{72} & 2D-E_1-E_3-E_4-E_5-E_6 & 1 & -1 & 0 & 0 & 0 & 0 & 2/3 \\ 
X_{73} & 2D-E_1-E_2-E_4-E_5-E_6 & 0 & 1 & -1 & 0 & 0 & 0 & 2/3 \\ 
X_{48} & 2D-E_1-E_2-E_3-E_5-E_6 & 0 & 0 & 1 & -1 & 0 & -1 & 2/3 \\ 
X_{58} & 2D-E_1-E_2-E_3-E_4-E_6 & 0 & 0 & 0 & 1 & -1 & -1 & 2/3 \\ 
X_{68} & 2D-E_1-E_2-E_3-E_4-E_5 & 0 & 0 & 0 & 0 & 1 & -1 & 2/3 
\end{array}

\end{array}
\label{divisors_dP6}
\eeq

}

\bigskip

\begin{figure}[htb]
\begin{minipage}[b]{0.3\textwidth}
\centering
\includegraphics[width=0.7\textwidth]{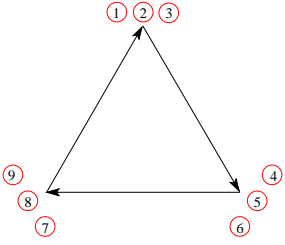}
\caption{Quiver diagram for $dP_6$.}
\label{quiver_dP6}
\end{minipage}%
\begin{minipage}[b]{0.7\textwidth}
\centering
{\footnotesize
\beq
\begin{array}{c|c}
 \ \ \ \ \ \mbox{Fields} \ \ \ \ \ & \ \ \ \ \ E_6\ \ \ \ \ \\
\hline
\hline
\begin{array}{l}(X_{81},X_{82},X_{83},X_{47},X_{57},X_{67},X_{93},X_{92}, X_{14}\\
                \ X_{15},X_{16},X_{91},X_{24},X_{25},X_{26},X_{34},X_{35},X_{36}\\
                \ X_{69},X_{59},X_{49},X_{71},X_{72},X_{73},X_{48},X_{58},X_{68}) 
\end{array}& 27
\end{array}
\label{symmetries_dP6}
\eeq
}
\bigskip
\bigskip
\bigskip
\bigskip
\end{minipage}
\end{figure}

All fields are accommodated in the fundamental ${\bf 27}$ representation of $E_6$, as shown in \eref{symmetries_dP6}. The superpotential 
for this theory is simply

\beq
W=27^3
\eeq


\section{Partial representions}

\label{section_partial_representations}

We have discussed how the matter content of the quiver theories for each $dP_n$ can be arranged into irreducible 
representations of the corresponding $E_n$ group. The superpotential can therefore be expressed as the gauge 
invariant part of a combination of fields invariant under the global symmetry transformations. Our discussion of this 
classification will be extended in Section \ref{section_symmetries_seiberg}, where we will study how to relate the 
representation content of Seiberg dual theories.

In Section \ref{section_symmetries}, we encoutered some examples (Model II
of $dP_4$, and  Models II and III of $dP_5$) that seem to challenge the 
applicability of our classification strategy. We mentioned there that in these cases
we have to go a step further and consider {\it partial representations}, and postponed 
the explanation to this point. The purpose of this section is to give a detailed description 
of the concept of partial representations and to show that they are a natural construction that 
enables us to study all the toric del Pezzo quivers from the same unified perspective. We will 
devote this subsection to explaining the simple rules that can be derived for these theories from 
a field theory point of view. The next subsection will sharpen these concepts but bases the discussion 
on the geometry of partial representations.

Theories with {\it partial representations} are those in which
it is not possible to arrange matter fields so that the corresponding $E_n$ representations
are completely filled. Naively, it is not clear what are the transformation properties that should 
be assigned to fields that seem not to fit into representations in these cases. It is not even clear 
that they can be organized into irreducible representations at all. As we will discuss, this situation 
neither implies a loss of predictive power nor that these models are exceptionl cases outside of the 
scope of our techniques, since in order for partial representations to exist, 
very specific conditions have to be fulfilled.

The idea is to find those fields that seem to be absent from the quiver, and that would 
join the fields that are present to form irreducible $E_n$ representations. These bifundamental fields 
should appear in representations and have gauge charges such that, following the rules given in Section 
\ref{section_symmetries}, quadratic terms appear in the superpotential. That is they can form quadratic 
invariants of the global symmetry group and, in quiver language, they appear as bidirectional arrows. These 
symmetric terms give masses to the fields under consideration, removing them from the low energy effective 
description.

Following the previous reasoning, partial representations appear in such a way that the same 
number of fields are missing from those representations that form quadratic terms. In 
some cases it is possible for the missing fields to lie on the same representation, which combines with itself to form
a quadratic invariant. When this occurs, the number of missing fields is even. The R charges of fields in  
representations that combine into quadratic invariants and become partial representations add up to 2. Thus,
for the specific case of self-combining representations, they should have $R=1$ in order to be capable of 
becoming partial.

These general concepts are sufficient to classify the quivers into representations, but do not 
indicate which are the precise nodes that are connected by the missing fields. One possible
way to determine them uses the assignation of divisors to bifundamental fields and is the 
motive of the next subsection.

\subsection{The geometry of partial representations}

\label{section_geometry_partial}

An important question is what the location in the quiver of the fields that are needed in order to complete 
partial representations is. As we discussed in the previous section, fields missing from partial representations form 
bidirectional arrows and are combined into quadratic mass terms.

The lists in Section 3 summarize the baryonic $U(1)$ and R charges of the fields that are present in the quiver. For each 
arrow, these numbers indicate the intersection of its associated divisor with the $n+1$ curves in the non-orthogonal basis 
of \eref{simple_roots}. Thus, these charges define a set of $n+1$ equations from which the divisor associated to a given bifundamental
field can be deduced. Furthermore, as explained in Section 2, the $n$ flavor charges correspond to the Dynkin components 
of each state. Then, the Dynkin components of the missing fields can be inferred by looking for those that are absent from 
partial representations. Once these charges are determined, they can be used together with the R charge of the 
representation to follow the process explained above and establish the divisors for the missing fields.

Based on the divisors that correspond to each node, \eref{nodes_divisors} gives the divisors for every possible bifundamental field. Comparing them
with the ones for missing fields we determine where they are in the quiver. Let us remark that the examples in Section
\ref{section_symmetries} show us that different bifundamentals can have the same associated divisors. In the cases
we will study, it is straightforward to check that such ambiguity does not hold for the fields we are trying to identify.

Let us consider the example of Model II of $dP_4$. There are 19 fields in this theory. Some of them form a full ${\bf 10}$ 
and a full $\bar{{\bf 5}}$ representations. There are four remaining fields that cannot be arranged into full representations
of $E_4$. From the Dynkin components in \eref{divisors_dP4_II}, we conclude that $X_{27}$ and $X_{37}$ sit in 
an incomplete ${\bf 5}$, while $Y_{21}$ and $Y_{31}$ are part of a $\bar{{\bf 5}}$. There are six missing fields that 
should complete the ${\bf 5}$ and $\bar{{\bf 5}}$. The Dynkin components ($U(1)$ charges) that are needed to complete
the representations can be immediately determined. They are listed in the second column of \eref{partial_geometry_dP4}.
From those, the divisors in the third column are computed. The divisors for each node in the quiver appear in 
\eref{divisors_dP4_II}. Using them, we determine the nodes in the quiver that are connected by the missing fields.

\beq
\begin{array}{|c|c|c|c|}
\hline
\ \ \mbox{Representation} \ \ & \ \ \ \ (J_1,J_2,J_3,J_4,R) \ \ \ \ & \mbox{Divisor}     & \ \ \mbox{Bifundamental} \ \ \\
\hline 
                              &                                     &                    &                       \\
                              & (-1,1,0,0,4/5)                      & D-E_2              & Y_{16}                \\
\bar{5}                       & (0,0,-1,1,4/5)                      & D-E_4              & Y_{14}                \\
                              & (0,0,0,-1,4/5)                      & 2D-E_1-E_2-E_3-E_4 & Y_{15}                \\ 
                              &                                     &                    &                       \\ 
\hline 
                              &                                     &                    &                       \\
                              & (1,-1,0,0,6/5)                      & 2D-E_1-E_3-E_4     & X_{61}                \\
5                             & (0,0,1,-1,6/5)                      & 2D-E_1-E_2-E_3     & X_{41}                \\
                              & (0,0,0,1,6/5)                       & D                  & X_{51}                \\ 
                              &                                     &                    &                       \\
\hline
\end{array}
\label{partial_geometry_dP4}
\eeq

\begin{figure}[ht]
  \epsfxsize = 4cm
  \centerline{\epsfbox{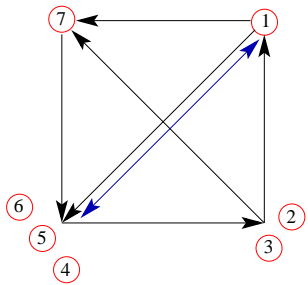}}
  \caption{Quiver diagram for Model II of $dP_4$ showing in blue the fields that are missing from partial representations.}
  \label{quiver_partial_dP4}
\end{figure}

The quiver with the addition of these extra fields is shown in \fref{quiver_partial_dP4}.

Now that we have identified the fields that are missing from the partial representations, we can rewrite the superpotential 
for this model in an expression that includes all the fields in the theory, both massless and massive. It becomes (note the 
mass term for the ${\bf 5}$ and ${\bf \bar{5}_b}$ representations)

\beq
W_{II}=[10 \otimes \bar{5}_a \otimes \bar{5}_a+10 \otimes 10 \otimes 5+10 \otimes \bar{5}_a \otimes \bar{5}_b]+
10 \otimes \bar{5}_b \otimes \bar{5}_b+5 \otimes \bar{5}_b
\eeq

The products of representations between brackets are already present in \eref{W_dP4_II}. Keeping in mind that some of the fields in the ${\bf 5}$ and 
${\bf \bar{5}_b}$ remain massless, it is straightforward to prove that the previous expression reduces to \eref{W_dP4_II} (which only includes massless 
fields) when the massive fields are integrated out.

Let us now consider a different example. In Model II of $dP_5$, fields within a single representation are combined to form quadratic terms. The procedure described
above can be applied without changes to this situation. The first step is to identify the Dynkin components and R charge of the missing fields. From them, the 
corresponding divisors are computed. Finally, the gauge charges of the missing fields are determined. This information is 
summarized in Table \eref{partial_geometry_dP5}. 

\beq
\begin{array}{|c|c|c|}
\hline
 \ \ \ \ (J_1,J_2,J_3,J_4,J_5,R) \ \ \ \ & \mbox{Divisor}     & \ \ \mbox{Bifundamental} \ \ \\
\hline 
                                     &                        &                       \\
  (-1,0,0,0,0,1)                      & \ 2D-E_2-E_3-E_4-E_5 \ & X_{14}                \\
  (1,-1,0,0,0,1)                     &   2D-E_1-E_3-E_4-E_5   & X_{15}                \\
  (0,1,-1,0,0,1)                     &   2D-E_1-E_2-E_4-E_5   & X_{16}                \\ 
                                     &                        &                       \\ 
\hline 
                                     &                        &                       \\
 (1,0,0,0,0,1)                         &   D-E_1                & X_{41}                \\
 (-1,1,0,0,0,1)                        &   D-E_2                & X_{51}                \\
 (0,-1,1,0,0,1)                        &   D-E_3                & X_{61}                \\ 
                                     &                        &                       \\
\hline
\end{array}
\label{partial_geometry_dP5}
\eeq

\begin{figure}[ht]
  \epsfxsize = 4cm
  \centerline{\epsfbox{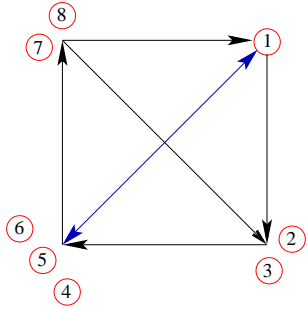}}
  \caption{Quiver diagram for Model II of $dP_5$ showing in blue the fields that are missing from partial representations.}
  \label{quiver_partial_dP5}
\end{figure}

\fref{quiver_partial_dP5} shows where the fields that are missing from the partial ${\bf 10}$ appear in the quiver for
 Model II of $dP_5$.

Including the massive fields, we can write the superpotential as
\beq
W_{II}=16^2 \otimes 10 + 10^2
\eeq
which reproduces \eref{W_dP5_II} when integrating out those fields in the ${\bf 10}$ representation that are massive.

\subsection{Partial representations and $E_n$ subgroups}

It is possible to make a further characterization of theories with partial representations using group theory. This is
attained by considering the transformation properties of fields, both present and missing from the quiver, under subgroups 
of $E_n$. These subgroups appear when some of the nodes in a quiver theory fall into blocks. For a block containing $n_i$ nodes, 
a subgroup $SU(n_i)$ of the enhanced global symmetry $E_n$ becomes manifest. In the general case, the manifest subgroup of the 
enhanced global symmetry will be a product of such $SU(n_i)$ factors \footnote{$dP_n$ quivers have $n+3$ nodes. For each block
of $n_i$ nodes, the associated $SU(n_i)$ factor has rank $n_i-1$. Thus, we see that for the specific case of 3-block quivers
the sum of the ranks of the three $SU(n_i)$ factors is $n$, corresponding to a maximal subgroup of the corresponding $E_n$.}. 
A matter field is charged under one of these factors if it is attached to one of the nodes in the corresponding block. Let us 
see how this works in the quivers where partial representations appear.
 
The first example is Model II of $dP_4$. Its quiver, shown in \fref{quiver_dP4_2}, features two
blocks, with two and three nodes each. The corresponding subgroup of $E_4$ is $SU(2)\times SU(3)$. There are two 
partial representations, a ${\bf 5}$ and a ${\bf \bar{5}}$. The way they decompose under this subgroup is
\beq
\begin{array}{l}
5\rightarrow (2,1) + (1,3)\\
\bar{5} \rightarrow (2,1) + (1,\bar{3}).
\end{array}
\eeq
The fields which do not appear in the low energy limit are the ones in the $(1,3)$ and $(1,\bar{3})$ and 
run from nodes 4, 5, 6 to node 1 and vice versa, as was derived in Section \ref{section_geometry_partial}.
These fields form a quadratic gauge invariant which is a mass term and 
is the singlet in the product $(1,3)\otimes (1,\bar{3})$. The singlet in  $(2,1)\otimes (2,1)$ is not gauge 
invariant so these fields are massless and make up the partial representations.
The important observation is that fields present and missing from the quiver can be organized in representations of these 
relatively small subgroups of $E_n$, their classification becomes simpler. 


A similar situation occurs for Model II of $dP_5$, depicted in \fref{quiver_dP5_2}. The subgroup here 
is $SU(2)\times SU(2)\times SU(3)$ and we have a partial ${\bf 10}$ representation which decomposes as
\beq
10\rightarrow (2,2,1) + (1,1,3) + (1,1,\bar{3}).
\eeq
The missing fields are the ones in  $(1,1,3)$ and $(1,1,\bar{3})$, extending from node $1$ to nodes $4,5,6$
and vice-versa respectively in the quiver. As before, the singlet of $(1,1,3)\otimes (1,1,\bar{3})$ is a 
quadratic gauge invariant that makes these fields massive.

The last example where partial representations occur is Model III of $dP_5$. This is a three block model and
the manifest subgroup of $E_5$ is $SU(2)\times SU(2)\times SU(4)$, which is maximal. The ${\bf 10}$ of $SO(10)$
breaks as
\beq
10\rightarrow (2,2,1) + (1,1,6).
\eeq
The four fields that appear in each of the two partial ${\bf 10}$'s are the ones in $(2,2,1)$, running from 
nodes 5, 6 to nodes 7, 8. The matter fields in $(1,1,6)$ are charged only under the $SU(4)$ factor and thus 
cannot extend between two blocks of nodes. These fields run between pairs of nodes in the $SU(4)$ block 
(there are exactly six distinct pairs). The fields coming from the two partial representations run in opposite 
directions making the arrows between the nodes bidirectional. The corresponding superpotential terms are given 
by the singlet of $(1,1,6)\otimes (1,1,6)$ and make the fields massive.

\section{Higgsing}

\label{section_higgsing}

It is interesting to understand how the gauge theories for different del Pezzos are related to each other.
The transition from $dP_n$ to $dP_{n-1}$ involves the blow-down of a 2-cycle. This operation appears
in the gauge theory as a higgsing, by turning a non-zero VEV for a suitable bifundamental field.
The determination of possible choices of this field has been worked out case by case in the literature. 

The $(p,q)$ web techniques introduced in \cite{Franco:2002ae} provide us with a systematic approach
to the higgsing problem. In these diagrams, finite segments represent compact 2-cycles. The blow down
of a 2-cycle is represented by shrinking a segment in the web and the subsequent combination of the
external legs attached to it. The bifundamental field that acquires a non-zero VEV corresponds to the 
one charged under the gauge groups associated to the legs that are combined. The reader is referred to \cite{Franco:2002ae} 
to a detailed explanation of the construction, interpretation and applications of $(p,q)$ webs in the 
context of four dimensional gauge theories on D3-branes on singularities. $(p,q)$ webs are 
traditionally associated to toric geometries, since they represent the reciprocal lattice of a toric diagram (see
\cite{Feng:2004uq} for a recent investigation of the precise relation). 
Nevertheless, their range of applicability can be extended to the determination of quivers \cite{Hanany:2001py} 
and higgsings of non-toric del Pezzos.

In this section, we will derive all possible higgsings from $dP_n$ down
to $dP_{n-1}$ using $(p,q)$ webs. The passage from $dP_3$ to $dP_2$ will be discussed in detail, and the presentation
of the results for other del Pezzos will be more schematic. After studying these examples, it will be clear that this 
determination becomes trivial when using the information about global symmetries summarized in Section 3. The 
problem can be rephrased as looking for how to higgs the global symmetry group from $E_n$ to
$E_{n-1}$ by giving a non-zero VEV to a field that transform as a non-singlet under $E_n$. We will 
conclude this section by writing down the simple group theoretic explanation of our findings.

\subsection{Del Pezzo 3}

As our first example, we proceed now to determine all possible higgsings from the four phases of $dP_3$ down to 
the two phases of $dP_2$.

\subsubsection*{Model I}

\begin{figure}[ht]
  \epsfxsize = 12cm
  \centerline{\epsfbox{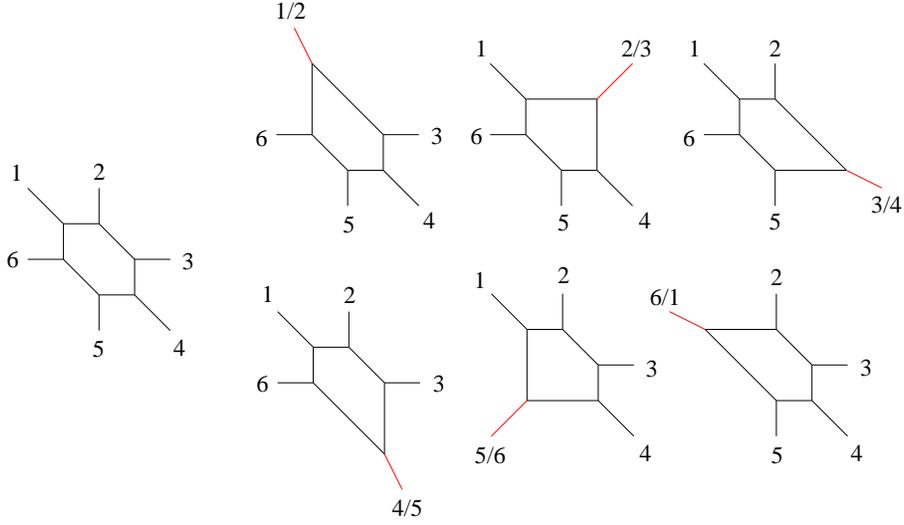}}
  \caption{$(p,q)$ webs describing all possible higgsings from Model I of $dP_3$ down to $dP_2$.}
  \label{higgs_dP3_1}
\end{figure}

The $(p,q)$ webs representing the higgsing of this phase down to $dP_2$ are presented
in \fref{higgs_dP3_1}. We have indicated in red, the combined external legs that result from blowing down  
2-cycles. All the resulting webs in \fref{higgs_dP3_1} are related by $SL(2,\IZ)$ transformations, implying 
that in this case it is only possible to obtain Model II of $dP_2$ by higgsing.

From the external legs that have to be combined in the $(p,q)$ webs in \fref{higgs_dP3_1}, we conclude that the lowest 
component of the following bifundamental chiral superfields should get a non-zero VEV in order to produce the higgsing

\beq
\begin{array}{cc}
\left. 
\begin{array}{lcl}
X_{12}, X_{23}, X_{34}, X_{45}, X_{56}, X_{61} & \rightarrow & \mbox{Model II} 
\end{array} \right\} &
\rightarrow (2,3)
\end{array}
\eeq

We notice that these fields form precisely a $(2,3)$, i.e. a fundamental representation, of $E_3=SU(2) \times SU(3)$. 
We will see that the same happens for all the del Pezzos.

\subsubsection*{Model II}

\begin{figure}[ht]
  \epsfxsize = 9cm
  \centerline{\epsfbox{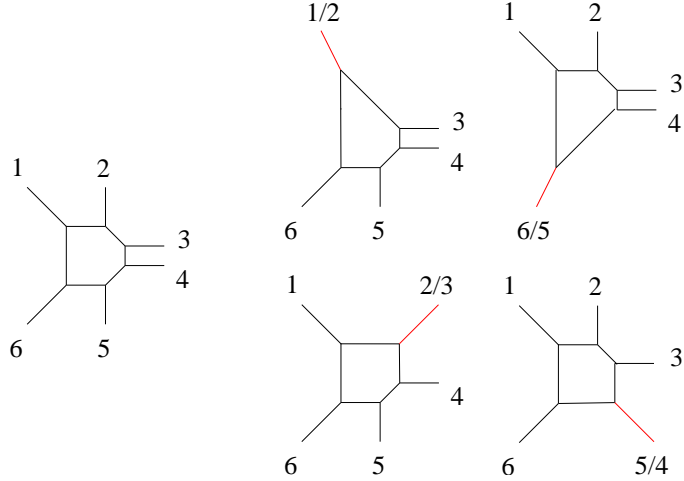}}
  \caption{$(p,q)$ webs describing all possible higgsings from Model II of $dP_3$ down to $dP_2$.}
  \label{higgs_dP3_2}
\end{figure}

The $(p,q)$ web diagrams for this model are shown in \fref{higgs_dP3_2}, along with the possible higgsings.

It is important to remind the reader of the node symmetries that appear in the gauge theory
when parallel external legs are present \cite{Franco:2002ae}. This situation appears in this 
example, and we have drawn only one representative of each family of $(p,q)$ webs related by this
kind of symmetries. In the language of exceptional collections, each set of parallel external legs
corresponds to a block of sheaves. The full list of bifundamental fields associated to the higgsings
in \fref{higgs_dP3_2} is summarized in the following list

\beq
\begin{array}{cc}
\left. 
\begin{array}{lcl}
X_{21}, X_{65}                 & \rightarrow & \mbox{Model I} \\
X_{32}, X_{42}, X_{53}, X_{54} & \rightarrow & \mbox{Model II}
\end{array} \right\} &
\rightarrow (2,3)
\end{array}
\eeq

\subsubsection*{Model III}

\begin{figure}[ht]
  \epsfxsize = 6.2cm
  \centerline{\epsfbox{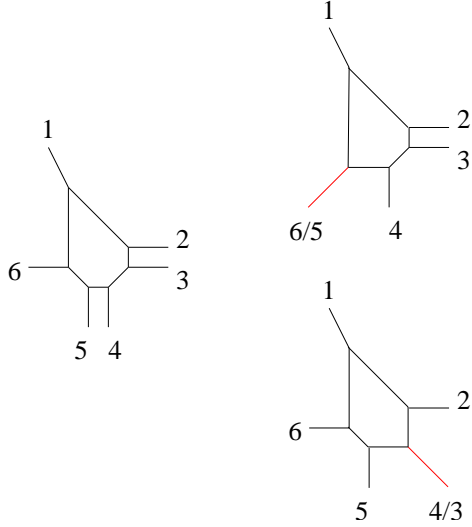}}
  \caption{$(p,q)$ webs describing all possible higgsings from Model III of $dP_3$ down to $dP_2$.}
  \label{higgs_dP3_3}
\end{figure}

\fref{higgs_dP3_3} shows the $(p,q)$ webs describing the higgsings of this phase. Once again, we have
included only one representative of each set of webs related by node symmetries. 

Then, we have the following set of fields that produce a higgsing to $dP_2$.

\beq
\begin{array}{cc}
\left. 
\begin{array}{lcl}
X_{64}, X_{65}                 & \rightarrow & \mbox{Model I} \\
X_{42}, X_{43}, X_{52}, X_{53} & \rightarrow & \mbox{Model II}
\end{array} \right\} &
\rightarrow (2,3)
\end{array}
\eeq

\subsubsection*{Model IV}

\begin{figure}[ht]
  \epsfxsize = 6cm
  \centerline{\epsfbox{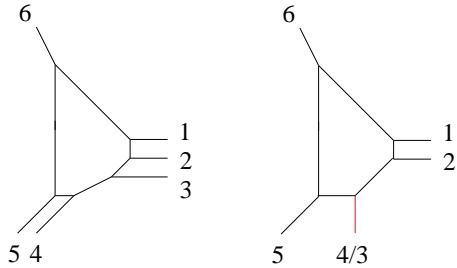}}
  \caption{$(p,q)$ webs describing all possible higgsings from Model IV of $dP_3$ down to $dP_2$.}
  \label{higgs_dP3_4}
\end{figure}

Finally, for Model IV of $dP_3$, we have the webs shown in \fref{higgs_dP3_4}

Indicating that the following fields can take us from this phase to $dP_2$.

\beq
\begin{array}{cc}
\left. 
\begin{array}{lcl}
X_{41}, X_{42}, X_{43}, X_{51}, X_{52}, X_{53} & \rightarrow & \mbox{Model I} 
\end{array} \right\} &
\rightarrow (2,3)
\end{array}
\eeq

\bigskip

Having studied the four toric phases of $dP_3$, we see that for all of them the fields that 
produce a higgsing down to $dP_2$ are those in the fundamental $(2,3)$ representation of 
$E_3=SU(2) \times SU(3)$. We will show below how the higgsing of all other del Pezzos 
is also attained by a non-zero VEV of any field in the fundamental representation of
$E_n$.

\subsection{Del Pezzo 4}

Let us now analyze the two toric phases of $dP_4$.

\subsubsection*{Model I}

\begin{figure}[ht]
  \epsfxsize = 3cm
  \centerline{\epsfbox{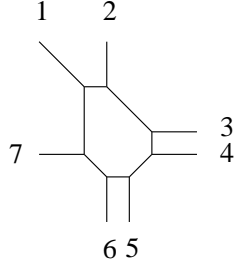}}
  \caption{$(p,q)$ web for Model I of $dP_4$.}
  \label{dP4_1}
\end{figure}

A possible $(p,q)$ web for this model is the one in \fref{dP4_1}. From now on, for simplicity, we will only 
present the original webs but not the higgsed ones. From \fref{dP4_1}, we determine the following higgsings

\beq
\begin{array}{cc}
\left. 
\begin{array}{lcl}
X_{35},X_{36},X_{45},X_{46} & \rightarrow & \mbox{Model I} \\
X_{23},X_{24},X_{57},X_{67} & \rightarrow & \mbox{Model II} \\
X_{12},X_{71}               & \rightarrow & \mbox{Model III} 
\end{array} \right\} &
\rightarrow 10
\end{array}
\eeq

As expected, the fields that higgs the model down to some $dP_3$ phase form the fundamental
${\bf 10}$ representation of $E_4$.

\subsubsection*{Model II}

\begin{figure}[ht]
  \epsfxsize = 3cm
  \centerline{\epsfbox{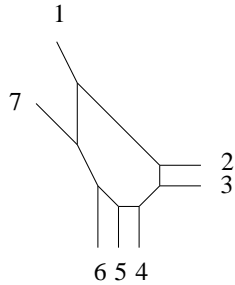}}
  \caption{$(p,q)$ web for Model II of $dP_4$.}
  \label{dP4_2}
\end{figure}

The $(p,q)$ web for Model II is presented in \eref{dP4_2}.

The higgsings in this case become

\beq
\begin{array}{cc}
\left. 
\begin{array}{lcl}
X_{42},X_{52},X_{62},X_{43},X_{53},X_{63} & \rightarrow & \mbox{Model II} \\
X_{74},X_{75},X_{76}                      & \rightarrow & \mbox{Model III} \\
X_{17}                                    & \rightarrow & \mbox{Model IV} 
\end{array} \right\} &
\rightarrow 10
\end{array}
\eeq

\subsection{Del Pezzo 5}

There are three toric phases for $dP_5$. We study now their higgsing down to $dP_4$.

\subsubsection*{Model I}

\begin{figure}[ht]
  \epsfxsize = 3cm
  \centerline{\epsfbox{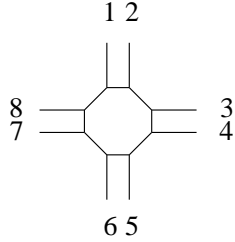}}
  \caption{$(p,q)$ web for Model I of $dP_5$.}
  \label{dP5_1}
\end{figure}

The web diagram corresponding to this theory is shown in \fref{dP5_1}. From it, we see that the
fields that can take us to $dP_4$ by getting a non-zero VEV are

\beq
\begin{array}{cc}
\left. 
\begin{array}{lcl}
\begin{array}{c}
X_{13},X_{14},X_{23},X_{24},X_{35},X_{36},X_{45},X_{46} \\ 
X_{57},X_{58},X_{67},X_{68},X_{71},X_{72},X_{81},X_{82}  
\end{array} & \rightarrow & \mbox{Model I} 
\end{array} \right\} &
\rightarrow 16
\end{array}
\eeq

\subsubsection*{Model II}

\begin{figure}[ht]
  \epsfxsize = 3cm
  \centerline{\epsfbox{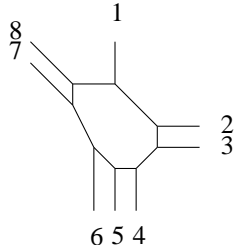}}
  \caption{$(p,q)$ web for Model II of $dP_5$.}
  \label{dP5_2}
\end{figure}

The web in this case is shown in \fref{dP5_2}.

The fields that higgs the theory down to $dP_4$ are

\beq
\begin{array}{cc}
\left. 
\begin{array}{lcl}
X_{24},X_{25},X_{26},X_{34},X_{35},X_{36},X_{47},X_{57},X_{67},X_{48},X_{58},X_{68}  & \rightarrow & \mbox{Model I} \\
X_{12},X_{13},X_{71},X_{81}                                                          & \rightarrow & \mbox{Model II} 
\end{array} \right\} &
\rightarrow 16
\end{array}
\eeq

\subsubsection*{Model III}

\begin{figure}[ht]
  \epsfxsize = 3cm
  \centerline{\epsfbox{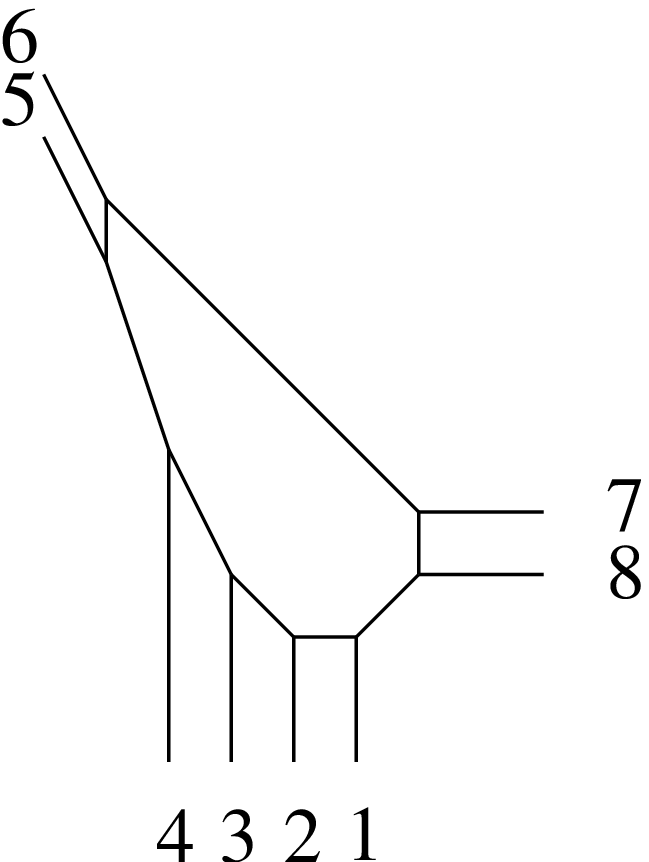}}
  \caption{$(p,q)$ web for Model III of $dP_5$.}
  \label{dP5_3}
\end{figure}

From the $(p,q)$ web in \fref{dP5_3}, we see that the following fields higgs the theory down to $dP_4$

\beq
\begin{array}{cc}
\left. 
\begin{array}{lcl}
\begin{array}{c}
X_{71},X_{72},X_{73},X_{74},X_{81},X_{82},X_{83},X_{84} \\ 
X_{15},X_{25},X_{35},X_{45},X_{16},X_{26},X_{36},X_{46}  
\end{array} & \rightarrow & \mbox{Model II} 
\end{array} \right\} &
\rightarrow 16
\end{array}
\eeq

\subsection{Del Pezzo 6}

\begin{figure}[ht]
  \epsfxsize = 3cm
  \centerline{\epsfbox{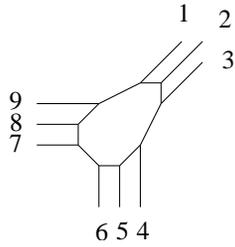}}
  \caption{$(p,q)$ web for $dP_6$.}
  \label{dP6}
\end{figure}

Finally, we present the web for $dP_6$ in \fref{dP6}, from where we read the following higgsings

\beq
\begin{array}{cc}
\left. 
\begin{array}{lcl}
\begin{array}{c}
X_{14},X_{15},X_{16},X_{24},X_{25},X_{26},X_{34},X_{35},X_{36} \\ 
X_{47},X_{48},X_{49},X_{57},X_{58},X_{59},X_{67},X_{68},X_{59} \\ 
X_{71},X_{72},X_{73},X_{81},X_{82},X_{83},X_{91},X_{92},X_{93} 
\end{array} & \rightarrow & \mbox{Model II} 
\end{array} \right\} &
\rightarrow 27
\end{array}
\eeq


\subsection{Higgsing global symmetry groups}

\label{subsection_higgsing_groups}

The results above show that, once the we classify quiver theories using their global symmetry groups,
the identification of higgsings that correspond to blow-downs of the geometry becomes straightforward. In
particular, we have seen for each $dP_n$ quiver theories that turning on a non-zero VEV for any component 
field of the fundamental representation reduces the global symmetry from $E_n$ to 
$E_{n-1}$ and produces a toric phase of $dP_{n-1}$. 

We will now present a comprehensive discussion on the group theory considerations that lead to the 
choice of the appropriate bifundamental fields that acquire a non-zero VEV. We will also present a  
description of how the blow-down process corresponds to a Higgs mechanism for the relevant gauge
groups, both from four and five dimensional perspectives.

In the previous section, we established which fields produce the desired higgsing with the aid of $(p,q)$ webs, and
discovered that in all cases they form the fundamental representation of the corresponding $E_n$ group. 
In fact, it is possible to determine which representation to choose for higgsing using solely group theoretic 
considerations. Generically, more than one representation are present in a given quiver theory. The representations 
that appear are always basic, meaning that their highest weight vectors are simple roots of the algebra. The clue to 
the right representation is provided by the Dynkin diagrams for the $E_n$ Lie algebras. The higgsing, and the corresponding 
enhanced symmetry breaking $E_n\rightarrow E_{n-1}$ can be depicted as the removal of a certain node from the $E_n$ Dynkin 
diagram. The representation that must be used for the higgsing is the one corresponding to the removed node, in the
sense that its highest weight vector is equal to the root corresponding to this node. Practically this means that 
the higgsed representation must contain a matter field with only one non-zero $U(1)$ charge in the same 
position as the removed node in the Dynkin diagram\footnote{The case $E_3\rightarrow E_2$ is 
slightly special, in that we have to remove two nodes from the (disconnected) Dynkin diagram. An obvious 
generalization of the described procedure shows that the representation we have to higgs is the (2,3).}.
The numbering of the nodes in the Dynkin diagram is unique in the basis of $U(1)$'s that we have chosen, 
leaving no space for ambiguity.

Let us now discuss the blow-down as a Higgs mechanism in four dimensions. As explained in previous sections, we go down 
from $dP_n$ to $dP_{n-1}$ by blowing down one of the exceptional divisors $E_i$ ($i=1 \ldots n-1$) of the del Pezzo. In 
fact, by acting on the $E_i$'s with the Weyl group of $E_n$, other possible choices of cycles to blow down are generated. 
As we have seen, all these divisors form the fundamental representation of $E_n$ and give precisely the divisors 
$L_{\alpha \beta}$ of the bifundamentals $X_{\alpha \beta}$ that can be used to appropriately higgs the gauge theory to one for 
$dP_{n-1}$. The bifundamental fields $X_{\alpha \beta}$ transform in the $(\bar{N},N)$ representation of the 
four dimensional gauge groups $U(N)_{(\alpha)} \times U(N)_{(\beta)}$ of the nodes they connect. Blowing down 
$L_{\alpha \beta}$ corresponds to $X_{\alpha \beta}$ getting a VEV proportional to the $N \times N$ identity matrix, higgsing 
$U(N)_{(\alpha)} \times U(N)_{(\beta)}$ to the diagonal subgroup. The non-zero VEV introduces a scale in the otherwise conformal 
field theory, and the new quiver will correspond to a new fixed point at the IR limit of the renormalization group flow.

On the other hand, bifundamental fields are also charged under the baryonic $U(1)$ global symmetries, which 
correspond to gauge symmetries in $AdS_5$. Then, the blow-down can also be interpreted as Higgs mechanism of a 
different gauge group, this time in five dimensions. The original five dimensional gauge group is in this case 
$U(1)^n$ and is higgsed down to $U(1)^{(n-1)}$. The simplest case corresponds to $X_{\alpha \beta}$ having
charge $q_1$ under a factor $U(1)_{(1)}$ and $q_2$ under $U(1)_{(2)}$. Then, the VEV of $X_{\alpha \beta}$ higgses
$U(1)_{(1)} \times U(1)_{(2)}$ to $q_2 U(1)_{(1)} - q_1 U(1)_{(2)}$, while the orthogonal combination becomes massive. The general 
case, in which the bifundamental field is charged under more than two $U(1)$'s is analogous, and simply amounts to 
a different combination of the original $U(1)'s$ becoming massive. 

Let us conclude this section with a few words describing how the enhanced $E_n$ global symmetry group of $dP_n$ is related 
to the $E_{n-1}$ group of $dP_{n-1}$. BPS states correspond to certain limits of D3-branes wrapping 3-cycles with 
$S^2 \times S^1$ topology in the non-spherical horizon $H_5$. In particular, these wrapped branes give rise to the charged
gauge bosons in $AdS_5$ that, together with the generators in the Cartan subalgebra that come from the reduction over 
3-cycles of the RR $C_4$, generate an $E_n$ gauge symmetry on $AdS_5$. The masses of the five dimensional $W$ bosons, as
well as the four dimensional gauge couplings, are K\"ahler moduli. The $E_n$ enhancement only appears at infinite
gauge coupling. Finite couplings produce an adjoint higgsing of $E_n$ down to $U(1)^n$, giving mass to the $W$ 
bosons. We have discussed above how $U(1)^n$ is connected to $U(1)^{n-1}$ by higgsing with a bifundamental field in the 
quiver. Then, we see that the global symmetry groups of $dP_n$ and $dP_{n-1}$ at finite and infinite 
coupling are connected as follows

\beq
\begin{array}{ccccc}
 &     & \begin{array}{c} \mbox{\footnotesize{bifundamental}} \\ \mbox{\footnotesize{higgsing}} \end{array} & & \\
 & E_n & \longrightarrow & E_{n-1} & \\
\begin{array}{c} \mbox{\footnotesize{finite coupling}} \\ \mbox{\footnotesize{(adjoint higgsing)}} \end{array}
 & \downarrow &  & \downarrow & 
\begin{array}{c} \mbox{\footnotesize{finite coupling}} \\ \mbox{\footnotesize{(adjoint higgsing)}} \end{array} \\
 & U(1)^n & \longrightarrow & U(1)^{n-1} &  \\
 &        & \mbox{\footnotesize{bifundamental}} & & \\
 &        & \mbox{\footnotesize{higgsing}} & & 
\end{array}
\eeq
where we have included the possibility of connecting the exceptional groups at infinite coupling directly.


\section{Global symmetries and Seiberg duality}

\label{section_symmetries_seiberg}

Section 3 explained how to determine the divisors associated to bifundamental fields when
blowing-up or blowing-down 2-cycles or when performing a Seiberg duality. We also saw there 
how the intersections with the divisors generating the $U(1)$ flavor symmetries determine
the $E_n$ Dynkin labels for each bifundamental field. Therefore, given the representation structure
of an original gauge theory, it is possible to determine how a Seiberg dual quiver 
is organized into $E_n$ representations, by carefully following this algorithm.  

The purpose of this section is to give a straightforward alternative procedure to determine the transformation properties 
under global symmetries of fields in a Seiberg dual theory. It consists of three simple rules, which have their origin in 
how $E_n$ symmetries are realized in the quivers, and also admits a geometric interpretation. Before going on, it is 
important to remind the reader the key fact that the bifundamental fields transforming in an irreducible 
representation of the $E_n$ global symmetry group do not necessarily have the same gauge quantum numbers (i.e. they 
can be charged under different pairs of gauge groups).

The three steps to deduce the representation structure of a theory based on that of a Seiberg dual are:

\begin{itemize}

\item
{\bf \underline{Step 1}:} Fields that are neutral under the dualized gauge group remain in the same representations. 
Some places in those representations might be left empty by the fields (otherwise known as dual quarks) that are conjugated (their transformation 
properties are yet to be determined) and by fields that become massive. If the representation is such that it cannot
appear in partial form (i.e. it is not possible to combine it with other representation or with itself to form
a quadratic invariant), these places will be completed either by meson fields or the conjugated ones (dual quarks).

\item
{\bf \underline{Step 2}:} Seiberg mesons appear in the product of the representations of the constituent fields. 
The precise representation is chosen from all the ones appearing in the product by requiring that those superpotential
terms that include the mesons are singlets under the global symmetry group. The geometric interpretation of this
step is that the divisor associated to mesons is given by the composition of the two divisors corresponding to
the component fields. As we studied above, the requirement that the superpotential terms are
$E_n$ singlets is translated to the geometric condition that the associated divisor lays in the canonical class.

\item
{\bf \underline{Step 3}:} The representations for the conjugated fields are determined by requiring that the cubic meson 
terms added to the superpotential are singlets of the global symmetry group. When doing so, it is very useful to choose these representations 
based on the entries that were left vacant at step {\bf (1)}, if this is possible. Once again, the geometric perspective
is that the sum of the divisors appearing in a superpotential term should be in the canonical class.

\end{itemize}

We will now use this technique to determine the global symmetries for all toric phases starting from
one of them, for del Pezzo surfaces from $dP_2$ to $dP_5$. As we will see, the results obtained this way are consistent 
with the Dynkin components assignations listed in Section \ref{section_symmetries}, obtained using the geometric 
prescription.

\subsection{Del Pezzo 2}

\subsection*{Model II of $dP_2$}

Let us start from Model I and dualize on node 1. In this case, we will have

\beq
\begin{array}{ccc}
X_{41} \rightarrow X_{14} & \ \ \ \ \ & X_{51} \rightarrow X_{15} \\
X_{13} \rightarrow X_{31} & \ \ \ \ \ & Y_{13} \rightarrow Y_{31} 
\end{array}
\eeq
and the following Seiberg mesons have to be added

\beq
\begin{array}{ccc}
M_{43}=X_{41} X_{13} & \ \ \ \ \ & \tilde{M}_{43}=X_{41} Y_{13} \\
M_{53}=X_{51} X_{13} & \ \ \ \ \ & \tilde{M}_{53}=X_{51} Y_{13} \\
\end{array}
\eeq

When dualizing, the fields which become massive, and have to be integrated out using
their equations of motion are

\beq
X_{34} \ \ M_{43} \ \  Y_{35} \ \ Z_{35} \ \ M_{53} \ \ \tilde{M}_{53}
\eeq
Next we demonstrate the three steps.

\bigskip

{\bf Step 1:} Fields that are neutral under the dualized gauge group remain invariant.

\beq
\begin{array}{ccc}
           & \ \ \ \ \ & SU(2) \times U(1) \\ 
(*,X_{42}) & & 2_3    \\
(X_{52},*) & & 2_{-1} \\
(*,X_{23}) & & 2_{-1} \\
(*,Y_{23}) & & 2_{-1} \\
X_{35}     & & 1_2    \\
X_{54}     & & 1_{-4} 
\end{array}
\eeq
where we have left an empty space for each conjugated or massive field.

\bigskip

{\bf Step 2:} Seiberg mesons are composite fields, and thus transform in an irreducible
representation in the product of their constituents. The right representation is chosen
from the existent terms in the superpotential

\beq
\begin{array} {ccccrclcr}
M_{43}&=&X_{41} X_{13}&=&2_3 \otimes 2_{-1}&=&(1 \oplus 3)_2 & \rightarrow & 1_2  \\ 
\tilde{M}_{43}&=&X_{41} Y_{13}&=&2_3 \otimes 2_{-1}&=&(1 \oplus 3)_2 & \rightarrow & 1_2 \\
M_{53}&=&X_{51} X_{13}&=&2_{-1} \otimes 2_{-1}&=&(1 \oplus 3)_{-2} & \rightarrow & 1_{-2} \\ 
\tilde{M}_{53}&=&X_{51} Y_{13}&=&2_{-1} \otimes 2_{-1}&=&(1 \oplus 3)_{-2} & \rightarrow & 1_{-2}
\end{array}
\eeq

\bigskip

{\bf Step 3:} We still have to determine the transformation properties
of $X_{14}$, $X_{15}$, $X_{31}$ and $Y_{31}$. This is done by requiring that the meson
terms added to the superpotential are invariant under the $SU(2) \times U(1)$ transformations  

\beq
M_{43} X_{31} X_{14}+\tilde{M}_{43} Y_{31} X_{14}+M_{53} X_{31} X_{15}+\tilde{M}_{53} Y_{31} X_{15}
\eeq
from were we conclude that the fields transform according to

\beq
\begin{array}{lclclcl}
X_{14} & \in & 2_{-1} & \ \ \ \ \ & X_{15} & \in & 2_3 \\
X_{31} & \in & 2_{-1} & \ \ \ \ \ & Y_{31} & \in & 2_{-1}
\end{array}
\eeq
resulting in the following arrangement of the bifundamental chiral fields

\beq
\begin{array}{ccc}
           & \ \ \ \ \ & SU(2) \times U(1) \\ 
(X_{15},X_{42}) & & 2_3 \\
(X_{52},X_{14}) & & 2_{-1} \\
(X_{31},X_{23}) & & 2_{-1} \\
(Y_{31},Y_{23}) & & 2_{-1} \\
X_{35} & & 1_2 \\
\tilde{M}_{43} & & 1_2 \\
X_{54} & & 1_{-4} 
\end{array}
\eeq

After the renaming of nodes indicated by the quiver diagram $(1,2,3,4,5) \rightarrow (2,5,1,4,3)$ we see
that we recover the results in \eref{symmetries_dP2_2}, derived using geometric methods.

\subsection{Del Pezzo 3}

Let us repeat the program we used for $dP_2$ in the case of $dP_3$, deriving the symmetry properties of Models II, III
and IV from Model I. 

\subsection*{Model II of $dP_3$}

Model II is obtained from Model I by dualizing node 1. 
We get

\beq
\begin{array}{ccc}
X_{12} \rightarrow X_{21} & \ \ \ \ \ & X_{13} \rightarrow X_{31} \\
X_{51} \rightarrow X_{15} & \ \ \ \ \ & X_{61} \rightarrow X_{16} 
\end{array}
\eeq
and the following mesons are added

\beq
\begin{array}{ccc}
M_{52}=X_{51} X_{12} & \ \ \ \ \ & M_{53}=X_{51} X_{13} \\
M_{62}=X_{61} X_{12} & \ \ \ \ \ & M_{63}=X_{61} X_{13} 
\end{array}
\eeq

Proceeding as before

\bigskip

{\bf Step 1:}

\beq
\begin{array}{ccc}
           & \ \ \ \ \ & SU(2) \times SU(3) \\
(*,X_{23},X_{34},X_{45},X_{56},*) & & (2,3) \\
(X_{24},X_{46},X_{62}) & & (1,\bar{3})
\end{array}
\eeq

One of the $({\bf 1},{\bf \bar{3}})$ representations has completely disappeared, while the other one stays unchanged. Note that $M_{53}$ and $X_{35}$ become massive and are in complex conjugate representations.

\bigskip

{\bf Step 2:}

\beq
\begin{array} {ccccrcrcr}
M_{52}&=&X_{51} X_{12}&=&(1,\bar{3}) \otimes (2,3)&=&(2,1 \oplus 8) & \rightarrow & (2,1) \\ 
M_{53}&=&X_{51} X_{13}&=&(1,\bar{3}) \otimes (1,\bar{3})&=&(1,3 \oplus \bar{6}) & \rightarrow & (1,3) \\ 
M_{62}&=&X_{61} X_{12}&=&(2,3) \otimes (2,3)&=&(1 \oplus 3,\bar{3} \oplus 6) & \rightarrow & (1,\bar{3})  \\ 
M_{63}&=&X_{61} X_{13}&=&(2,3) \otimes (1,\bar{3})&=&(2,1 \oplus 8) & \rightarrow & (2,1)
\end{array}
\eeq

\bigskip

{\bf Step 3:}

In order to determine the transformation properties of $X_{21}$, $X_{31}$, $X_{15}$ and $X_{16}$, we study
the meson terms in the superpotential

\beq
M_{52} X_{21} X_{15}+M_{53} X_{31} X_{15}+M_{62} X_{21} X_{16}+M_{63} X_{31} X_{16}
\eeq

Thus, we conclude that

\beq
\begin{array}{lcl}
X_{21} \in (1,\bar{3}) & \ \ \ \ \ & X_{31} \in (2,3) \\
X_{15} \in (2,3)& \ \ \ \ \ & X_{16} \in (1,\bar{3})
\end{array}
\eeq

Putting all these results together

\beq
\begin{array}{ccc}
           & \ \ \ \ \ & SU(2) \times SU(3) \\
(X_{31},X_{23},X_{34},X_{45},X_{56},X_{15}) & & (2,3) \\
(X_{24},X_{46},X_{62}) & & (1,\bar{3}) \\
(M_{62},X_{21},X_{16}) & & (1,\bar{3}) \\
(M_{52},M_{63}) & & (2,1) 
\end{array}
\eeq
which becomes \eref{symmetries_dP3_2} after relabeling the gauge groups according to $(1,2,3,4,5,6) \rightarrow (4,6,5,3,2,1)$.

\subsection*{Model III of $dP_3$}

Model III is obtained by dualizing node 5 of Model II. The dual quarks are

\beq
\begin{array}{ccc}
X_{15} \rightarrow X_{51} & \ \ \ \ \ & X_{65} \rightarrow X_{56} \\
X_{54} \rightarrow X_{45} & \ \ \ \ \ & X_{53} \rightarrow X_{35}
\end{array}
\eeq

The Seiberg mesons are

\beq
\begin{array}{ccc}
M_{13}=X_{15} X_{53} & \ \ \ \ \ & M_{14}=X_{15} X_{54} \\
M_{63}=X_{65} X_{53} & \ \ \ \ \ & M_{64}=X_{65} X_{54} 
\end{array}
\eeq

The following fields become massive and are integrated out: $X_{31}$, $M_{13}$, $X_{41}$ and $M_{14}$. Let us now
apply our set of rules.

\bigskip

{\bf Step 1:}

\beq
\begin{array}{ccc}
           & \ \ \ \ \ & SU(2) \times SU(3) \\
(*,*,*,X_{32},X_{21},X_{42}) & & (2,3) \\
(X_{63},*,X_{16}) & & (1,\bar{3}) \\
(X_{64},*,Y_{16}) & & (1,\bar{3}) \\
(X_{26},*) & & (2,1) 
\end{array}
\eeq

\bigskip

{\bf Step 2:}

\beq
\begin{array} {ccccrcrcr}
M_{13}&=&X_{15} X_{53}&=&(2,1) \otimes (2,3)&=&(1 \oplus 3,3) & \rightarrow & (1,3) \\ 
M_{14}&=&X_{15} X_{54}&=&(2,1) \otimes (2,3)&=&(1 \oplus 3,3) & \rightarrow & (1,3) \\ 
M_{63}&=&X_{65} X_{53}&=&(2,3) \otimes (2,3)&=&(1 \oplus 3,\bar{3} \oplus 6) & \rightarrow & (1,\bar{3})  \\ 
M_{64}&=&X_{65} X_{54}&=&(2,3) \otimes (2,3)&=&(1 \oplus 3,\bar{3} \oplus 6) & \rightarrow & (1,\bar{3})
\end{array}
\eeq

\bigskip

{\bf Step 3:}

We now determine the representations for $X_{51}$, $X_{56}$, $X_{35}$ and $X_{45}$. The new
terms in the superpotential are

\beq
M_{13} X_{35} X_{51}+M_{14} X_{45} X_{51}+M_{63} X_{35} X_{56}+M_{64} X_{45} X_{56}
\eeq

Then,

\beq
\begin{array}{lcl}
X_{51} \in (2,3) & \ \ \ \ \ & X_{56} \in (2,1) \\
X_{35} \in (2,3) & \ \ \ \ \ & X_{45} \in (2,3)
\end{array}
\eeq

Putting all the fields together we have

\beq
\begin{array}{ccc}
           & \ \ \ \ \ & SU(2) \times SU(3) \\
(X_{51},X_{35},X_{45},X_{32},X_{21},X_{42}) & & (2,3) \\
(X_{63},M_{64},X_{16}) & & (1,\bar{3}) \\
(M_{63},X_{64},Y_{16}) & & (1,\bar{3}) \\
(X_{26},X_{56}) & & (2,1) 
\end{array}
\eeq
which becomes \eref{symmetries_dP3_3} after renaming the gauge groups according to $(1,2,3,4,5,6)\rightarrow (6,2,3,1,4,5)$.

\bigskip

\subsection*{Model IV of $dP_3$}

Model IV can be obtained for example by dualizing node 6 of Model III. Then,

\beq
\begin{array}{ccc}
X_{16} \rightarrow X_{61} & \ \ \ \ \ & Y_{16} \rightarrow Y_{61} \\
X_{64} \rightarrow X_{46} & \ \ \ \ \ & X_{65} \rightarrow X_{56} 
\end{array}
\eeq

The following mesons are added

\beq
\begin{array}{ccc}
M_{14}=X_{16} X_{64} & \ \ \ \ \ & \tilde{M}_{14}=Y_{16} X_{64} \\
M_{15}=X_{16} X_{65} & \ \ \ \ \ & \tilde{M}_{15}=Y_{16} X_{65} 
\end{array}
\eeq

There are no fields that become massive in this case.

\bigskip

{\bf Step 1:}

\beq
\begin{array}{ccc}
          & \ \ \ \ \ & SU(2) \times SU(3) \\
(X_{42},*,X_{43},X_{53},*,X_{52}) & & (2,3) \\
(X_{21},*,X_{31}) & & (1,\bar{3}) \\
(Y_{21},*,Y_{31}) & & (1,\bar{3}) \\
(X_{15},X_{14}) & & (2,1) 
\end{array}
\eeq

\bigskip

{\bf Step 2:} 

Meson fields transform according to

\beq
\begin{array} {ccccrcrcr}
M_{14}&=&X_{16} X_{64}&=&(1,\bar{3}) \otimes (2,3)&=&(2,1 \oplus 8) & \rightarrow & (2,1) \\ 
\tilde{M}_{14}&=&Y_{16} X_{64}&=&(1,\bar{3}) \otimes (2,3)&=&(2,1 \oplus 8) & \rightarrow & (2,1) \\ 
M_{15}&=&X_{16} X_{65}&=&(1,\bar{3}) \otimes (2,3)&=&(2,1 \oplus 8) & \rightarrow & (2,1) \\ 
\tilde{M}_{15}&=&Y_{16} X_{65}&=&(1,\bar{3}) \otimes (2,3)&=&(2,1 \oplus 8) & \rightarrow & (2,1) 
\end{array}
\eeq

\bigskip

{\bf Step 3:}

We determine the representations for $X_{46}$, $X_{56}$, $X_{61}$ and $Y_{61}$ by requiring
the meson superpotential terms to be invariant

\beq
M_{14} X_{46} X_{61}+\tilde{M}_{14} X_{46} Y_{61}+M_{15} X_{56} X_{61}+\tilde{M}_{15} X_{56} Y_{61}
\eeq

And we see that

\beq
\begin{array}{lcl}
X_{46} \in (2,3) & \ \ \ \ \ & X_{56} \in (2,3) \\
X_{61} \in (1,\bar{3}) & \ \ \ \ \ & Y_{61} \in (1,\bar{3})
\end{array}
\eeq
leading to

\beq
\begin{array}{ccc}
          & \ \ \ \ \ & SU(2) \times SU(3) \\
(X_{42},X_{46},X_{43},X_{53},X_{56},X_{52}) & & (2,3) \\
(X_{21},X_{61},X_{31}) & & (1,\bar{3}) \\
(Y_{21},Y_{61},Y_{31}) & & (1,\bar{3}) \\
(X_{15},X_{14}) & & (2,1) \\
(M_{15},M_{14}) & & (2,1) \\
(\tilde{M}_{15},\tilde{M}_{14}) & & (2,1) 
\end{array}
\eeq
that reduces to \eref{symmetries_dP3_4} by renaming $(1,2,3,4,5,6) \rightarrow (6,1,3,5,4,2)$.

\subsection{Del Pezzo 4}

We will now derive the global symmetry properties of Model II of $dP_4$ from those of Model I.

\subsection*{Model II of $dP_4$}

The preceding examples show in detail how to operate with the rules in Section \ref{section_symmetries_seiberg}
and classify the matter content of dual theories according to their global symmetry properties. We will now
move on and apply our program to Model II of $dP_4$. This example is of particular interest because,
as we mentioned in Sections \ref{section_symmetries} and \ref{section_partial_representations}, it is the first 
one to exhibit {\it partial representations}. 
  
We obtain Model II by dualizing Model I on node 7. As usual, those bifundamental fields that are
charged under the dualized gauge group reverse their orientation 

\beq
\begin{array}{ccc}
X_{71} \rightarrow X_{17} & \ \ \ \ \ & X_{72} \rightarrow X_{27} \\
X_{57} \rightarrow X_{75} & \ \ \ \ \ & X_{67} \rightarrow X_{76} 
\end{array}
\eeq

The following meson fields have to be incorporated

\beq
\begin{array}{ccc}
M_{51}=X_{57} X_{71} & \ \ \ \ \ & M_{52}=X_{57} X_{72} \\
M_{61}=X_{67} X_{71} & \ \ \ \ \ & M_{62}=X_{67} X_{72}
\end{array}
\eeq

There are no fields in this theory that become massive, so we end up in a theory with 19 fields.
At this point, we see the first indications that this model is rather peculiar since, as long
as singlets are not used, it seems impossible to arrange these 19 fields into a combination 
of $SU(5)$ representations. Let us apply the three step program as before.  

\bigskip

{\bf Step 1:}

\beq
\begin{array}{ccc}
 \ \ \ \ \ \mbox{Fields} \ \ \ \ \ & \ \ \ \ \ SU(5) \ \ \ \ \ \\
(X_{45},X_{23},X_{46},*,X_{36},X_{24},*,X_{35},X_{12},*) & 10 \\
(X_{51},*,X_{61},X_{13},X_{14}) & \bar{5}
\end{array}
\eeq

\bigskip

{\bf Step 2:}

\beq
\begin{array} {ccccrcrcr}
M_{51}&=&X_{57} X_{71}&=&10 \otimes 10     &=& \bar{5} \oplus \bar{45} \oplus \bar{50} & \rightarrow & \bar{5} \\ 
M_{52}&=&X_{57} X_{72}&=&10 \otimes \bar{5}&=& 5 \oplus 45 & \rightarrow & 5 \\ 
M_{61}&=&X_{67} X_{71}&=&10 \otimes 10     &=& \bar{5} \oplus \bar{45} \oplus \bar{50} & \rightarrow & \bar{5} \\
M_{62}&=&X_{67} X_{72}&=&10 \otimes \bar{5}&=& 5 \oplus 45 & \rightarrow & 5 
\end{array}
\eeq

Both the ${\bf 5}$ and ${\bf \bar{5}}$ representations will be partially filled with two fields each. The same
number of fields are missing in both representations, as explained in 
Section \ref{section_partial_representations}.

\bigskip

{\bf Step 3:}

Looking at the superpotential terms 

\beq
M_{51} X_{17} X_{75}+M_{52} X_{27} X_{75}+M_{61} X_{17} X_{76}+M_{62} X_{27} X_{76}
\eeq
we see that

\beq
\begin{array}{lcl}
X_{27} \in 10      & \ \ \ \ \ & X_{17} \in \bar{5} \\
X_{75} \in \bar{5} & \ \ \ \ \ & X_{76} \in 10
\end{array}
\eeq
and we obtain \eref{symmetries_dP4_2}. We thus see that partial representations appear naturally when we
study the transformation of theories under Seiberg duality.

\subsection{Del Pezzo 5}

We will obtain in this section the global symmetry structure of Models II and III of $dP_5$
from Model I.

\subsection*{Model II of $dP_5$}

Model II is obtained by dualizing Model I on node 2.

\beq
\begin{array}{ccc}
X_{23} \rightarrow X_{32} & \ \ \ \ \ & X_{24} \rightarrow X_{42} \\
X_{72} \rightarrow X_{27} & \ \ \ \ \ & X_{82} \rightarrow X_{28} 
\end{array}
\eeq

The following mesons appear

\beq
\begin{array}{ccc}
M_{73}=X_{72} X_{23} & \ \ \ \ \ & M_{74}=X_{72} X_{24} \\
M_{83}=X_{82} X_{23} & \ \ \ \ \ & M_{84}=X_{82} X_{84} 
\end{array}
\eeq

No fields become massive.

\bigskip

{\bf Step 1:}

\beq
\begin{array}{ccc}
          & \ \ \ \ \ & SO(10) \\
\begin{array}{l}(X_{81},X_{57},*,X_{36},X_{46},*,X_{58},X_{13}, \\
                \ X_{14},X_{71},X_{45},X_{35},*,*,X_{67},X_{68}) 
\end{array}& & 16 
\end{array}
\eeq

\bigskip

{\bf Step 2:}

\beq
\begin{array} {ccccrcrcr}
M_{73}&=&X_{72} X_{23}&=&16 \otimes 16&=& 10 \oplus 120 \oplus 126 & \rightarrow & 10 \\ 
M_{74}&=&X_{72} X_{24}&=&16 \otimes 16&=& 10 \oplus 120 \oplus 126 & \rightarrow & 10 \\ 
M_{83}&=&X_{82} X_{23}&=&16 \otimes 16&=& 10 \oplus 120 \oplus 126 & \rightarrow & 10 \\ 
M_{84}&=&X_{82} X_{24}&=&16 \otimes 16&=& 10 \oplus 120 \oplus 126 & \rightarrow & 10 
\end{array}
\eeq

\bigskip

{\bf Step 3:}

The representations for $X_{32}$, $X_{42}$, $X_{27}$ and $X_{28}$ are determined by considering
the superpotential terms

\beq
M_{73} X_{32} X_{27}+M_{74} X_{42} X_{27}+M_{83} X_{32} X_{28}+M_{84} X_{42} X_{28}
\eeq

Thus,

\beq
X_{32}, X_{42}, X_{27}, X_{28} \in 16
\eeq
and we conclude that the matter is arranged as in \eref{symmetries_dP5_2}.

\subsection*{Model III of $dP_5$}

Dualizing Model II on node 1 we get Model III.

\beq
\begin{array}{ccc}
X_{71} \rightarrow X_{17} & \ \ \ \ \ & X_{81} \rightarrow X_{18} \\
X_{12} \rightarrow X_{21} & \ \ \ \ \ & X_{13} \rightarrow X_{31} 
\end{array}
\eeq

The Seiberg mesons are

\beq
\begin{array}{ccc}
M_{72}=X_{71} X_{12} & \ \ \ \ \ & M_{82}=X_{81} X_{12} \\
M_{73}=X_{71} X_{13} & \ \ \ \ \ & M_{83}=X_{81} X_{13} 
\end{array}
\eeq

There are no massive fields.

\bigskip

{\bf Step 1:}

\beq
\begin{array}{ccc}
          & \ \ \ \ \ & SO(10) \\
\begin{array}{l}(*,X_{25},X_{24},X_{67},X_{68},X_{34},X_{35},*, \\
                \ *,*,X_{58},X_{57},X_{48},X_{47},X_{26},X_{36}) 
\end{array}& & 16 \\
(X_{73},X_{83},X_{72},X_{82},*,*,*,*,*,*) & & \mbox{partial } 10 
\end{array}
\eeq

\bigskip

{\bf Step 2:}

\beq
\begin{array} {ccccrcrcr}
M_{72}&=&X_{71} X_{12}&=&16 \otimes 16&=& 10 \oplus 120 \oplus 126 & \rightarrow & 10 \\ 
M_{82}&=&X_{81} X_{12}&=&16 \otimes 16&=& 10 \oplus 120 \oplus 126 & \rightarrow & 10 \\ 
M_{73}&=&X_{71} X_{13}&=&16 \otimes 16&=& 10 \oplus 120 \oplus 126 & \rightarrow & 10 \\ 
M_{83}&=&X_{81} X_{13}&=&16 \otimes 16&=& 10 \oplus 120 \oplus 126 & \rightarrow & 10 
\end{array}
\eeq

\bigskip

{\bf Step 3:}

Looking at the following superpotential terms, we determine the representations for
$X_{17}$, $X_{18}$, $X_{21}$ and $X_{31}$.

\beq
M_{72} X_{21} X_{17}+M_{82} X_{21} X_{18}+M_{73} X_{31} X_{17}+M_{83} X_{31} X_{18}
\eeq

Then,

\beq
X_{17}, X_{18}, X_{21}, X_{31} \in 16
\eeq
and we recover \eref{symmetries_dP5_3}.

\section{Dibaryon operators}

\label{section_dibaryon_operators}

Section \ref{section_dibaryon_counting} will be devoted to another application of our $E_n$ classification of quivers, the 
counting of dibaryon operators. These operators were introduced in Section \ref{section_En_and_geometry}, where we also 
discussed their AdS realization as D3-branes wrapping 3-cycles in $H_5$. Equation \eref{simple_dibaryons} presented a 
special type of dibaryons, formed by antisymmetrizing $N$ copies of a single kind of bifundamental field. Quiver theories 
admit a larger variety of dibaryons, which can be constructed from more complicated paths or subquivers \cite{Herzog:2003dj}. 
Therefore, generic quivers can have dibaryons that correspond to bifurcated paths. This situation complicates the direct application 
of our techniques. We will restrict ourselves to two types of quivers, in which computations are relatively simple.

\begin{figure}[ht]
  \epsfxsize = 6cm
  \centerline{\epsfbox{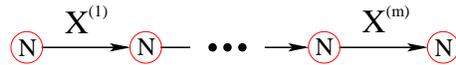}}
  \caption{Path associated to a dibaryon in a toric quiver.}
  \label{dibaryon_toric}
\end{figure}

The first class corresponds to the toric quivers we have studied so far. In these quivers, all gauge groups are identical. This
prevents bifurcations in the paths that represent dibaryons, leaving us with linear paths as the one shown in 
\fref{dibaryon_toric}

Dibaryons in these models take the form  

\beq
\epsilon_{\alpha_1 \cdots \alpha_N} \epsilon^{\beta_1 \cdots \beta_N} X^{\alpha_1 (1)}_{i_1} \cdots X^{\alpha_N (1)}_{i_N}
                    X^{i_1 (2)}_{j_1} \cdots X^{i_N (2)}_{j_N} \cdots X^{k_1 (m)}_{\beta_1} \cdots X^{k_N (m)}_{\beta_N}
\eeq

There are exactly $N$ copies of each $X^{(i)}$. In this sense, these dibaryons are analogous to the simple ones in 
\eref{simple_dibaryons} and, for the purposes of assigning divisors or representation under the Weyl group of $E_n$, 
we can consider just one representation for each of the arrows in \fref{dibaryon_toric}. 

\begin{figure}[ht]
  \epsfxsize = 6cm
  \centerline{\epsfbox{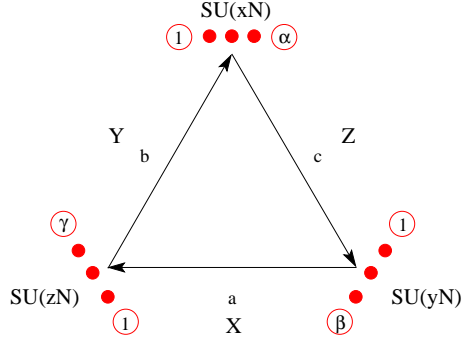}}
  \caption{Quiver diagram for a generic 3-block collection.}
  \label{quiver_3block}
\end{figure}

The second class of theories that we will consider are 3-block quivers. Some of the examples that we will study are non-toric quivers.  
They have the general structure shown in \fref{quiver_3block}, with no bifundamental fields connecting nodes in the same 
block.

These quivers are obtained as solutions of a Diophantine (Markov type) equation of the form \cite{Karpov:1}

\beq
\alpha x ^2+\beta y^2+\gamma z^2=\sqrt{K_n^2 \alpha \beta \gamma} x y z
\label{diophantine_3block}
\eeq
where $\alpha$, $\beta$ and $\gamma$ are the number of gauge groups in each of the three blocks, $K_n$ denotes the 
canonical class as before and $K_n^2=9-n$. The number of bifundamental fields between the nodes are computed as

\beq
\begin{array}{ccccc}
a=\alpha k x & \ \ \ \ & b=\beta k y & \ \ \ \ & c=\gamma k z
\end{array}
\label{anomaly_3block}
\eeq
with

\beq
k=\sqrt{{K_n^2 \over \alpha \beta \gamma}}=\sqrt{{9-n \over \alpha \beta \gamma}}.
\eeq

As a result, the arrow numbers satisfy the following Markov type equation.
\beq
\frac{a ^2}{\alpha}+\frac{b^2}{\beta}+\frac{c^2}{\gamma}= a b c
\label{markov_3block}
\eeq

This equation is a reduction of the general Diophantine equation which was derived for a generic del Pezzo quiver 
in \cite{Feng:2002kk} to the case in which the corresponding $(p,q)$ branes are grouped into 3 sets of parallel branes, 
of multiplicities $\alpha, \beta, \gamma,$ respectively. It is interesting to point out that equations 
\eref{diophantine_3block} and \eref{markov_3block} are derived by demanding that the quiver theory of 
\fref{quiver_3block} will be conformally invariant. Equation \eref{anomaly_3block} has the interpretation of the anomaly 
cancellation condition for the quiver theories. The positive integers $\alpha$, $\beta$ and $\gamma$ are solutions to
\beq
\alpha+\beta+\gamma=n+3,\qquad \alpha\beta\gamma(9-n)=(integer)^2.
\label{abc}
\eeq

In addition these numbers represent all maximal subgroups of $E_n$, of A type,
\beq
SU(\alpha)\times SU(\beta)\times SU(\gamma)\subset E_n,
\eeq
with the convention that $SU(1)$ is null. As an example, for $E_8$ there are 4 such triples: $(1,1,9)$, $(1,2,8)$, 
$(1,5,5)$, $(2,3,6)$.

\begin{figure}[ht]
  \epsfxsize = 4cm
  \centerline{\epsfbox{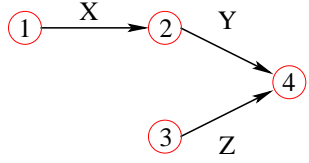}}
  \caption{An example of a possible path in a quiver associated to a dibaryon. Nodes $2$ and $3$ can actually be the same.}
  \label{path_dibaryon}
\end{figure}

We will study examples of non-toric 3-block quivers. An elegant discussion of the construction of dibaryons in general 
quivers can be found in \cite{Herzog:2003dj}. There, a vector space $V_i$ of dimension $d_i N$ is associated to each node, and 
every bifundamental field represents a linear map between its tail and its head. These maps can be multiplied 
to form linear maps $A$ for arbitrary paths in the quiver associated to a given dibaryon, forming an algebra. Generically, these 
paths can have bifurcations. We show an example in \fref{path_dibaryon}. 

A useful quantity that can be associated to a given path $A$ is its rank

\beq
r(A)=\sum_i N d_i (h_i-t_i) \ ,
\eeq
where $h_i$ and $d_i$ are respectively the number of arrow heads and tails at each node. The rank of $A$ counts the number
of uncontracted fundamental minus antifundamental $SU(d_i N)$ indices at each node. When $r(A)=0$, the antisymmetrized product
that gives rise to the dibaryon is simply given by the determinant of the linear map $A$ between the vector spaces of the path's 
tail and head.

We have discussed how for non-toric quivers the baryonic $U(1)$ charges are generically fractional, and cannot be interpreted as
Dynkin coefficients. We will see in Section \ref{section_dP7_and_dP8} how in these cases it is still possible to exploit the decomposition 
of the global symmetry into subgroups to efficiently organize the counting of dibaryons.





The operators defined above, both for toric and general 3-block quivers do not have definite transformation properties 
under the global symmetry group. In order to achieve that, we have to construct linear combinations of them using 
appropriate Clebsch-Gordon coefficients. The techniques described in the next section will help us to directly organize 
dibaryons of different dimensions into representations of the Weyl groups of $E_n$, based on the classification of 
bifundamental fields.

\section{Dibaryon counting}

\label{section_dibaryon_counting}

We have already seen how the classification into an $E_n$ symmetric language has proved useful in simplifying and 
organizing some problems in $dP_n$ quivers. It has reduced the computation of superpotentials to the construction of $E_n$ invariants and the 
determination of possible higgsings from $dP_n$ to $dP_{n-1}$ to the problem of higgsing the global
symmetry group by a non-zero VEV for a field in the fundamental representation. We will explore in this section
an additional application of this machinery, using it to count dibaryon operators in the gauge theories. Matching the 
counting performed in the gauge theory with the one on the gravity side is another check of the AdS/CFT correspondence.

\subsection{Geometric counting}

\label{subsection_geometric_counting}

The {\it geometric counting} of dibaryons corresponds to the enumeration of holomorphic curves in the del Pezzo surface under 
consideration. The R charge (equivalently the dimension $\Delta$) of the dibaryons that we want to count determines the degree of the 
associated curve $\mathcal{C}$ by

\beq
R_\mathcal{C}=k {2 N \over (9-n)}
\label{R_charge}
\eeq
which is analogous to the corresponding expression for bifundamental fields \eref{R-charge_formula}. As explained in equation \eref{equation_degree}, 
the degree $k$ is computed as minus the intersection with the canonical class $K_n$

\beq
k=-(K_n \cdot \mathcal{C})=-(3 H-\sum_i E_i) \cdot \mathcal{C}
\label{degree}
\eeq

The genus of $g$ of $\mathcal{C}$ is related to its degree and self-intersection by the adjunction formula \cite{Griffiths:1}

\beq
\mathcal{C} \cdot \mathcal{C}=k-2+2g
\label{adjunction}
\eeq

Then, in order to have genus greater or equal to zero, we have

\beq
\mathcal{C} \cdot \mathcal{C} \geq (k-2)
\eeq

For computational purposes, it is useful to rewrite this equation in terms of the numbers of D's and $E_i$'s

\beq
N_D^2-\sum_i N_{E_i}^2 \geq (k-2)
\label{genus}
\eeq

Equations \eref{R_charge} and \eref{genus} will be our main tools for the geometric counting.
It is important to note that the geometric method can be used not only to provide the multiplicity of states, but also 
the specific representation of the Weyl group of $E_n$ under which dibaryons transform. This is accomplished by computing their 
Dynkin components as the intersections of the corresponding curves with the $U(1)$ generators in \eref{simple_roots}. 
This information can also be contrasted against the results of the algebraic procedure, which is explained in the next 
section.



\subsection{Algebraic counting}

The counting of dibaryon operators in the gauge theory can be simplified by exploiting the global symmetry representations 
assigned to the fields in the quiver. We will refer to this approach as {\it algebraic counting}. Our discussion applies to 
the class of toric quivers. Similar ideas for non-toric 3-block quivers will be presented in Section \ref{section_dP7_and_dP8}. 
The starting point is to look for all possible multiplications of fields that give the appropriate $R$ charge of the dibaryons 
that we want to consider. In fact, there is no need to take into account all possible combinations, although it is useful to do 
so in order to check results. The simplicity of this method resides in that it is not necessary to be concerned about the details 
of how these fields are combined, the proper number of states will be the result of the group structure of the quiver. 

Following the discussion in the previous section, we consider one representation for each type of fields and multiply them, 
yielding for each combination a sum of candidate representations in which dibaryons can in principle transform. The appropriate
representation can be distinguished by supplementing the group theory discussion with a simple input consisting of an upper
bound on the number of dibaryons coming from inspection of the quiver (some representations are simply too large to correspond to
dibaryons formed with fields in the quiver). This representation will in general be generated by some of the alternative 
combinations considered.

In the following sections, we will perform the geometric counting of dibaryons on the gravity side and compare the results with 
the ones obtained using the algebraic procedure described above. We will do that for the specific cases of $dP_4$ and $dP_5$. 
These are interesting theories for various reasons. The quiver theories
in these cases are fairly non-trivial, with seven and eight gauge groups respectively under which bifundamental chiral multiplets
are charged in very different ways. Furthermore, for both geometries, it is possible to perform a cross check between the 
counting done on different dual gauge theories using the $E_n$ representation structure presented in Section 3. 
Finally, some of these models also contain partial representations, constituing a more subtle check of our 
techniques. 


\subsection{Dibaryons in del Pezzo 4}

\subsubsection*{Geometric counting}

Let us count the dibaryons of degree $k=2$ to $5$, by using equations \eref{degree} and \eref{genus}. Degree $1$ dibaryons
form the fundamental ${\bf 10}$ representation of $E_4$ and had been considered in \cite{Herzog:2003wt}. The results 
in this case are

\beq
\begin{array} [t] {|cc|cc|cc|cc|}
\hline
\mbox{k=2} & \ \ \ \ & \mbox{k=3} & \ \ \ \ & \mbox{k=4} & \ \ \ \ & \mbox{k=5} & \ \ \ \ \\
\hline 
                         &   &                             &   &                            &   &                             &    \\   
(1,\underline{-1,0,0,0}) & 4 & (1,0,0,0,0)                & 1 & (2,\underline{-1,-1,0,0})   & 6 & (2,\underline{-1,0,0,0})    & 4  \\
(2,-1,-1,-1,-1)          & 1 & (2,\underline{-1,-1,-1,0}) & 4 & (3,\underline{-2,-1,-1,-1}) & 4 & (3,\underline{-2,-1,-1,0})  & 12 \\
                         &   &                            &   &                             &   & (4,\underline{-2,-2,-2,-1}) & 4 \\
                         &   &                            &   &                             &   &                             &    \\
\hline
                         & {\bf \bar{5}} &                & {\bf 5} &                       & {\bf \bar{10}}  &               & {\bf 24}-4 \\
\hline
\end{array}
\label{geometric_table_dP4}
\eeq
where we have used the vector notation of \eref{vector_notation} to indicate divisors, and the convention is that all 
possible permutations of the underlined entries have to be considered. This notation will be used along the rest
of the paper. We list in this table the multiplicities of divisors and the representation in which
the curves of each degree, and hence the dibaryons, transform.

The dibaryons at $k=5$ form a ${\bf 20}$ Weyl orbit of $E_4$. We have indicated it as ${\bf 24}-4$ in
\eref{geometric_table_dP4} to make contact with the more familiar $E_4$ representations. We see that ${\bf 20}$
corresponds to the adjoint ${\bf 24}$ of $E_4$ minus the four Cartan generators, which correspond to dibaryons whose
associated divisor would be the canonical class of $dP_4$.

As explained in Section \ref{subsection_geometric_counting}, we have been able not only to count the multiplicity of 
states at each $k$, but also to determine the specific representations, by computing the Dynkin components that 
correspond to the divisors.



\subsubsection*{Algebraic counting}

The quiver diagrams for the two toric phases of $dP_4$ were presented in \fref{quiver_dP4_1} and \fref{quiver_dP4_2}. Let us 
now compute the number of dibaryons of each degree using the representation structure derived in Section 3.

The $R$ charges of the bifundamental fields are, for Model I,

\beq
\begin{array}{c|c}
\mbox{Representation} & \ \ \ \ R \ \ \ \ \\
\hline
10                    & 2/5 \\
\bar{5}               & 4/5
\end{array}
\eeq
and, for Model II,

\beq
\begin{array}{c|c}
\mbox{Representation} & \ \ \ \ R \ \ \ \ \\
\hline
10                    & 2/5 \\
\bar{5}_a               & 4/5 \\
\mbox{partial } \bar{5}_b       & 4/5 \\
\mbox{partial } 5               & 6/5 
\end{array}
\eeq

As we discussed in Section \ref{subsection_Weyl_group}, basic representations of $E_n$ coincide with Weyl orbits (with the 
exception of the adjoint). These are the representations in which
dibaryons of small degree are organized. For this reason, whenever we should multiply Weyl orbits in the coming sections, we will be in fact
computing products of $E_n$ representations. This is intended to simplfy the computations, since the products of $E_n$ representations 
can be found in standard references such as \cite{Slansky:yr}. This will generally be sufficient for our purposes in all the cases we will 
study, since we will only be interested in small representations emerging from such products. We hope the readers will not be mislead by this 
assumption and will keep it in mind.

\subsubsection*{k=2}

According to \eref{R_charge}, these states have $R=4N/5$. Let us start by considering Model I. Taking into account how the fields forming 
the representations above are distributed in the quiver, we see that we can construct these dibaryons in the following ways

\beq
\begin{array}{ll}
\left. \begin{array}{l}
10 \otimes 10=\underline{\bar{5}} \oplus \bar{45} \oplus \bar{50} \\ \\
\underline{\bar{5}}
\end{array} \right \} & \rightarrow \bar{5}
\end{array}
\label{algebraic_dP4_k2_1}
\eeq
It is immediate to discard the ${\bf \bar{45}}$ and ${\bf \bar{50}}$ representations because of their large dimensions.
The resulting one is then the ${\bf \bar{5}}$. We have underlined it in \eref{algebraic_dP4_k2_1} to indicate how it 
appears in both alternative ways of constructing these dibaryons.

We arrive at the same result when we look at how $R=4N/5$ dibaryons can be formed in Model II, a dual theory.

\beq
\begin{array}{ll}
\left. \begin{array}{l}
10 \otimes 10=\underline{\bar{5}} \oplus \bar{45} \oplus \bar{50} \\ \\
\underline{\bar{5}_a} \\ \\
\underline{\bar{5}_b}
\end{array} \right \} & \rightarrow \bar{5}
\end{array}
\eeq

\subsubsection*{k=3}

For $k=3$, we have $R=6N/5$. Then the possible ways of constructing these dibaryons are

\beq
\begin{array}{ll}
\left. \begin{array}{l}
10 \otimes 10 \otimes 10=\underline{5} \oplus \bar{5} \oplus 2 \ 45 \oplus \bar{45} \oplus 50 \oplus 70 \oplus 157'' \oplus 2 \ 280 \\ \\
10 \otimes \bar{5}=\underline{5} \oplus 45
\end{array} \right \} & \rightarrow 5
\end{array}
\eeq

By comparing the two possible constructions, one can conclude that the representation will be either a ${\bf 5}$, a 
${\bf 45}$ or a direct sum of both of them. Quick inspection of the quiver reveals that it is not possible to form at 
least 45 dibaryons, leaving us with the ${\bf 5}$ representation. The same result
can be obtained in Model II, where we have

\beq
\begin{array}{ll}
\left. \begin{array}{l}
10 \otimes \bar{5}_a=\underline{5} \oplus 45 \\ \\
10 \otimes \bar{5}_b=\underline{5} \oplus 45 \\ \\
\underline{5}
\end{array} \right \} & \rightarrow 5
\end{array}
\eeq

\subsubsection*{k=4}

We have now $R=8N/5$, and the following possibilities in Model I

\beq
\begin{array}{ll}
\left. \begin{array}{l}
10 \otimes 10 \otimes 10 \otimes 10 = \underline{\bar{10}} \oplus \bar{15} \oplus ... \\ \\
10 \otimes 10 \otimes \bar{5}=\underline{\bar{10}} \oplus \bar{15} \oplus ... \\ \\
\bar{5} \otimes \bar{5}=\underline{\bar{10}}+\bar{15}
\end{array} \right \} & \rightarrow \bar{10}
\end{array}
\eeq
where the dots indicate large representations that are not relevant for our discussion.
As in the previous case, both the ${\bf \bar{10}}$ and the ${\bf \bar{15}}$ seem possible, but the $\bar{15}$ is discarded 
because of its large multiplicity. Repeating the calculation for Model II, we get

\beq
\begin{array}{ll}
\left. \begin{array}{l}
10 \otimes 10 \otimes \bar{5}_a = \underline{\bar{10}} \oplus \bar{15} \oplus ... \\ \\
10 \otimes 10 \otimes \bar{5}_b = \underline{\bar{10}} \oplus \bar{15} \oplus ... \\ \\
\bar{5}_a \otimes \bar{5}_a=\underline{\bar{10}} \oplus \bar{15} \\ \\
\bar{5}_b \otimes \bar{5}_a=\underline{\bar{10}} \oplus \bar{15} \\ \\
10 \otimes 5=\underline{\bar{10}} \oplus \bar{40}
\end{array} \right \} & \rightarrow \bar{10}
\end{array}
\eeq
in coincidence with the computation in Model I.

\subsubsection*{k=5}

These dibaryons have $R=2N$. In Model I, we have the following alternatives to construct them

\beq
\begin{array}{ll}
\left. \begin{array}{l}
\bar{5} \otimes \bar{5} \otimes 10 = 1 \oplus \underline{24} \oplus 
\bar{
24
}
\oplus 2 \ 75 \oplus 126 \oplus 175' \\ \\
10 \otimes 10 \otimes 10 \otimes \bar{5}=1 \oplus 2 \ \underline{24} + \bar{24} \oplus ... \\ \\
10 \otimes 10 \otimes 10 \otimes 10 \otimes 10
\end{array} \right \} & \rightarrow 24
\end{array}
\eeq

Applying the same arguments as before, we obtain the ${\bf 24}$ representation. We now remember that we multiplied $E_n$ representations
for simplicity, but the objects we are actually interested in, are Weyl orbits. Then, we conclude that $k=5$ dibaryons form the ${\bf 20}$
Weyl orbit, which is obtained from the ${\bf 24}$ adjoint representation of $E_n$ by removing the Cartan generators. This is again in 
agreement with the geometric counting. We get the same result in Model II, by considering

\beq
\begin{array}{ll}
\left. \begin{array}{l}
10 \otimes 10 \otimes 5 = 1 \oplus 2 \ \underline{24} \oplus 75 \oplus 126 \\ \\
10 \otimes \bar{5}_a \otimes \bar{5}_a= 1 \oplus \underline{24} \oplus \bar{24} \oplus 2 \ 75 \oplus 126 \oplus 175'
\end{array} \right \} & \rightarrow 24
\end{array}
\eeq

\subsection{Dibaryons in del Pezzo 5}

There are three toric phases for $dP_5$, whose quivers are shown in Figures \ref{quiver_dP5_1}, \ref{quiver_dP5_2} and \ref{quiver_dP5_3}. 
We will perform the algebraic counting for Model I and Model II. Model III differs from Model II simply by the presence 
of an additional copy of a partial ${\bf 10}$ representation, and thus the counting works identically.

\subsubsection*{Geometric counting}

Degree 1 dibaryons were studied in \cite{Herzog:2003wt} and they form a ${\bf 16}$ of $E_5$. Based on equations 
\eref{degree} and \eref{genus}, the geometric procedure determines the following representations for dibaryons of degree 
2 to 4

\beq
\begin{array}{|cc|cc|cc|}
\hline
\mbox{k=2} & & \mbox{k=3} & & \mbox{k=4} & \\
\hline 
                              &    &                                &    &                                &    \\
(1,\underline{-1,0,0,0,0})    &  5 & (1,0,0,0,0,0)                  &  1 & (2,\underline{-1,-1,0,0,0})    & 10 \\
(2,\underline{-1,-1,-1,-1,0}) &  5 & (2,\underline{-1,-1,-1,0,0})   & 10 & (3,\underline{-2,-1,-1,-1,0})  & 20 \\
                              &    & (3,\underline{-2,-1,-1,-1,-1}) &  5 & (4,\underline{-2,-2,-2,-1,-1}) & 10 \\
                              &    &                                &    &                                &    \\
\hline
                              & {\bf 10} &                          & {\bf \bar{16}} &                    & {\bf 45}-5 \\
\hline
\end{array}
\eeq

\subsubsection*{Algebraic counting}

The quivers for Models I and II are presented in \fref{quiver_dP5_1} and \fref{quiver_dP5_2}. The R charge assignations are  

\beq
\begin{array}{c|c}
\mbox{Representation} & \ \ \ \ R \ \ \ \ \\
\hline
16                    & 1/2 
\end{array}
\eeq
for Model I, and

\beq
\begin{array}{c|c}
\mbox{Representation} & \ \ \ \ R \ \ \ \ \\
\hline
16                    & 1/2 \\
\mbox{partial } 10    &   1 
\end{array}
\eeq
for Model II.

\subsubsection*{k=2}

These dibaryons have $R=N$. In Model I, all possible combinations that produce them take the form

\beq
\left. 16 \otimes 16=\underline{10} \oplus 120 \oplus 126 \right\} \rightarrow 10
\eeq

The ${\bf 10}$ representation has been chosen because it is already clear from the quiver that is not possible to form 
enough dibaryons to fill the other representations

Moving to Model II, we get two possible ways of forming these dibaryons

\beq
\begin{array}{ll}
\left. \begin{array}{l}
16 \otimes 16=\underline{10} \oplus 120 \oplus 126 \\ \\
\underline{10}
\end{array} \right\} &
\rightarrow 10
\end{array}
\eeq

Reproducing the result obtained with Model I. Notice that although we have included all the possible combinations of 
bifundamental fields giving rise to $k=2$ dibaryons, the computation actually reduces in this case to noticing that it is possible 
to construct them by antisymmetrizing fields of a single kind (the ones with $R=1$) and thus the resulting dibaryons should fill the 
representation in which these fields transform, i.e. a ${\bf 10}$.

\subsubsection*{k=3}

The $k=3$ dibaryons have $R=3N/2$. Their construction in Model I corresponds to the combination

\beq
\left. 16 \otimes 16 \otimes 16=2 \ \underline{\bar{16}} \oplus 3 \ \bar{144} \oplus \bar{560} \oplus \bar{672} \oplus 2 \ \bar{1200} 
\right\} \rightarrow \bar{16}
\eeq

Proceeding as in previous sections by considering an upper bound in the number of dibaryons, we conclude that they form
a ${\bf \bar{16}}$ representation. We can arrive at the same conclusion by studying Model II, where

\beq
\begin{array}{ll}
\left. \begin{array}{l} 
10 \otimes 16=\underline{\bar{16}} \oplus \bar{144} \\ \\
16 \otimes 16 \otimes 16=2 \ \underline{\bar{16}} \oplus ... 
\end{array} \right\} &
\rightarrow \bar{16}
\end{array}
\eeq

\subsubsection*{k=4}

These operators have $R=2N$ and can be built in Model I in the following way

\beq
\left. 16 \otimes 16 \otimes 16 \otimes 16 =1 \oplus 6\ \underline{45} \oplus ... \right\} \rightarrow 45
\eeq

The extra terms in the product correspond extra representations that cannot correspond to dibaryons due to their
large dimensions, leaving us with the ${\bf 45}$. Once again, remember that the relevant objects are Weyl orbits, so we 
conclude that $k=4$ dibaryons form the ${\bf 40}$ Weyl orbit that results when removing the Cartan generators from the
adjoint ${\bf 45}$. This result agrees with the geometric counting. Moving to Model II, we get the same result.

\beq
\begin{array}{ll}
\left. \begin{array}{l}
10 \otimes 16 \otimes 16 =1 \oplus 2 \ \underline{45} \oplus ... \\ \\
16 \otimes 16 \otimes 16 \otimes 16 =1 \oplus 6\ \underline{45} \oplus ... 
\end{array} \right\} &
\rightarrow 45
\end{array}
\eeq

We have been able to observe a complete match between the geometric and algebraic enumeration of dibaryon operators.
The algebraic counting appears as a useful tool for organizing the counting. We will study in the next section a related
idea that applies to non-toric quivers.



\section{Del Pezzo 7 and 8}

\label{section_dP7_and_dP8}

The methods we have developed for classifying bifundamental fields in del Pezzo theories based on 
their relation to the $E_n$ Lie algebras cease to work in a simple way for models in which the gauge 
group factors have different ranks. The main reason was pointed out in Section \ref{section_En_and_geometry}, where we
noted that equations \eref{nodes_divisors} and \eref{R-charge_formula} predict in these theories fractional $U(1)$ charges 
that cannot be identified with Dynkin coefficients of $E_n$ representations.

Dibaryon operators correspond to the antisymmetrized product of a large number of bifundamental fields and 
do transform into $E_n$ representations. Moreover, we discussed in Section \ref{section_dibaryon_operators} 
how, for the special case of three block models, a maximal $SU(\alpha)\times SU(\beta)\times SU(\gamma)$ subgroup of $E_n$ becomes 
manifest. This subgroup will play a similar role to the one of $E_n$ in the algebraic counting of dibaryons in
3-block non-toric quivers. Bifundamental fields can be organized into representations of the maximal subgroup which
can be used to determine the representations of dibaryons. 

Contrary to what happens in toric quivers, the representations of the maximal subgroup in which dibaryons transform are not determined 
by straigthforward multiplication of represenations of constituent bifundamental fields, but result from the application of the following 
procedure. For each of the factors $SU(\alpha)\times SU(\beta)\times SU(\gamma)$, the appropriate representation is determined by the nodes 
that are contracted with antisymmetric tensors (i.e. nodes that are at heads or tails of the quiver path defining the dibaryon, 
and that are not contracted with fundamental or antifundamental indices of other bifundamental fields). We have to multiply one 
$SU(\alpha)$ representation for each of these `free nodes', and take the antisymmetric combination in the product. 
This is the main difference with traditional multiplication of representations, since the number of `free nodes' 
for each $SU(\alpha)$ can be different.


\subsection*{Del Pezzo 7}

\begin{figure}[ht]
  \epsfxsize = 6cm
  \centerline{\epsfbox{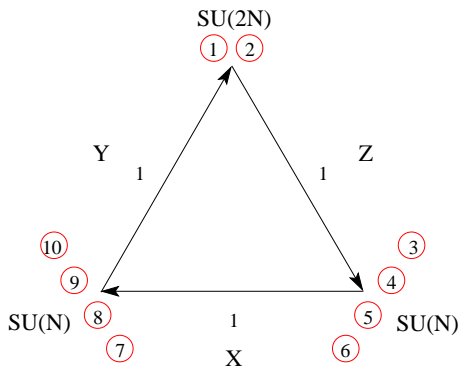}}
  \caption{One of the possible 3-block quivers for $dP_7$.}
  \label{quiver_dP7_2}
\end{figure}

In the case of $dP_7$, there are three minimal solutions to the Diophantine equation \eref{diophantine_3block}, from which all others can 
be generated through Seiberg dualities: $(\alpha,\beta,\gamma)=(1,1,8)$, $(2,4,4)$ and $(1,3,6)$ \cite{Karpov:1}. We will examine now the 
$(2,4,4)$ model. \fref{quiver_dP7_2} shows the quiver diagram for this theory.

The gauge groups for nodes in each of the three blocks are $U(2N)$, $U(N)$ and $U(N)$. The maximal subgroup of $E_7$ is in this case 
$SU(2)\times SU(4)\times SU(4)$, under which bifundamental fields have the following transformation properties

\beq
\begin{array}{lr}
X: & (1,4,\bar{4}) \\
Y: & (2,1,4)       \\
Z: & (2,\bar{4},1)
\end{array}
\eeq

Their R-charges are 
\beq
\begin{array}{ccc}
R_X=1 & \ \ \ \ \ &  R_Y=R_Z=1/2 .
\end{array}
\eeq

Before moving on to count dibaryons in this theory, let us consider in detail two explicit examples, in
order to gain familiarity with the application of the rules described above. The first example corresponds 
to the level one dibaryons constructed by antisymmetrizing  
$2N$ $Y$ fields. Suppressing the $N$ factor in the multiplicity, which will be common to all dibaryons, we will 
refer to them as $Y^2$ dibaryons.

\begin{figure}[ht]
  \epsfxsize = 4.5cm
  \centerline{\epsfbox{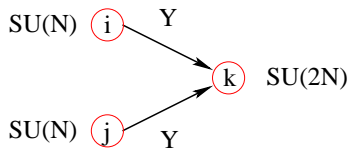}}
  \caption{Level one $Y^2$ dibaryons in the $(2,4,4)$ $dP_7$ 3-block model.}
  \label{path_dibaryon_dP7_1}
\end{figure}

In the figure, $k \in \{1,2 \}$ and $i,j \in \{7,8,9,10 \}$. The $Y$ fields transform trivially under the first $SU(4)$.
We have only one node in the $SU(2)$ block, and thus the $Y^2$ dibaryons will transform under it in the ${\bf 2}$ representation. 
Nodes $i$ and $j$ transform in the ${\bf 4}$ of the second $SU(4)$ and thus we need to take the antisymmetric combination in the product 
of two ${\bf 4}$'s, giving the ${\bf 6}$. In summary, $Y^2$ dibaryons form the $(2,1,6)$ representation of $SU(2) \times SU(4) \times SU(4)$. 

Let us now consider the level two dibaryons constructed from $2N$ $Z$ and $N$ $X$ fields. As before, we refer to them as $Z^2 X$. 
They correspond to paths in the quiver of the form

\begin{figure}[ht]
  \epsfxsize = 5.0cm
  \centerline{\epsfbox{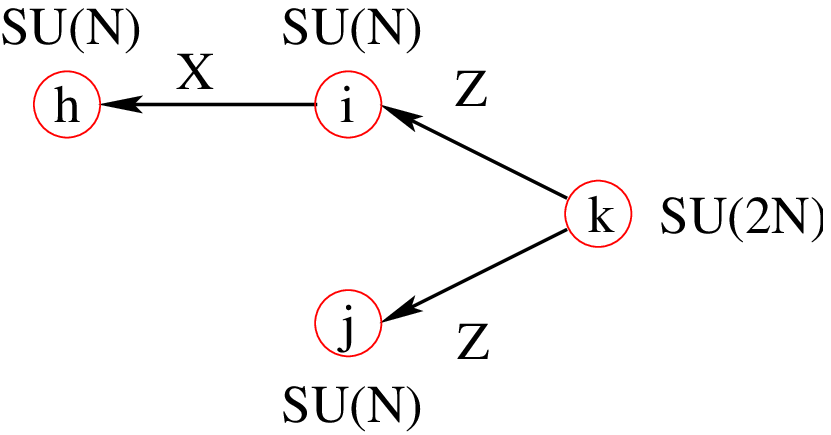}}
  \caption{Level two $Z^2 X$ dibaryons in the $(2,4,4)$ $dP_7$ 3-block model.}
  \label{path_dibaryon_dP7_2}
\end{figure}

Here $h \in \{7,8,9,10 \}$, $i,j \in \{3,4,5,6 \}$ and $k \in \{1,2 \}$.
There is only one node for each $SU(\alpha)$ that is contracted with epsilon tensors. For $SU(2)$ this corresponds to head of $Z$ fields,
the first $SU(4)$ corresponds to the tail of $Z$ and the second $SU(4)$, to the tail of $X$. Using that $X$ transforms as $(1,4,\bar{4})$
and $Z$ as $(2,\bar{4},1)$ we immediately conclude that the $Z^2X$ dibaryons form the $(2,\bar{4},4)$ representation.

Having illustrated the practical details of the classification into maximal subgroups with two examples, we can move on 
and write down the combinations of fields in the gauge theory that give rise to dibaryons of different levels explicitly.
Level one dibaryons have R-charge $N$ and fill the fundamental {\bf 56 } of $E_7$ \cite{Herzog:2003wt}. They can be organized 
as shown in the table below.

 \beq
\begin{array}{|c|c|c|}
\hline
\mbox{Dibaryon} & \mbox{Number of states} & SU(2)\times SU(4)\times SU(4) \\
\hline  
                                     &  &  \\
\mbox{X}    & 16 & (1,4,\bar{4})  \\
\mbox{YZ}   & 16 &  (1,\bar{4},4) \\
\mbox{Y}^2  & 12 & (2,1,6)  \\
\mbox{Z}^2  & 12 & (2,6,1)   \\
                                     & &    \\
\hline
\end{array}
\eeq

Our results correspond to the branching of the ${\bf 56}$ representation under the maximal subalgebra 

\beq 
56 \rightarrow (1,4,\bar{4}) + (1,\bar{4},4) + (2,1,6) + (2,6,1) .
\label{56_branching}
\eeq

At level two, the dibaryons have R-charge $2N$. The geometric counting gives 

\beq
\begin{array}{|c|c|}
\hline
\mbox{Divisor} & \mbox{Number of states} \\
\hline  
                                     &    \\
(1,\underline{-1,0,0,0,0,0,0})       & 7  \\
(2,\underline{-1,-1,-1,-1,0,0,0})    & 35 \\
(3,\underline{-2,-1,-1,-1,-1,-1,0})  & 42 \\
(4,\underline{-2,-2,-2,-1,-1,-1,-1}) & 35 \\
(5,\underline{-2,-2,-2,-2,-2,-2,-1}) & 7  \\ 
                                     &    \\
\hline
\mbox{{\bf Total}}                   & {\bf 133}-7 \\
\hline
\end{array}
\eeq
corresponding to  part of the adjoint representation of $E_7$, which has dimension 133. The remaining seven elements
correspond to the Cartan generators of $E_7$. The counting on the gauge theory is shown in the next table.

\beq
\begin{array}{|c|c|c|}
\hline
\mbox{Dibaryon} & \mbox{Number of states} & SU(2)\times SU(4)\times SU(4) \\
\hline  
                                     &  &  \\
\mbox{ZXY}    & 2 & (3,1,1) - 1 \\
\mbox{XYZ}    & 12 &  (1,15,1) - 3  \\
\mbox{YXZ}    & 12 &  (1,1,15) - 3 \\
\mbox{X}\mbox{Y}^2  & 32 &  (2,4,4)   \\
\mbox{Z}^2\mbox{X}   & 32 &  (2,\bar{4},\bar{4}) \\ 
\mbox{Y}^2\mbox{Z}^2 & 36 & (1,6,6) \\
                                   & &    \\
\hline
\mbox{{\bf Total}}       & 126   & {\bf 133}-7 \\
\hline
\end{array}
\eeq
where we have removed the Cartan generators from each of the adjoint representations. The result is in complete
agreement with the geometric counting. We see that dibaryons are classified under the $SU(2)\times SU(4)\times SU(4)$
maximal subgroup according to the branching of the ${\bf 133 }$ $E_7$ representation

\beq 
133\rightarrow (3,1,1) + (1,15,1) + (1,1,15) + (2,4,4) + (2,\bar{4},\bar{4}) + (1,6,6)
\label{133_branching}
\eeq


\subsection*{Del Pezzo 8}

\begin{figure}[ht]
  \epsfxsize = 6.5cm
  \centerline{\epsfbox{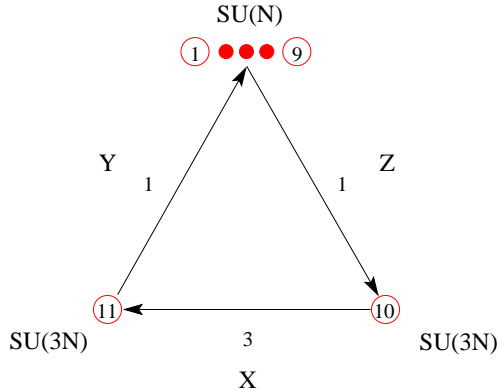}}
  \caption{One of the possible 3-block quivers for $dP_8$.}
  \label{quiver_dP8_1}
\end{figure}

The eigth del Pezzo can be studied using the same methodology. The 3-block Diophantine equation for $dP_8$ has four minimal 
solutions: $(\alpha,\beta,\gamma)=(1,1,9)$, $(1,2,8)$, $(2,3,6)$ and $(1,5,5)$. The $(1,2,8)$ case has been studied in \cite{Herzog:2003dj}, 
where the 240 level one dibaryons were written explicitly in the gauge theory. Let us apply maximal subgroup decomposition to classify
level one dibaryons in the $(1,1,9)$ model. The quiver for this theory is shown in \fref{quiver_dP8_1}.

The maximal subgroup of $E_8$ is in this case $SU(9)$. All bifundamental fields in this model have R-charge $2/3$ \footnote{As a curiosity note that this quiver is precisely that of the $C^3/\Delta(27)$ orbifold \cite{Hanany:1998sd,Feng:2000mw}.}.
Proceeding as before, we present in next table the $240$ level one dibaryons, along with their tranformation properties 
under $SU(9)$.

\beq
\begin{array}{|c|c|c|}
\hline
\mbox{Dibaryon} & \mbox{Number of states} & SU(9) \\
\hline  
                                     &  &  \\
\mbox{ZXY}    & 72 & 80 - 8 \\
\mbox{Y}^3    & 84 &  \bar{84}  \\
\mbox{Z}^3    & 84 &  84 \\
                                  & &    \\
\hline
\mbox{{\bf Total}}       & 240   & {\bf 248} - 8 \\
\hline
\end{array}
\eeq


\section{Conclusions}

In this paper we have identified the global symmetries of the gauge theories on D3-branes probing complex cones over del 
Pezzo surfaces, which have their origin in the automorphism of the underlying geometry. This has been possible due to the 
association of divisors in the del Pezzo surface to every bifundamental field in the quiver. The correspondence between bifundamental
fields and divisors follows from studies of a special class of dibaryon operators. Each of them is constructed by 
antisymmetrizing various copies of a single field in the quiver. For each $dP_n$, the bifundamental 
matter of the theories has been explicitly organized in irreducible representations of $E_n$. We presented
the results in Section \ref{section_symmetries}.
This classification has been obscure in the past due to the fact that, in general, irreducible representation of the global 
symmetry group are formed by bifundamental fields charged under different pairs of gauge groups.

We encountered some theories in which the matter content seems, at a first glance, insufficient to complete representations. 
We discussed how all the models can be naturally studied within the same framework. The fields that appear to be absent 
from partial representations sit in bidirectional arrows in the quiver (i.e. quadratic gauge invariants) and that also 
form quadratic invariants under the global symmetry group. Thus, following the same rules that apply to all other cases, 
mass terms for these fields are present in the superpotential, and they are integrated out when considering the low energy 
physics. The geometric origin of partial representations was discussed in Section \ref{section_geometry_partial}, where we 
explained how to determine the location in the quiver of the massive fields. 

The $E_n$ classification of the models becomes particularly helpful in writing down superpotentials, both for toric and
non-toric del Pezzos. The basic elements of their construction are the gauge invariant projections of the singlets under 
the action of the global symmetry group. We have seen how superpotentials become completly determined by this principle
(and, in a few cases, information about the higgsing from a higher $dP_n$).

The blow-down of a 2-cycle takes us from $dP_n$ to $dP_{n-1}$. This geometric action translates on the gauge theory side to a 
non-zero VEV for a bifundamental field that higgses the quiver. We have shown in Section \ref{section_higgsing} how to use 
the group theory classification of the quiver to identify the bifundamental fields that do the correct job. By turning on a 
VEV for any field in the fundamental representation of $E_n$, it is higgsed down to one of $E_{n-1}$ 
and a $dP_{n-1}$ quiver is produced. In this way, we have presented a clear systematic prescription that identifies
all possible ways in which a $dP_n$ quiver can be higgsed to obtain another quiver that corresponds to $dP_{n-1}$.

It would be interesting to extend the discussion of this paper to gauge theories on D3-branes probing different 
singularities, in which other groups of automorphisms will point towards global symmetries of the corresponding field theories. 
In the case that these symmetry groups include or are included in the ones for del Pezzo theories, the group theory 
concepts of Section \ref{subsection_higgsing_groups} would indicate how to derive those theories by (un)higgsing.

In Section \ref{section_symmetries_seiberg}, we presented a simple set of rules that determine how the representation structure 
of a gauge theory transform under Seiberg duality. This was used to rederive the classification based on divisors obtained in 
Section 3.

We showed in Section \ref{section_dibaryon_counting} how the gauge theory counting of dibaryon operators is organized and 
simplified by using the $E_n$ structure. We further verified that the algebraic counting in the field 
theory match the geometric counting of curves of different degrees, in agreement with the AdS/CFT correspondence.  

Finally, in Section \ref{section_dP7_and_dP8}, we used maximal subgroups of $E_n$ to classify and count dibaryons in 
non-toric 3-block quivers for $dP_7$ and $dP_8$.

\section*{Acknowledgements}

The authors would like to thank Y. H. He, C. Herzog and J. Walcher for various insightful discussions and comments on early versions of the manuscript. S. F. also thanks M. Schvellinger for helplful 
conversations. A. H. appreciates the hospitality of the Institute for Advanced Study during the preparation 
of this work. This Research was supported in part by the CTP and LNS of MIT and the U.S. Department of Energy 
under cooperative research agreement \# DE-FC02-94ER40818. A. H. is also indebted to the Reed Fund Award and 
a DOE OJI ward.


 \bibliographystyle{JHEP}

\end{document}